\documentclass[graphics, floatfix, footinbib, tightenlines, nobibnotes, aps, prb, twocolumn]{revtex4-1}
\usepackage{amsmath}
\usepackage{amssymb}
\usepackage{graphicx}
\usepackage[utf8]{inputenc}
\usepackage{comment}
\usepackage{xcolor}
\usepackage{braket}
\usepackage{bbold}
\usepackage[colorlinks=true ,urlcolor=blue,urlbordercolor={0 1 1}]{hyperref}
\usepackage{ragged2e}
\usepackage{caption}
\usepackage{ulem}

\renewcommand{\v}[1]{\mathbf{#1}} 

\newcommand{\be}{\begin{equation}}
\newcommand{\ba}{\begin{align}}
\newcommand{\ee}{\end{equation}}
\newcommand{\bea}{\begin{eqnarray}}
\newcommand{\eea}{\end{eqnarray}}
\newcommand{\beq}{\begin{equation}}
\newcommand{\eeq}{\end{equation}}
\newcommand{\beqn}{\begin{eqnarray}}
\newcommand{\eeqn}{\end{eqnarray}}

\usepackage{bm}
\usepackage{subcaption}

\begin{document}

\title{Excitonic Chern insulator and kinetic ferromagnetism in MoTe$_2$/WSe$_2$ moir\'e bilayer}
\author {Zhihuan Dong$^1$}
\author{ Ya-Hui Zhang$^2$}
\affiliation{$^1$Department of Physics, Massachusetts Institute of Technology, Massachusetts 02139, USA}
\affiliation{$^2$Department of Physics and Astronomy, Johns Hopkins University, Baltimore, Maryland 21218, USA
}

\date{\today}
\begin{abstract}
We propose a new mechanism for the quantum anomalous Hall (QAH) effect in the AB stacked MoTe$_2$/WSe$_2$ system.  Based on the observation that the inter-layer tunneling is suppressed in the AB stacking, we consider a model with two layers coupled through the Coulomb interaction. The moir\'e lattices of the two layers are shifted to form a honeycomb lattice. Initially, the system is in a layer-polarized Mott insulator at $\nu_T=1$. But with a displacement field, an equal number of holes and electrons are doped into the two layers respectively, forming inter-layer exciton condensation. Through mean field theory, we find $p\pm ip$ exciton condensation in a certain parameter regime, which leads to a Chern insulator with Chern number $C=\pm 1$. The valleys are polarized due to the kinetic energy instead of interaction. But the polarization in the two layers can be either the same or opposite.  Especially, an inter-valley-coherent (IVC) Chern insulator phase is possible, in agreement with the recent  magnetic circular dichroism (MCD) measurement. Our work opens a new direction to search for a topologically non-trivial excitonic insulator with high angular momentum  exciton pairing symmetry.
\end{abstract}

\maketitle

\textbf{Introduction} Chern insulator is one of the famous topologically non-trivial states, which supports the quantum anomalous Hall(QAH) effect with a quantized Hall conductivity at zero magnetic field\cite{liu2016quantum}. Although the  Chern insulator phase and QAH  effect was theoretically proposed to be realized in the Haldane model in 1988\cite{haldane1988model}, its experimental realization appeared only after 2013. The first realization is on the  surface of the three-dimensional topological insulator\cite{chang2013experimental}. The observations of the QAH effect\cite{sharpe2019emergent,serlin2020intrinsic,chen2019tunable,chen2021electrically,polshyn2020electrical,polshyn2020nonvolatile,polshyn2022topological} in truly two-dimensional systems were achieved only a few years ago in the moir\'e systems\cite{cao2018correlated,Wang2019Signatures,Wang2019Evidence,yankowitz2019tuning,Wang2019Signatures,chen2019tunable,lu2019superconductors,Cao2019Electric,Liu2019Spin,Shen2019observation,polshyn2020nonvolatile,chen2020electrically,sharpe2019emergent,serlin2020intrinsic,tang2020simulation,regan2019optical,wang2019magic,gu2021dipolar,zhang2021correlated,li2021continuous,ghiotto2021quantum,li2021quantum}.

QAH was first predicted theoretically for the graphene moir\'e systems\cite{zhang2019nearly}. The moir\'e superlattice folds the original band into narrow moir\'e bands in a mini Brillouin Zone (MBZ). Especially the two valleys K and K$^\prime$  form their own moir\'e bands, which can have non-zero but opposite Chern numbers\cite{zhang2019nearly}. Then the interaction can polarize the valley at filling $\nu_T=1$ due to the Stoner ferromagnetism mechanism. After that, the Chern band of one valley is fully occupied, leading to a Chern insulator phase. After the experimental observations in several graphene systems\cite{sharpe2019emergent,serlin2020intrinsic,chen2019tunable,chen2021electrically,polshyn2020electrical,polshyn2020nonvolatile,polshyn2022topological}, many other theoretical studies\cite{zhang2019twisted,bultinck2020mechanism,repellin2020ferro,po2018origin,wu2020collective,wu2020quantum,liu2021theories,alavirad2020ferromagnetism,shi2021moire,liu2020anomalous,mao2021quasiperiodicity} confirm this simple picture as the mechanism for the QAH phase in graphene-based moir\'e systems. In this scenario, interaction is needed for valley ordering, but the band topology already exists at the single particle level. One natural question is then whether  a Chern insulator phase is possible without single-particle band topology.  We will propose a new mechanism towards QAH with band topology generated by exciton condensation, motivated by the recent experimental measurements of a QAH phase in a moir\'e system based on the AB stacked transition metal dichalcogenide (TMD) MoTe$_2$/WSe$_2$ bilayer\cite{li2021quantum,tao2022valley}.

 In the AB stacked MoTe$_2$/WSe$_2$ hetero-bilayer\cite{li2021quantum,tao2022valley}, a Mott insulator evolves to a Chern insulator phase at density $\nu_T=1$ per moir\'e unit cell  when tuning the displacement field. There is evidence that the Chern insulator phase is in the regime where particles are transferred from the MoTe$_2$ layer to the WSe$_2$ layer, so the densities at the two layers are $1-x$ and $x$ respectively. Previous theories\cite{xie2022valley,zhang2021spin,rademaker2022spin,pan2021topological,devakul2022quantum,chang2022theory,su2022massive} all predict a phase with aligned valley polarizations in the two layers. Because these theories rely on a single particle tunneling term to couple the two layers to form a band, the valleys naturally align because the hopping term must be between the same valley due to momentum conservation. However, the recent experimental measurement of magnetic circular dichroism (MCD) indicates that the valleys in the two layers within the QAH phase are anti-aligned\cite{tao2022valley}. It is clear that all of the previous theories are inconsistent with this measurement and a new theoretical explanation is needed for this system.

\begin{figure}[ht]
\centering
\includegraphics[width=0.5\textwidth]{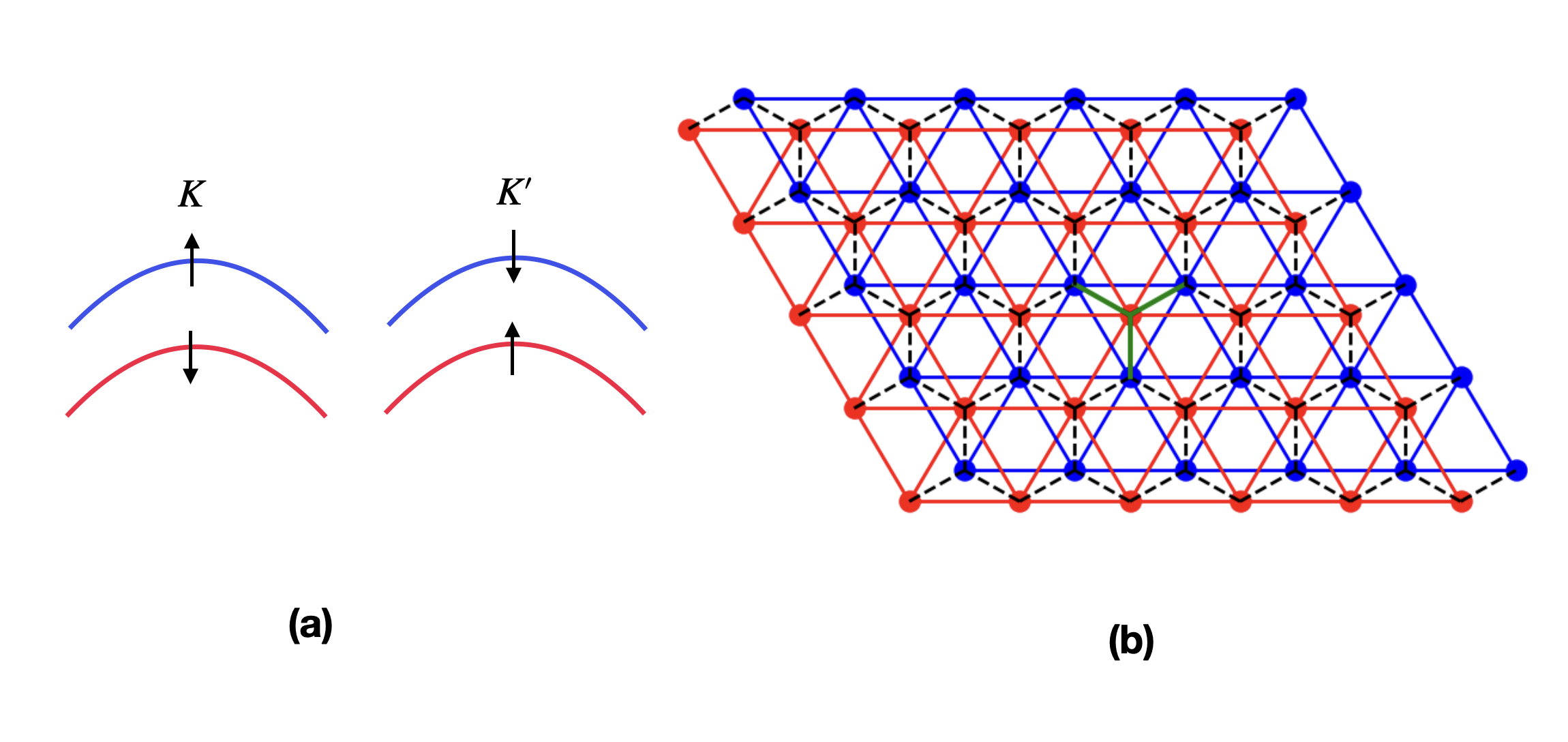}
\caption{(a) Illustration of the spin valley locking in MoTe$_2$/WSe$_2$ hetero-bilayer. Red and blue lines are the top valence band for the WSe$_2$ and MoTe$_2$ layer respectively. In AB stacking, one layer is rotated by $180^\circ$ compared to AA stacking, which flips the spin-valley locking in this layer. (b) Lattice model on the honeycomb lattice. Red and blue labels A and B sublattice sites, corresponding to WSe$_2$ and MoTe$_2$ layer. The solid line is the nearest neighbor bond within each layer. The dashed line is the AB bond connecting the sites of the two layers. In our model, the microscopic hopping $t_{AB}$ is assumed to be zero, but there can be effective inter-layer hopping spontaneously generated from exciton condensation. The exciton condensation order parameter is $\chi_{ij;\alpha \beta}=\langle c^\dagger_{i;\alpha}c_{j;\beta}\rangle$, where $\langle ij \rangle$ is illustrated as the green bond in (b). }
\label{fig:moire_bilayer}
\end{figure}

In this paper, we propose a completely new route towards QAH in AB-stacked TMD hetero-bilayer. Based on the observation that the inter-layer tunneling is suppressed in the AB stacking because the spin $S_z$ in the two layers for the same valley are opposite (see Fig.~\ref{fig:moire_bilayer}(a)),  we take the view that the most important coupling between the MoTe$_2$ and WSe$_2$ layer is the Coulomb interaction.  The physics is then similar to the recently proposed\cite{zhang20214,zhang2022doping} and experimentally fabricated\cite{zhang2021correlated,gu2021dipolar} Coulomb coupled moire bilayer. The major difference now is that the lattices of the two layers are shifted to form a honeycomb lattice together instead of being aligned. Then the electron and hole pockets in the two layers can form exciton condensation due to the inter-layer Coulomb interaction. In certain parameter regions, we find that the pairing symmetry of the exciton condensation is $p\pm ip$, similar to the $p \pm ip$ superconductor\cite{read2000paired} after a particle-hole transformation only for one layer. This leads to a topological phase we dub excitonic Chern insulator (ECI). In this case, we can have either intra-valley exciton condensation or inter-valley exciton condensation, which are nearly degenerate. With small valley contrasting flux of appropriate sign, inter-valley exciton condensation is selected and we reach an ECI phase with inter-valley-coherence (IVC), consistent with the MCD measurement\cite{tao2022valley}. Our theory points out the possibility of a topological excitonic insulator phase with higher angular momentum exciton pairing symmetry, which was not well explored before.

\textbf{Model} For single-layer TMD, due to the Ising spin-orbit coupling, the spin and valley degrees of freedom are locked. For AB stacked MoTe$_2$/WSe$_2$ system, the spin-valley locking is opposite in the two layers, as shown in Fig.~\ref{fig:moire_bilayer}(a).  For the WSe$_2$ or A layer, we have $(\tau^z_A,s_z)\in \{(K_A,\downarrow),(K'_A,\uparrow)\}$. For the MoTe$_2$ or B layer, we have  $(\tau^z_B,s_z)\in \{(K_B,\uparrow),(K'_B,\downarrow)\}$. Here $K_a, K'_a$ labels the two valleys for the layer $a=A, B$. $\tau_i, s_i$ with $i=x,y,z$ label Pauli matrices in the valley and spin space respectively. We also use $\sigma_i$ with $i=x,y,z$ to label Pauli matrices in the layer space.  In the following, the spin index will be ignored and the two valleys $K, K'$ will be labeled as $+,-$. WSe$_2$ and MoTe$_2$ layers will be labeled as $A, B$.

We  use a honeycomb lattice model for the MoTe$_2$/WSe$_2$ bilayer similar to  Ref.~\onlinecite{zhang2021spin}, where Wannier orbitals from WSe$_2$ and MoTe$_2$ live on A and B sublattice sites (see Fig.~\ref{fig:moire_bilayer}(b)). The essential physics is captured by a simple lattice model:

\begin{align}
    H& =-t_{AB} \sum_{\langle ij \rangle}(c^\dagger_{i;\alpha}c_{j;\alpha}+h.c.)+D\sum_i  (-1)^{\sigma_z(i)}n_i \notag \\
    &-t_A \sum_{\langle \langle ij \rangle \rangle_A}(e^{i \phi^A_{ij} \tau_z}c^\dagger_{i;\alpha}c_{j;\alpha}+h.c.) \notag \\
   &-t_B \sum_{\langle \langle ij \rangle \rangle_B}(e^{i \phi^B_{ij} \tau_z}c^\dagger_{i;\alpha}c_{j;\alpha}+h.c.) \notag \\ 
    &+ \frac{U}{2} \sum_{i}n_i(n_i-1)+ V\sum_{\langle ij\rangle}n_in_j + V'\sum_{\langle\langle ij\rangle\rangle}n_in_j
\label{Ham}
\end{align}
where $\alpha=+,-$ labels the valley index and we assume Einstein summation convention for $\alpha$. $n_i=\sum_{\alpha=+,-}c^\dagger_{i;\alpha}c_{i;\alpha}$ is the density at site $i$.  $\sigma_z(i)=1,-1$ for A and B layer respectively.   The two unit vectors of the lattice are $\mathbf a_1=(1,0)$ and $\mathbf a_2=(-\frac{1}{2},\frac{\sqrt{3}}{2})$. $\phi_{i;i+\mathbf{a_1}}^a=-\phi_{i;i-\mathbf a_1}=\phi_a$ is the valley contrasting hopping term for layer $a=A,B$.  $\phi_{ij}^a$ of the other bonds are generated by $C_3$ symmetry.  $\Phi_a=3\phi_a$ is the valley contrasting flux through each triangle of the layer $a$. In the above $\langle ij \rangle$ labels the nearest neighbor AB bond. $\langle \langle ij \rangle \rangle_a$ labels the intra-layer nearest neighbor bond of the triangular lattice in the layer $a=A, B$.   We expect $\phi_A \approx \frac{2\pi}{3}$ and $\phi_B \approx 0$ so that the band bottom of the two layers are at $\v K_M$ ($\v K'_M$) and $\v \Gamma_M$ in the mini Brillouin zone (MBZ). Time reversal symmetry flips the valley $\tau_z$ for both layers.  Despite the superficial similarity to the model in Ref.~\onlinecite{zhang2021spin} and Ref.~\onlinecite{devakul2022quantum}, we will consider the limit $\frac{V}{t_{AB}}\rightarrow \infty$, while the inter-layer interaction $V$ term was ignored in Ref.~\onlinecite{zhang2021spin} and Ref.~\onlinecite{devakul2022quantum}.

In most of the paper, we assume $t_{AB}=0$ for simplicity and will discuss its effect later. The global symmetry is then $U(1)^4$ at $t_{AB}=0$, corresponding to the conservation of the four flavors combining layer and valley indices. If $\Phi_a=0$, then the symmetry is  enlarged to U(1) $\times$ U(2), with an SU(2) rotation symmetry in the valley space only for the layer $A$. When $\Phi_A=\Phi_B=0$, the symmetry is U(2) $\times$ U(2). When $t_{AB}=0$, we are free to do a gauge transformation for only one layer to shift $\phi_a$ by $\pm \frac{2\pi}{3}$. Thus we can focus on the regime with $\phi_A \approx \phi_B \approx 0$. On the other hand, a finite $t_{AB}$ reduces the symmetry to U(1) $\times$ U(1) with only total charge and total valley conservation for generic values of $\Phi_A,\Phi_B$. In this case, $\phi_A, \phi_B$ can not be gauged away.

Let us focus on the filling with $\nu_T=n_A+n_B=1$. Although in the experiment the valence band is hole-doped, here we use an electron picture after a particle-hole transformation.  We start from large D where all of the electrons are in the B layer and there is a Mott insulator due to large $U/t_B$.  When D is decreased, the B layer is hole-doped and the A layer has an equal number of electrons.  A convenient way to deal with the large $U$ term is to project to the Hilbert space without double occupancy and consider a bilayer t-J model:

\begin{align}
 H&=-t_A \sum_{\langle \langle ij \rangle \rangle_A}(e^{i \phi^A_{ij} \tau_z}c^\dagger_{i;\alpha}c_{j;\alpha}+h.c.)+D\sum_{i \in A}n_i \notag \\
 &~~~-t_B \sum_{\langle \langle ij \rangle \rangle_B}(e^{i \phi^B_{ij} \tau_z}Pc^\dagger_{i;\alpha}c_{j;\alpha}P+h.c.)-D \sum_{i \in B} n_i \notag \\
 &~~~ + J_B\sum_{\langle\langle ij\rangle\rangle_B} \tau^\mu_iR^{\mu\nu}_z(2\phi^B_{ij})\tau^\nu_j +(V'-\frac{J_B}{4}) \sum_{\langle \langle ij \rangle \rangle_B}n_i n_j\notag \\
   &~~~+V\sum_{\langle ij\rangle}n_in_j + V'\sum_{\langle\langle ij\rangle\rangle_A}n_in_j
\label{HamtJ}
\end{align}
where $P$ is the projection operator to forbid double occupancy.   $R^{\mu\nu}_z(\phi)$ is the matrix representing rotation around $z$ axis (see the supplementary \footnote{See Supplemental Material for details on Schwinger boson mean field theory}). For $U/t_B=50$\cite{devakul2022quantum}, $J_B=\frac{4t_B^2}{U}=0.08t_B$. The effect of  $J_B$ is negligible. For simplicity, we do not impose a double-occupancy constraint for the A layer which is appropriate in the small $n_A$ limit.

\textbf{Phase diagram}
We can obtain a phase diagram by Hartree-Fock calculation directly for the Hubbard model in Eq.~\ref{Ham}. We find valley polarization at an intermediate value of $D$ and an excitonic Chern insulator (ECI) in a certain parameter regime.  To make the mechanism of the magnetic order transparent, it is more convenient to work on the bilayer t-J model in Eq.~\ref{HamtJ} and use the Schwinger boson parton construction for the B layer: $c_{i;\alpha}=b_{i;\alpha} f_i$  with the constraint $\sum_{\alpha}b^\dagger_{i;\alpha}b_{i;\alpha}=f^\dagger_i f_i$ at each site $i\in B$. $b_{i;\alpha}$ is Schwinger boson carrying the valley degree of freedom while $f_i$ is the spinless holon operator.  The magnetic order at layer B is decided by the condensation of the Schwinger boson:$ \begin{pmatrix} \langle b_{i;+} \rangle \\ \langle b_{i;-} \rangle\end{pmatrix}=F_i$. For  valley ferromagnetic (FM) state with $+$ valley, we use  $ F_i  =\sqrt{n_a} \begin{pmatrix} 1 \\ 0 \end{pmatrix}$.  On the other hand, if there is a $120^\circ$ anti-ferromagnetic (AF) order, we use $F_i=-\frac{\sqrt{n_a}}{\sqrt{2}} e^{\frac{i}{2}(\v Q \cdot \v r_i) \tau_z} \begin{pmatrix} 1 \\ 1 \end{pmatrix}$, where $\v Q = \mathbf K_M \  \mathrm{or} -\mathbf K_M$.  For the A layer, we assume full valley polarization as found in the Hartree Fock calculation. Then there is only spinless charge degree of freedom which we also label as $f_i$ for $i\in A$.

 After the magnetic orders are decided for both layers, the charge degree of freedom is conveniently captured by the following spinless model:

\begin{align}
    H_f&=H^f_0+H^f_V \notag \\
    H^f_0&=-\sum_{a=A,B} \sum_{\langle \langle ij \rangle \rangle_a} t^f_{ij;a} f_i^\dagger f_j +D\sum_{i \in A}n_i-D\sum_{i \in B}n_i \notag \\
    H^f_V&=V \sum_{\langle ij \rangle} n_i n_j+\sum_{a=A,B}\sum_{\langle \langle ij \rangle \rangle_a}V' n_i n_j
\end{align}
where $t^f_{ij;B}=t_B (F_i^\dagger e^{i \phi_{ij;B} \tau_z} F_j)$ and $t^f_{ij;A}=t_Ae^{i \phi_{ij;A}\tau_z}$. $n_i=f_i^\dagger f_i$ is constrained to be the physical density. The total filling is $n_A+n_B=1$ so there are electron and hole pockets in the two layers respectively. Due to the $V$ term, we can decouple the following mean field:
\begin{equation}
    H^f_M=-V\sum_{\langle ij \rangle} \chi_{ji} f^\dagger_i f_j
\end{equation}
where $\chi_{ij}=\langle f^\dagger_i f_j \rangle$ is the exciton order parameter. The above term is basically a spontaneously generated hopping term on the nearest neighbor AB bond of the honeycomb lattice.

The obtained phase diagram is shown in Fig.~\ref{pd}(a), similar to the Hartree Fock results (see the supplementary). We find the following phases: (I) MI: layer polarized Mott insulator with $120^\circ $ AFM in layer B; (II) FL: Fermi liquid with electron and hole pockets in A and B layer; (III) AF EI: excitonic insulator with AFM order in the B layer and  $\tau_z$ FM order in the A layer. The Chern number is $C=0$; (IV) NEI: nematic excitonic insulator with $\tau_z$ FM in both layers. The Chern number is $C=0$; (V) ECI: excitonic Chern insulator with $|C|=1$ and $\tau_z$ FM order in both layers. The transition between the trivial AF-EI and  the topological ECI phase is first order with a small jump of the charge gap $\Delta_c$.

\begin{figure}[!ht]
  \centering
  \subfloat[][]{\includegraphics[width=.24\textwidth]{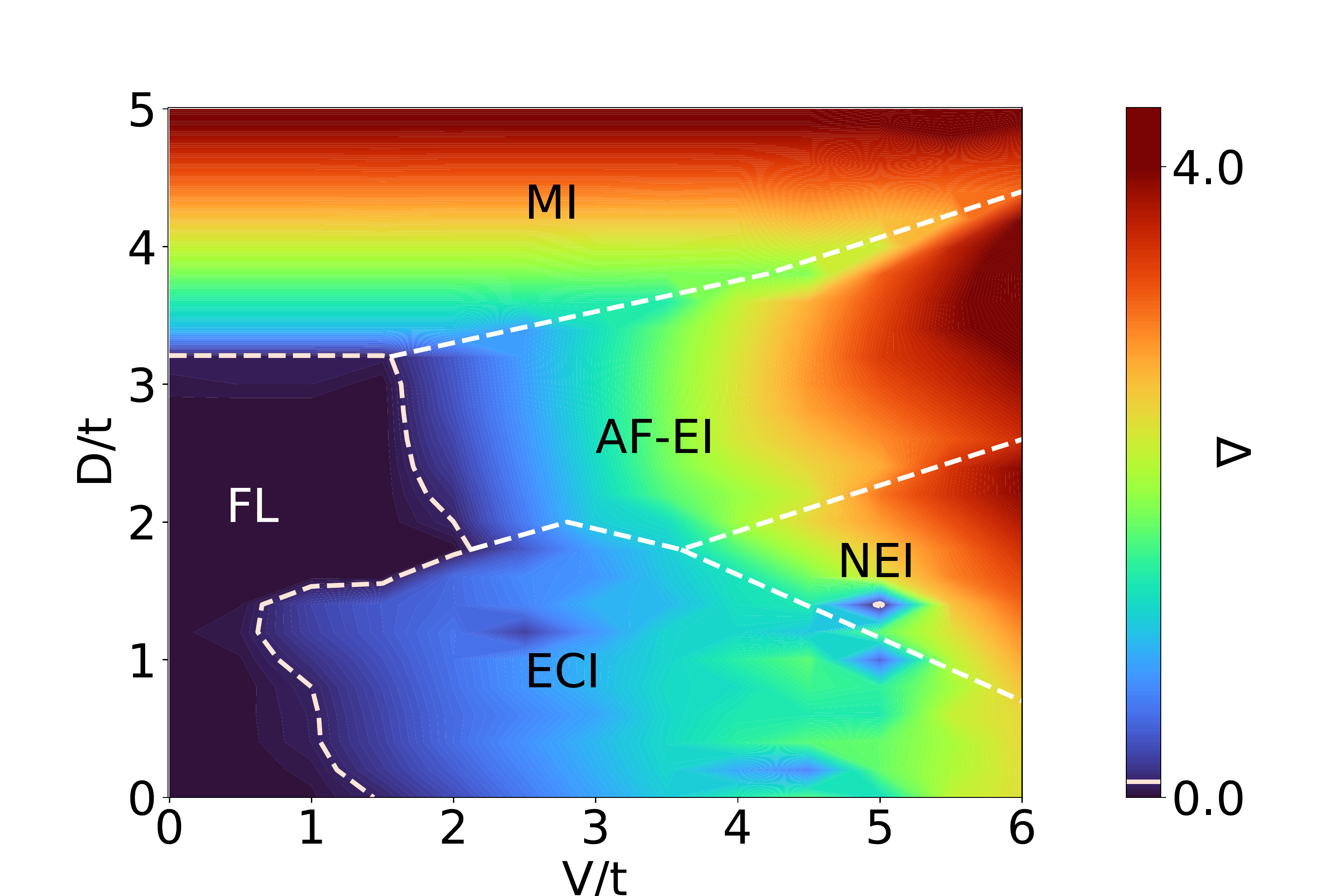}}
  \subfloat[][]{\includegraphics[width=.24\textwidth]{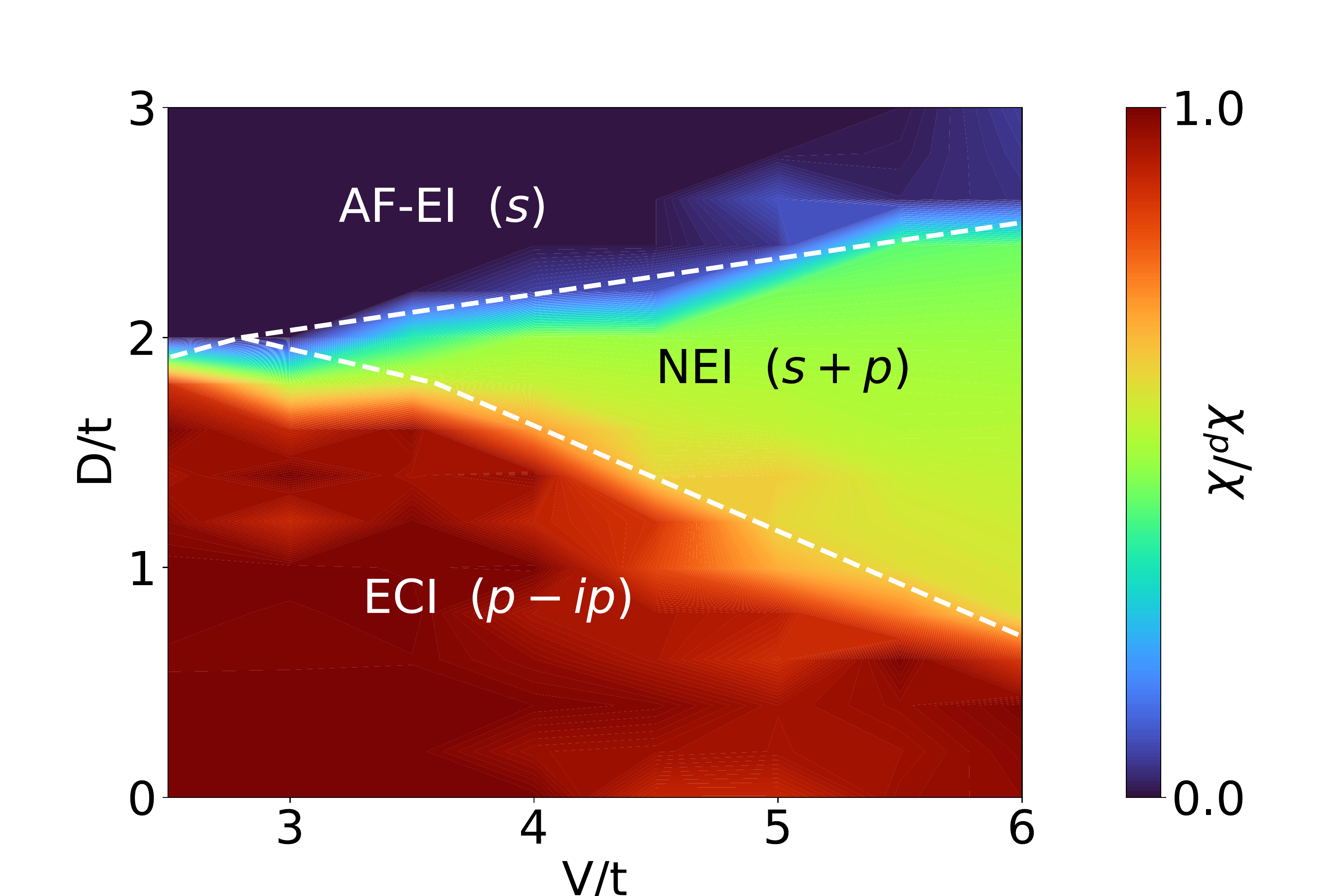}}\\
  \subfloat[][]{\includegraphics[width=.24\textwidth]{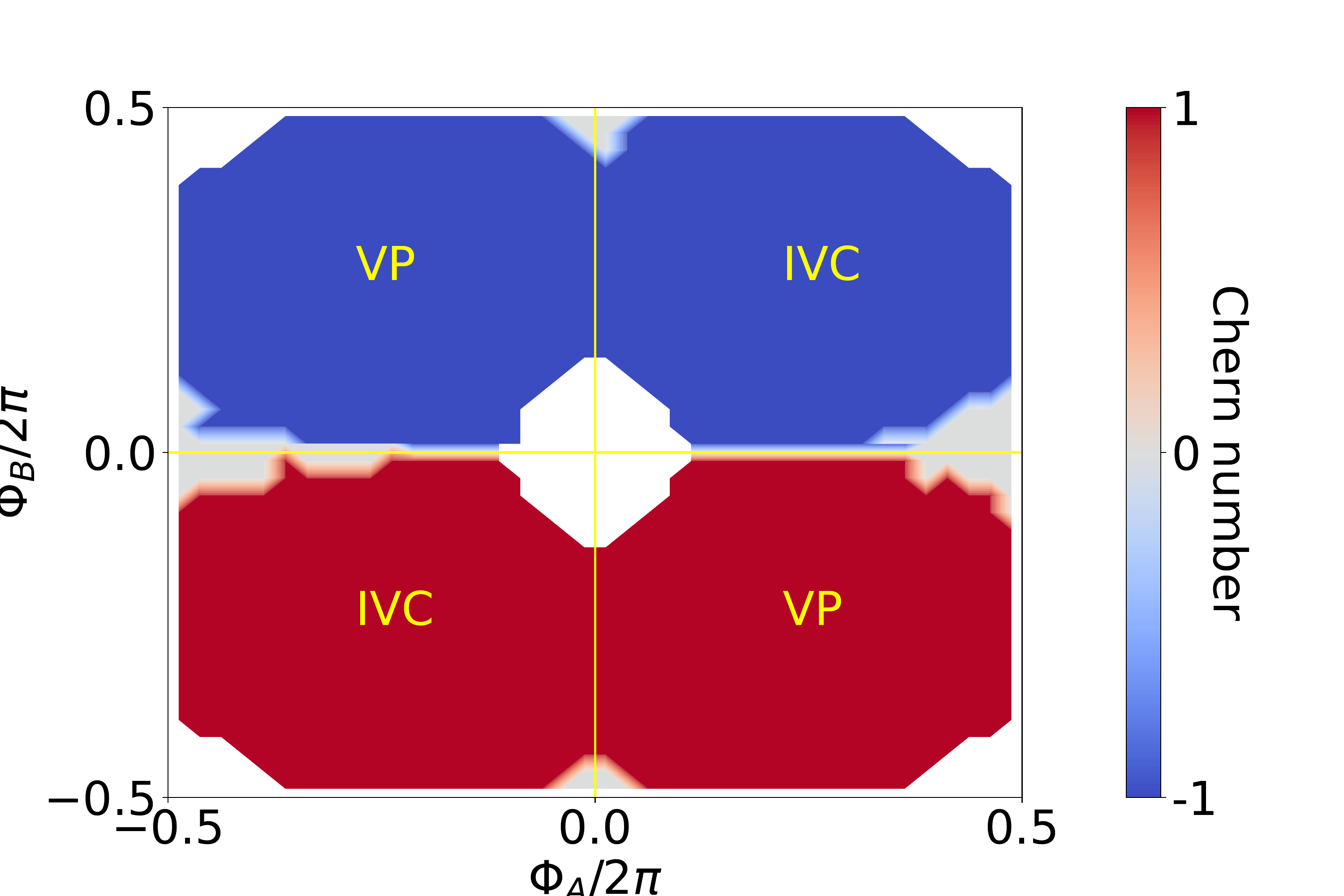}}
  \subfloat[][]{\includegraphics[width=.24\textwidth]{{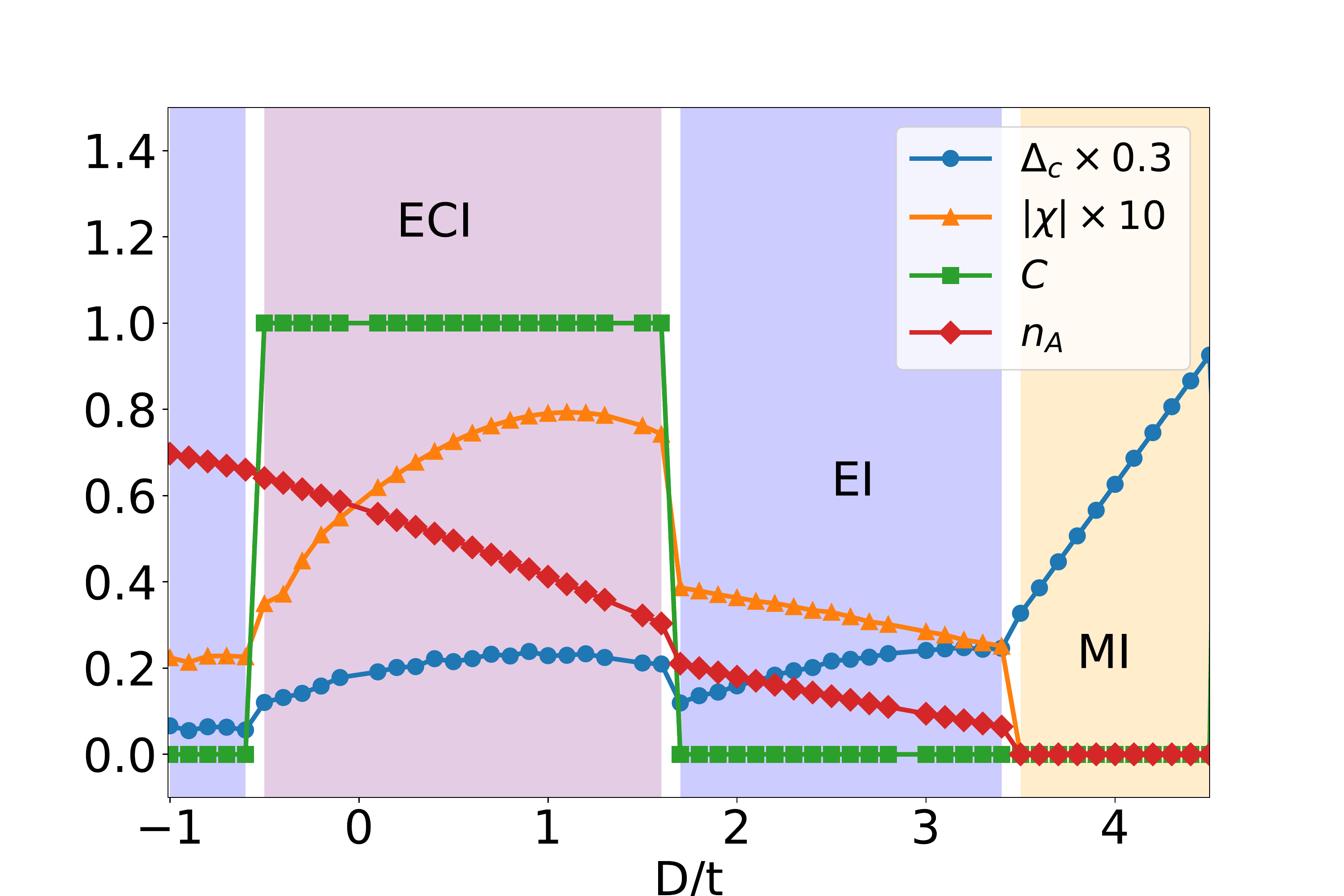}}}
   
  \caption{ (a) Schwinger boson mean field phase diagram in $D-V$ plane. The color bar shows the single electron charge gap $\Delta_c$. Various phases are marked: Mott Insulator(MI), Fermi liquid(FL), trivial exciton insulator with nematicity (NEI), trivial exciton insulator with $120^\circ$ order in B layer(AF-EI), excitonic Chern insulator with $\tau_z$ ferromagnetic order(ECI). In all of the three excitonic insulator phases, the A layer is valley polarized. (b) Evolution of exciton angular momentum. Colors show the weight of $p-ip$ (or $p+ip$) component. The ECI is dominated by $p\pm ip$ symmetry. NEI has a mixture of $s$ and $p-ip$ waves so it breaks $C_3$ symmetry.  (c) $\Phi_A-\Phi_B$ phase diagram for the Chern number. We used $D=0.5,V=4$. Metallic regions are whited out. The model has time-reversal symmetry. We have restricted to the state with $\langle \tau^z_A\rangle>0$. The corresponding valley-polarized (VP) and  inter-valley-coherent (IVC) regions are marked with text. (d) Displacement field driven evolution along a line cut with $V=2.5$. The transition from trivial to topological exciton insulating state is first order. Exciton order $\chi$ and charge gap $\Delta_c$ jumps on the phase boundary. The ECI phase onsets when $n_A$ is above a critical density around $0.25$.  In the calculation, $\tau^z_A$ is fixed to be polarized to the $+$ valley. We use the following parameters: $U=50$, $t_A=t_B=1$, $\phi_A = \pi/4$, $\phi_B = \pi/12$, $V=2V'=4$. The results in (a) and (b) are calculated using Schwinger boson mean field theory, while (c) and (d) are obtained from Hartree-Fock mean field theory.}
  \label{pd}
\end{figure}

\textbf{Exciton order parameter and Chern number} 
 We allow various translational symmetry breaking patterns in the $\sqrt{3}\times \sqrt{3}$ unit cell and the exciton order parameter is in the form:
\be
    \chi_{i j} = \sum_{Q,L \in Z} \chi_{Q,L} e^{iQ \v K_M \cdot \v r_i} e^{i L\theta(\v r_j-\v r_i)}
\ee
where $i\in A$ and $j \in B$. $\chi_{Q,L}$ is the amplitude.   $\theta(\v r_j -\v r_i)$ is the angle of $\vec r_j -\vec r_i$ relative to the $\hat x$ direction.  Here  $Q$ and $L$  are associated with the center of mass momentum and angular momentum of the exciton. 

Both $Q$ and $L$ are not gauge-invariant. One can simply shift the momentum of each layer by a gauge transformation $f_{i;a}\rightarrow f_{i;a}e^{\pm i \mathbf K_M \cdot \v r_i}$ with $a=A,B$ to shift $Q$ and $L$. We define $Q$ and $L$ by going to the gauge which shifts the electron and hole pockets of the two layers to the $\mathbf \Gamma_M$ point of the MBZ\footnote{This is a different gauge from the one we performed the Hartree Fock calculation, which shifts $\Phi_A,\Phi_B$ to be close to zero.}. In this gauge, the electron-hole problem can be mapped to the familiar Bardeen–Cooper–Schrieffer(BCS) theory of superconductor. Then we find there is only a $\mathbf Q=0$ component, meaning that the exciton order can efficiently hybridize the electron and hole pockets at $\mathbf \Gamma_M$ point.  The AF-EI has $L=0$ or $s$ wave exciton order. Its resulting band is trivial. The ECI phase is dominated by the $L=1$ or $L=-1$ component and the exciton order is of $p\pm ip$ symmetry. This leads to a Chern band with $C=\pm 1$ and a QAH effect.  In the NEI phase, the mixing between $s$ ($L=0$) and $p$ ($L=\pm 1$) wave for the exciton order breaks the C$_3$ symmetry. The Chern number is zero.  The exciton angular momentum is plotted in Fig.~\ref{pd}(b).

Let us also provide some intuitive explanations for why $p \pm ip$ exciton condensation is favored over the $s$ wave. In the supplementary, we show that the competition between $s$ wave and $p$ wave exciton instability is decided by the magnitude of the form factors $F_l(\mathbf k)=\sum_{j=0,1,2}e^{i  \frac{2\pi}{3} l j}e^{i \mathbf k \cdot \mathbf {\delta r_j}} $ near Fermi surfaces of the electron and hole pockets. Here $l=0,1,-1$ is the angular momentum of the exciton. $j=0,1,2$ is summed over three nearest neighbor bonds and $\mathbf{\delta r_{j}}$ is the displacement vector of the nearest neighbor bond. When $n_A>0.25$, we find that $|F_{l=\pm 1}(\mathbf k)|>|F_{l=0}(\mathbf k)|$ for $\mathbf k$ around the Fermi surfaces of the electron and hole pockets. This agrees with our phase diagram in Fig.~\ref{pd}(a) where the ECI phase onsets after a critical exciton density. We also find that a small valley flux $\Phi_A$ or $\Phi_B$ is needed to stabilize the ECI phase. For the $\Phi_A=\Phi_B=0$ point, the ECI phase shrinks to a quite small regime in $(D, V)$ parameter space (see the supplementary).

\textbf{Kinetic magnetism and symmetry selection}
The ECI phase in the above is fully valley polarized in both layers. Unlike the quantum Hall ferromagnetism, there is no Hund's coupling or band topology in our model. Instead, the ferromagnetism is kinetically driven: with the valley polarization, the hopping of the holon $f_i$ is larger, favored by the kinetic energy. This is similar to the celebrated Nagaoka ferromagnetism\cite{nagaoka1966ferromagnetism} in a hole-doped Mott insulator. 

\begin{figure}[ht]
  \centering
    \subfloat[][]{\includegraphics[width=.24\textwidth]{{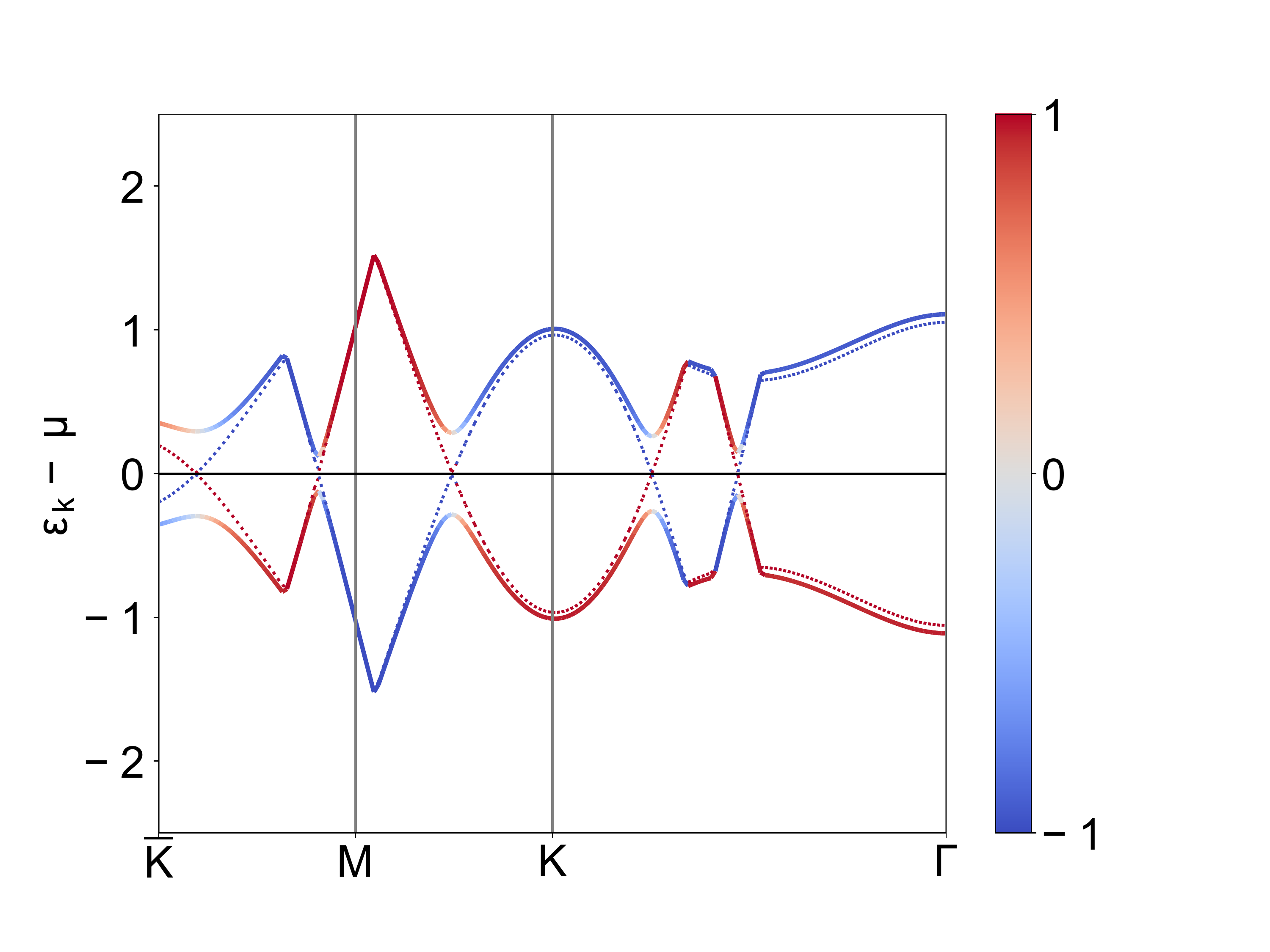}}}
    \subfloat[][]{\includegraphics[width=.24\textwidth]{{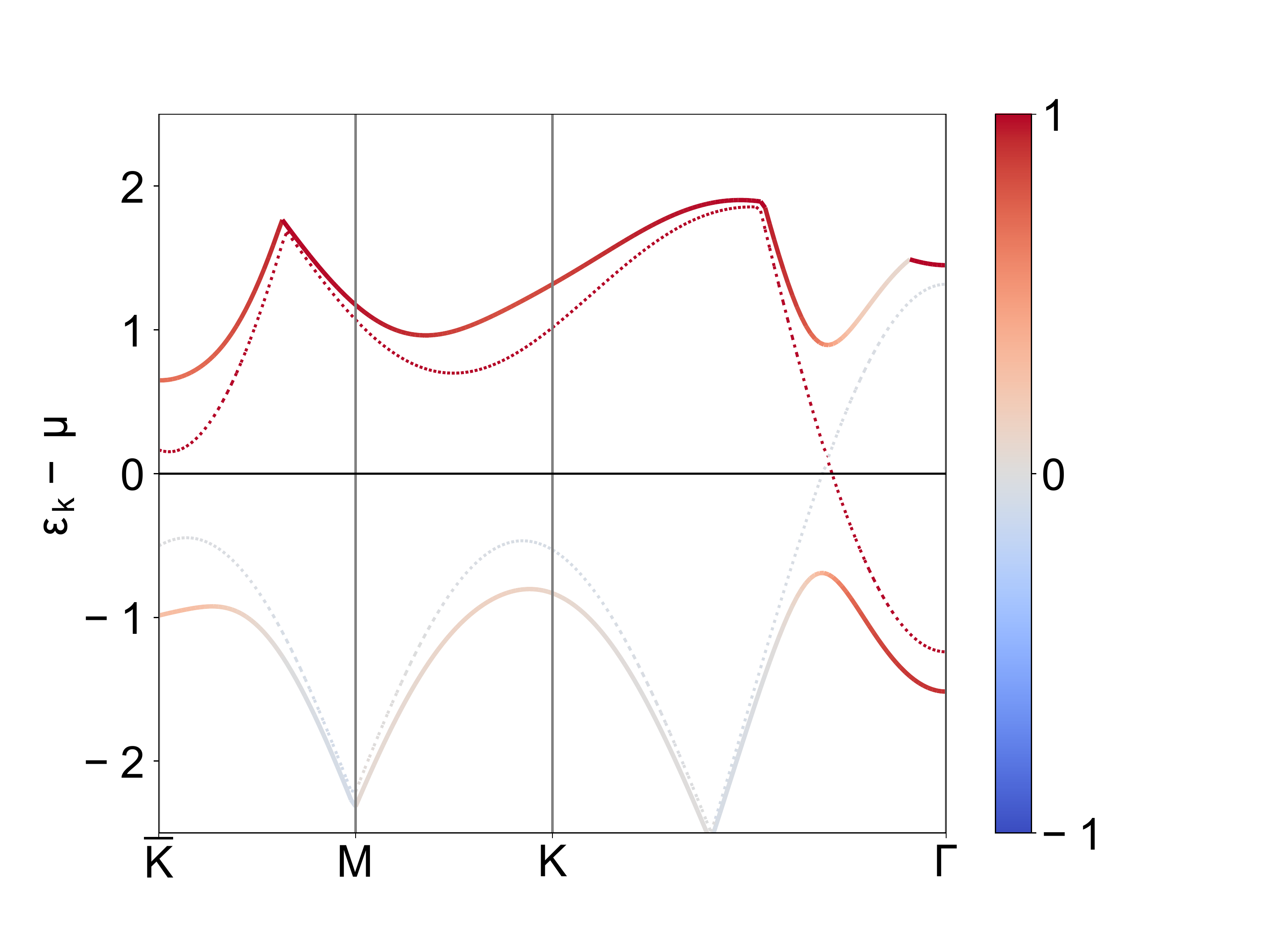}}}\\
    \subfloat[][]{\includegraphics[width=.24\textwidth]{{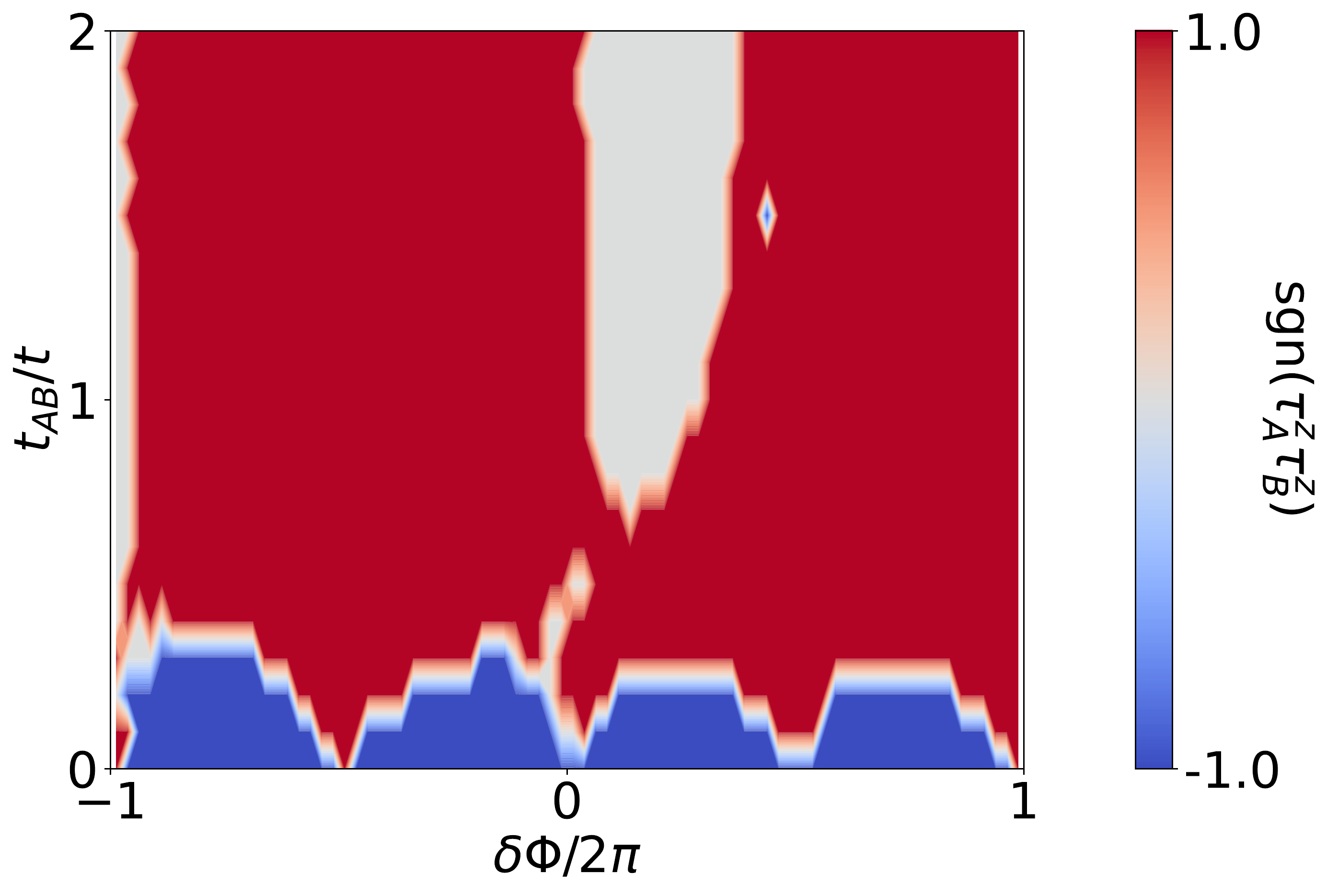}}}
    \subfloat[][]{\includegraphics[width=.24\textwidth]{{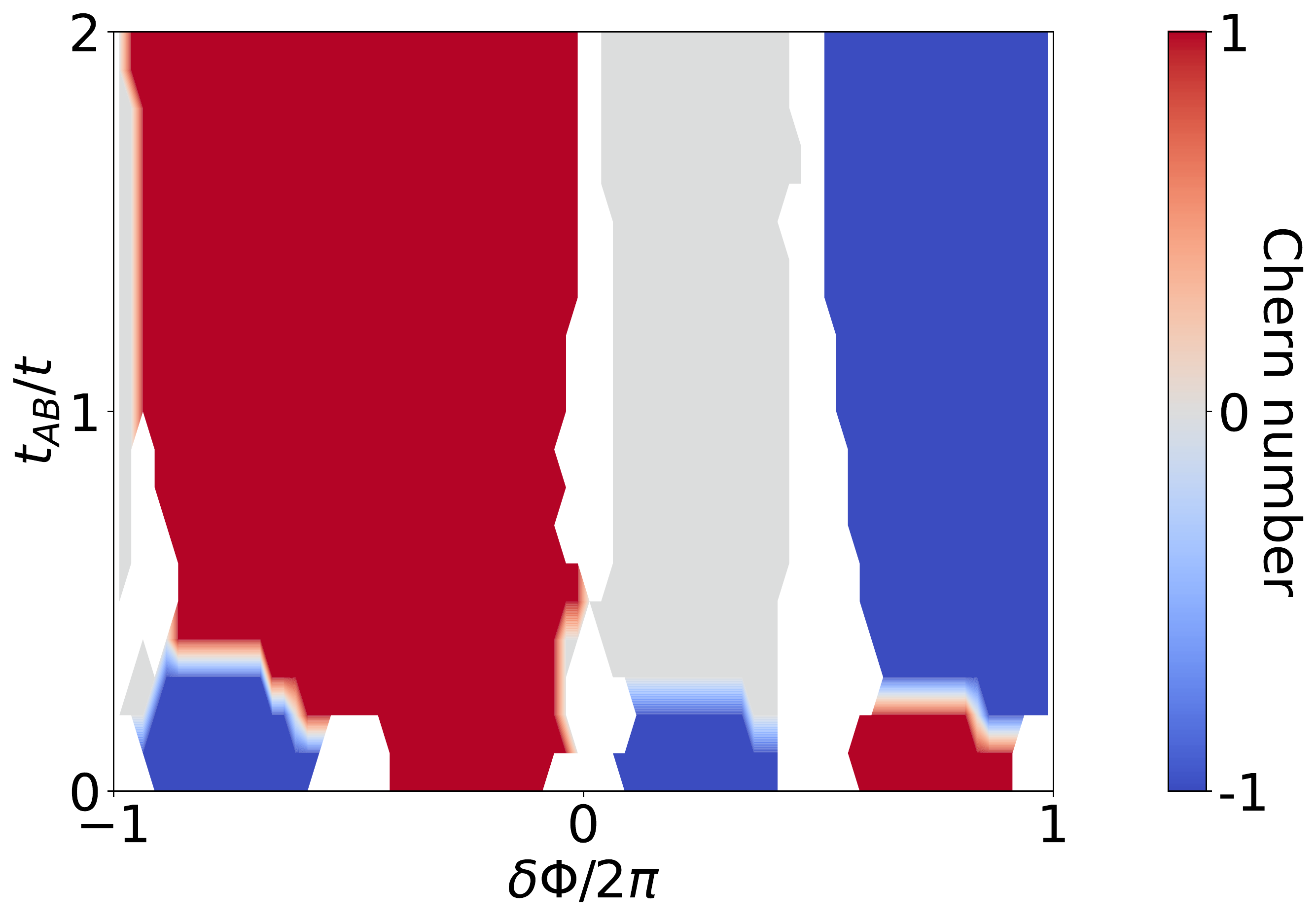}}}
    \caption{Exciton gap opening and phase diagrams in $t_{AB}-\delta\Phi$ plane. (a),(b) Mean-field electron band structure in ECI and EI phases. Dotted and solid curves stand for band structures before and after turning on exciton order parameters. Colors show the valley polarization. (a) and (b) are obtained with $V=1.5, D=0.5$ and $V=3, D=3$ respectively. (c) The sign of $\tau^z_A \tau^z_B$ when it is nonzero. (d) Chern number. The gapless region is whited out. In presence of $t_{AB}$, $\delta\Phi\rightarrow \delta\Phi+2\pi$ is no longer a gauge transform. These phase diagrams are calculated using Hartree-Fock mean field for $t_A=t_B=1$, $\Phi_A=2\pi+\delta\Phi$, $\Phi_B=\delta\Phi$, $U=50$, $V=2V'=4$ and $D=0.5$. }
    \label{tAB_effect_main}
\end{figure}

The valley polarizations in two layers can have the same or opposite signs. At $\Phi_B=0$ (or $\Phi_A=0$), the valley SU(2) symmetry of the B (or A) layer guarantees the degeneracy of intra-valley and inter-valley exciton condensation. With a finite $\Phi_B$, one of them is selected, while the other one is selected for an opposite sign of $\Phi_B$ because we can apply the transformation $i\tau_x$ only for layer $B$ to flip $\Phi_B$ and $\tau^z_B$ together. This selection is shown in Fig.~\ref{pd}(c). For $\Phi_A \Phi_B>0$, one finds that inter-valley exciton is favored and the QAH phase has inter-valley-coherence(IVC) with $\tau^z_A \tau^z_B<0$, consistent with the MCD measurement\cite{tao2022valley}.

 A finite $t_{AB}$ can also lift the degeneracy and select the $\tau^z_A \tau^z_B>0$ phase, as shown in Fig.~\ref{tAB_effect_main}(c).  We need a finite but small valley flux $\Phi_A$ or $\Phi_B$ (as small as $\frac{\pi}{12}$) to stabilize the ECI phase. Deep inside the region with $\Phi_A \Phi_B>0$, a sizable $t_{AB}$ as large as 0.35 t is consistent with an ECI phase with opposite valley polarizations. 
 
 Lastly, our Schwinger boson formulation has the nice feature of spin-charge separation.  In the supplementary, we show that increasing the superexchange $J_B$ term can lead to an excitonic Chern insulator with canted spin texture and only partial valley polarization.  This may be one explanation for the non-saturation of the MCD at a small magnetic field in the recent experiment\cite{tao2022valley}. However, we note that partial valley polarization can also be trivially explained by domain wall formation driven by disorder effects.

\textbf{Summary} In conclusion we propose a new mechanism for the QAH effect in the Coulomb coupled moire bilayer and apply it to the MoTe$_2$/WSe$_2$ system. In our model, the inter-layer hopping is spontaneously generated from inter-layer exciton condensation, which can have $p\pm ip$ symmetry and lead to an excitonic Chern insulator (ECI). The ECI phase has valley ferromagnetism arising from kinetic energy, in contrast to the quantum Hall ferromagnetism picture for the QAH effect in graphene moir\'e systems\cite{zhang2019nearly,repellin2020ferro}.  The valley polarizations in the two layers can be the same or opposite, depending on small perturbations away from a symmetric point where they are degenerate.  Our theory opens a new direction to search for interaction-driven topological phases without single-particle band topology, for example in moir\'e bilayer with an hBN barrier in the middle. Excitonic insulators with $p\pm ip$ pairing symmetry may also be realized in other systems outside of the moir\'e materials.

\textbf{Acknowledgement} We thank Trithep Devakul, Liang Fu, and  T. Senthil  for useful discussions. We thank Kin Fai Mak and Jie Shan for their communications on their unpublished experimental data. ZD is supported by NSF grant DMR-1911666. YHZ is supported by a startup fund from Johns Hopkins University.

\textit{Note added} When finalizing this manuscript, we become aware of a recent preprint\cite{KTLaw2022} which also proposes inter-valley $p\pm ip$ exciton order for MoTe$_2$/WSe$_2$ system. But the mechanism to select the inter-valley exciton order over intra-valley exciton order in Ref.~\onlinecite{KTLaw2022} relies on a large Hund's coupling and  is different from our theory.

\bibliographystyle{apsrev4-1}
\bibliography{LLL}

\newpage

\appendix
\onecolumngrid

\section{Details on Hartree-Fock calculation of the Hubbard model}

We start with the Hubbard model on the honeycomb lattice:

\begin{align}
    H& =-t_{AB} \sum_{\langle ij \rangle}(c^\dagger_{i;\alpha}c_{j;\alpha}+h.c.)+D\sum_{i\in A} n_i-D \sum_{i\in B}n_i \notag \\
    &-t_A \sum_{\langle \langle ij \rangle \rangle_A}(e^{i \phi^A_{ij} \tau_z}c^\dagger_{i;\alpha}c_{j;\alpha}+h.c.) -t_B \sum_{\langle \langle ij \rangle \rangle_B}(e^{i \phi^B_{ij} \tau_z}c^\dagger_{i;\alpha}c_{j;\alpha}+h.c.) \notag \\ 
    &+ \frac{U}{2} \sum_{i}n_i(n_i-1)+ V\sum_{\langle ij\rangle}n_in_j + V'\sum_{\langle\langle ij\rangle\rangle}n_in_j
\end{align}

We can decouple the interaction part through mean-field ansatz:

\begin{align}
    H_{HF}& =-t_{AB} \sum_{\langle ij \rangle}(c^\dagger_{i;\alpha}c_{j;\alpha}+h.c.)+D\sum_i  (-1)^{\sigma_z(i)}n_i \notag \\
    &-t_A \sum_{\langle \langle ij \rangle \rangle_A}(e^{i \phi^A_{ij} \tau_z}c^\dagger_{i;\alpha}c_{j;\alpha}+h.c.) -t_B \sum_{\langle \langle ij \rangle \rangle_B}(e^{i \phi^B_{ij} \tau_z}c^\dagger_{i;\alpha}c_{j;\alpha}+h.c.) \notag \\ 
    &+ U \sum_{i}(\rho_i c^\dagger_{i,\alpha}c_{i,\alpha} -\frac{1}{2}\rho_i^2) - U \sum_{i} (\tau^{\mu}_i c^\dagger_{i,\alpha}\tau^\mu_{\alpha\beta}c_{i,\beta} -\frac{1}{2} \tau_i^{\mu}\tau_i^{\mu}) \notag \\
    &+ 2V\sum_{\langle ij\rangle} (\rho_j c^\dagger_{i,\alpha}c_{i,\alpha} - \chi_{ji,\beta\alpha} c^\dagger_{i,\alpha}c_{j,\beta} -\frac{1}{2}\rho_i\rho_j +\frac{1}{2}|\chi_{ij,\alpha\beta}|^2) \notag \\
    &+ 2V'\sum_{\langle\langle ij\rangle\rangle} \rho_j c^\dagger_{i,\alpha}c_{i,\alpha} - \frac{1}{2}\rho_i\rho_j
\end{align}
with self-consistent equations:
\begin{align}
    \chi_{ij,\alpha\beta}&=\langle c^\dagger_{i,\alpha} c_{j,\beta} \rangle \;\mathrm{for\;} i,j\in \mathrm{NN} \notag \\
    \rho_{i} &= \langle c^\dagger_{i,\alpha} c_{j,\alpha} \rangle \notag \\
    \tau^{\mu}_{i} &= \langle c^\dagger_{i,\alpha} \tau^{\mu}_{\alpha,\beta}c_{j,\beta} \rangle
\end{align}

\begin{figure}[!ht]
  \centering
  \subfloat[][]{\includegraphics[width=.48\textwidth]{image/summary_phasediagram/M-V_pd_gap.pdf}}\quad
  \subfloat[][]{\includegraphics[width=.48\textwidth]{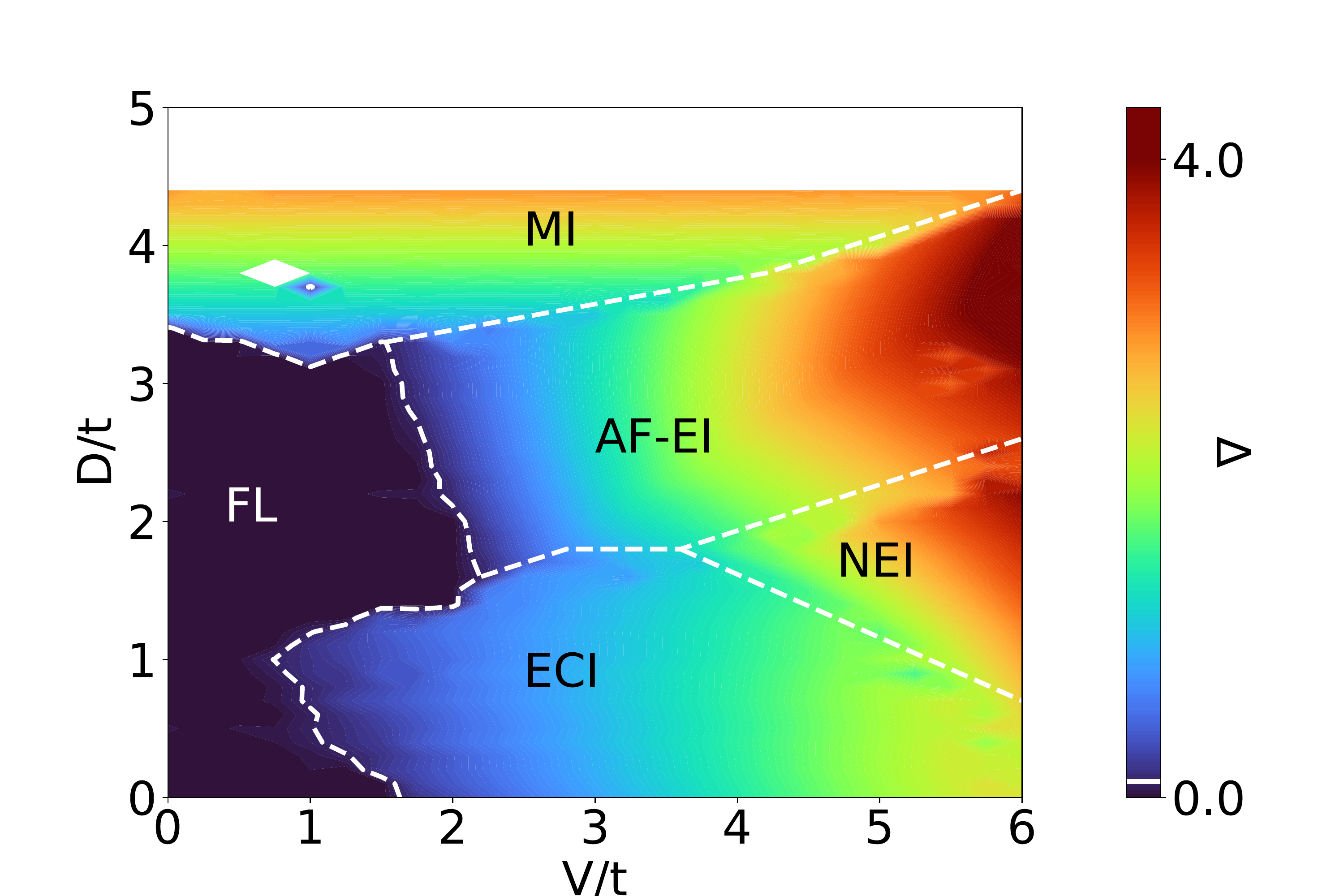}}
  \caption{Comparison between phase diagram from (a) t-J model treated with Schwinger boson mean field (b) Hubbard model treated with Hartree-Fock mean field. The color shows the charge gap. The two approaches are qualitatively consistent.  MI is the layer-polarized Mott insulator with $120^\circ$ anti-ferromagnetic order. FL is the Fermi liquid or electron-hole liquid. There are three different excitonic insulators: AF-EI, NEI, and ECI phase.  $\tau_z$ is polarized in the A layer for all these three phases. The magnetic order in the B layer is $120^\circ$ ordered in the AF-EI phase and $\tau_z$ polarized in the NEI and ECI phases.  ECI has a $p\pm ip$ exciton condensation symmetry and a non-zero Chern number $C=\pm 1$, while NEI has $s+p$ pairing symmetry for the exciton condensation and zero Chern number. }
  \label{compare}
\end{figure}

The mean field Hamiltonian is solved self-consistently. In addition, we use a $\sqrt{3}\times\sqrt{3}$ unit cell to allow states with translation symmetry breaking.  In Fig.\ref{compare}, we show that the phase diagram from the Hartree-Fock calculation is consistent with the Schwinger boson mean field theory of the t-J model detailed in the following section. In the Hartree-Fock calculation, magnetic orders in both layers are self-consistently generated. We find that the A layer is fully valley-polarized from this self-consistent calculation.

\section{Details on Schwinger boson calculation of the t-J model}
In the limit of large $U$, the Hubbard model is approximated by a $t-J$ model
\begin{align}
 H&=-t_A \sum_{\langle \langle ij \rangle \rangle_A}(e^{i \phi^A_{ij} \tau_z}c^\dagger_{i;\alpha}c_{j;\alpha}+h.c.)+D\sum_{i \in A}n_i \notag \\
 &~~~-t_B \sum_{\langle \langle ij \rangle \rangle_B}(e^{i \phi^B_{ij} \tau_z}Pc^\dagger_{i;\alpha}c_{j;\alpha}P+h.c.)-D \sum_{i \in B} n_i \notag \\
 &~~~ + J_B\sum_{\langle\langle ij\rangle\rangle_B} \tau^\mu_iR^{\mu\nu}_z(2\phi^B_{ij})\tau^\nu_j +(V'-\frac{J_B}{4}) \sum_{\langle \langle ij \rangle \rangle_B}n_i n_j\notag \\
   &~~~+V\sum_{\langle ij\rangle}n_in_j + V'\sum_{\langle\langle ij\rangle\rangle_A}n_in_j
\label{HamtJ2}
\end{align}
where $J=\frac{4t^2}{U}$, and $R^{\mu\nu}_z$ is the matrix representing rotation around $z$ axis, namely, $R(\phi) = \begin{pmatrix}
  \cos{\phi} & \sin{\phi} & 0\\
  -\sin{\phi} & \cos{\phi} & 0\\
  0 & 0& 1
\end{pmatrix}$ in the basis of $(\tau_x,\tau_y,\tau_z)$.  Note that since $A$ layer has a low density $x$, the double-occupancy projection $P$ is not expected to be relevant. So we have treated the A layer as free fermions instead of using the composite degrees of freedom. In the calculation, we will set the magnetization in the A layer to be polarized along $\tau^z$ direction, which is found to hold throughout the Hartree-Fock phase diagram.

We use a fermionic holon, bosonic spinon construction for the B layer to take care of the projected operator: $c_{i\alpha}=b_{i\alpha}f_i$, with a constraint $n^f_i=n^b_i$. $b_{i;\alpha}$ will be assumed to be condensed and just represents a magnetic order. This is just a convenient way to deal with the charge degree of freedom moving in the background of magnetic order. No exotic phase with fractionalization is targeted in our calculation. The Hamiltonian in Eq.\ref{HamtJ} is rewritten as
\be
    H = H_t + H_D + H_J + H_V
\ee
where
\be
\begin{split}
    H_t &=-t_A \sum_{\langle \langle ij \rangle \rangle_A}(e^{i \phi^A_{ij} \tau_z}c^\dagger_{i;\alpha}c_{j;\alpha}+h.c.)+D\sum_{i \in A}n_i \notag \\
 &~~~-t_B \sum_{\langle \langle ij \rangle \rangle_B}(e^{i \phi^B_{ij} \tau_z}b^\dagger_{i\alpha} b_{j\alpha}f^\dagger_{i}f_{j}+h.c.) \notag\\
    H_D &= D \sum_{i\in A}f_i^\dagger f_i-D \sum_{i\in B} f_i^\dagger f_i \notag \\
    H_J &= J \sum_{\langle\langle ij\rangle\rangle}b^\dagger_{i\alpha}\sigma^\mu_{\alpha\beta} b_{i\beta}R^{\mu\nu}_z(\phi_{ij})b^\dagger_{j\gamma}\sigma^\nu_{\gamma\delta} b_{j\delta}\\
    H_V &= V \sum_{\langle ij\rangle}f^\dagger_{i} f_{i}f^\dagger_{j}f_{j}+V' \sum_{a=A,B}\sum_{\langle \langle ij \rangle \rangle_a}n_i n_j
\end{split}
\label{tJham}
\ee

The exciton insulators found in Hartree-Fock always have $\tau_z$ polarization in the A layer. But the magnetic order of the B layer could be either $120^\circ$ order or $\tau_z$ polarized FM order. To study the competition between the two magnetic orders of the B phase, we adopt the following ansatz for the Schwinger boson part
\be
    \begin{pmatrix} \langle b_{i;+} \rangle \\ \langle b_{i;-} \rangle\end{pmatrix}=F_i 
\ee
with $ F^A_i  = \begin{pmatrix} 1 \\ 0 \end{pmatrix}$, $ F^B_i  =\sqrt{n_B} \begin{pmatrix} 1 \\ 0 \end{pmatrix}$ for the FM order.  or  $F^B_i=\sqrt{n_B} e^{\frac{i}{2}(\v Q \cdot \v r_i) \tau_z} \begin{pmatrix} \cos{\frac{\theta}{2}} \\ \sin{\frac{\theta}{2}} \end{pmatrix}$ for the canted AFM order, where $\v Q = \mathbf K_M \  \mathrm{or} -\mathbf K_M$.

After the magnetic orders are decided for both layers, the charge degree of freedom is captured by a spinless model. We also label the valley polarized operator $c_{i;+}$ in the A layer as $f_i$ with $i \in A$. The kinetic term is now
\begin{align}
    H_t^f&=-\sum_{a=A,B} \sum_{\langle \langle ij \rangle \rangle_a} t^f_{ij;a} f_i^\dagger f_j +D\sum_{i \in A} f_i^\dagger f_i -D \sum_{i \in B}f_i^\dagger f_i
    \label{eq:Ht_appendix}
\end{align}
where $t^f_{ij;a}=t_a (F_i^{a\dagger} e^{i \phi_{ij;a} \tau_z} F^a_j)$. $n_i=f_i^\dagger f_i$ is constrained to be the physical density. The total filling is $n_A+n_B=1$ so there are electron and hole pockets in the two layers respectively. 

The $H_J$ term simply contributes energy for each magnetic pattern:
\begin{equation}
    H_J = J \sum_{\langle\langle ij\rangle\rangle_B} n_B \bigg(\cos^2{\theta} + \sin^2{\theta}\cos{(2\phi_{ij}+\v Q\cdot\v r_{ij})}\bigg) 
\end{equation}
where $\v Q =\v \Gamma_M, \v K_M \;\mathrm{or} -\v K_M$. Here $\theta$ is the angle between $z$-axis and the spin vector $\vec{\tau}_B$, and $\phi$ is the valley-contrasting hopping phase. $H_V^{MF}$ term is taken care of by a mean-field decoupling
\begin{equation}
    H_V=-V\sum_{\langle ij \rangle} \chi_{ji} f^\dagger_i f_j + h.c. - |\chi_{ij}|^2
\end{equation}
with self-consistent equation
\begin{equation}
    \chi_{ij}=\langle f_i^\dagger f_j \rangle
\end{equation}
The mean field energy is
\begin{align}
    H_{MF} =&- \sum_{a=A,B} \sum_{\langle \langle ij \rangle \rangle_a} t_a (F_i^{a\dagger} e^{i \phi_{ij;a} \tau_z} F^a_j) \chi_{ij} + h.c. +D\sum_{i \in A} \langle n_i\rangle -D \sum_{j \in B} \langle n_j\rangle \\ \nonumber
    &- V\sum_{\langle ij \rangle} |\chi_{ij}|^2
    + J \sum_{\langle\langle jj'\rangle\rangle_B} n_B \bigg(\cos^2{\theta} + \sin^2{\theta}\cos{(2\phi_{jj'}+\v Q\cdot\v r_{jj'})}\bigg) 
\end{align}
The calculation proceeds in the following way. We target either FM or $120^\circ$ AFM in the B layer and assume $\tau_z$ FM in the A layer, after that for each magnetic order pattern we deal with $H_t^f+H_V$ term with $H_t^f$ in Eq.~\ref{eq:Ht_appendix}. We use a $\sqrt{3}\times \sqrt{3}$ unit cell, allowing the exciton order to break translation symmetry.  In the end, for each magnetic order pattern, we add the spin coupling energy from the $H_J$ term and compare their total energies. The phase diagram is shown in the main text.


\section{Exciton order parameter \label{appendix:exciton order parameter}}

\begin{figure}[!ht]
  \centering
  \subfloat[][]{\includegraphics[width=.2\textwidth]{{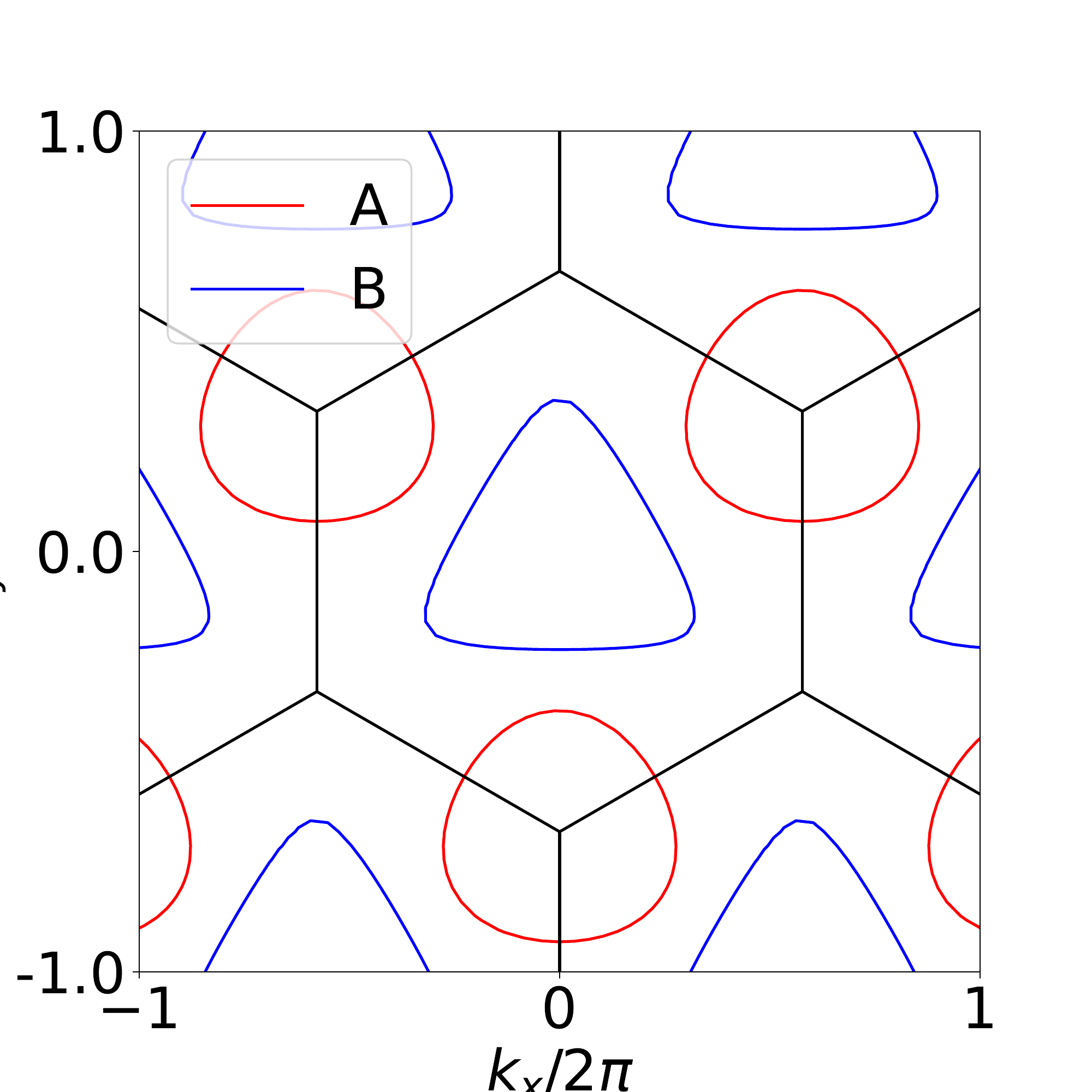}}}
  \subfloat[][]{\includegraphics[width=.2\textwidth]{{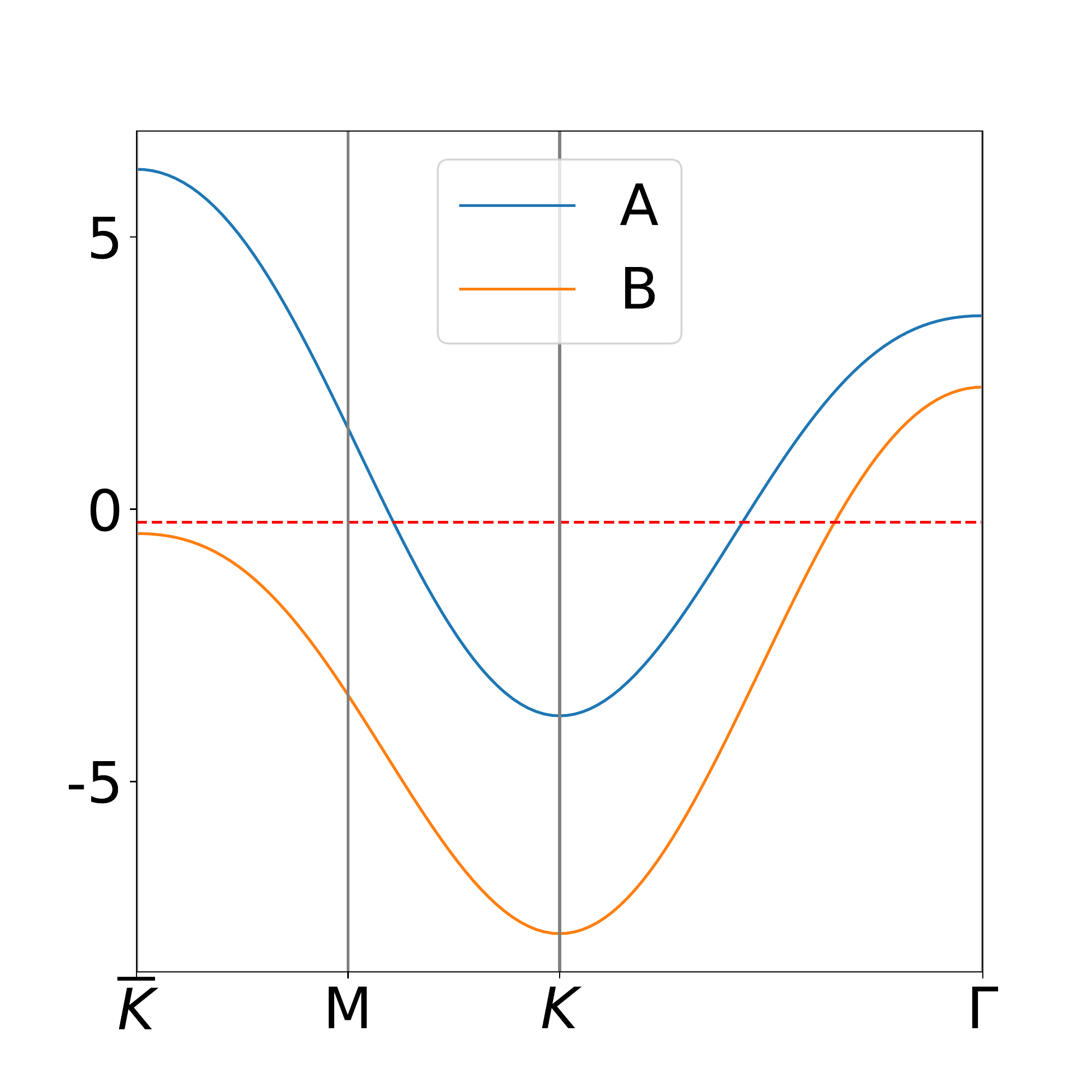}}}
  \subfloat[][]{\includegraphics[width=.2\textwidth]{{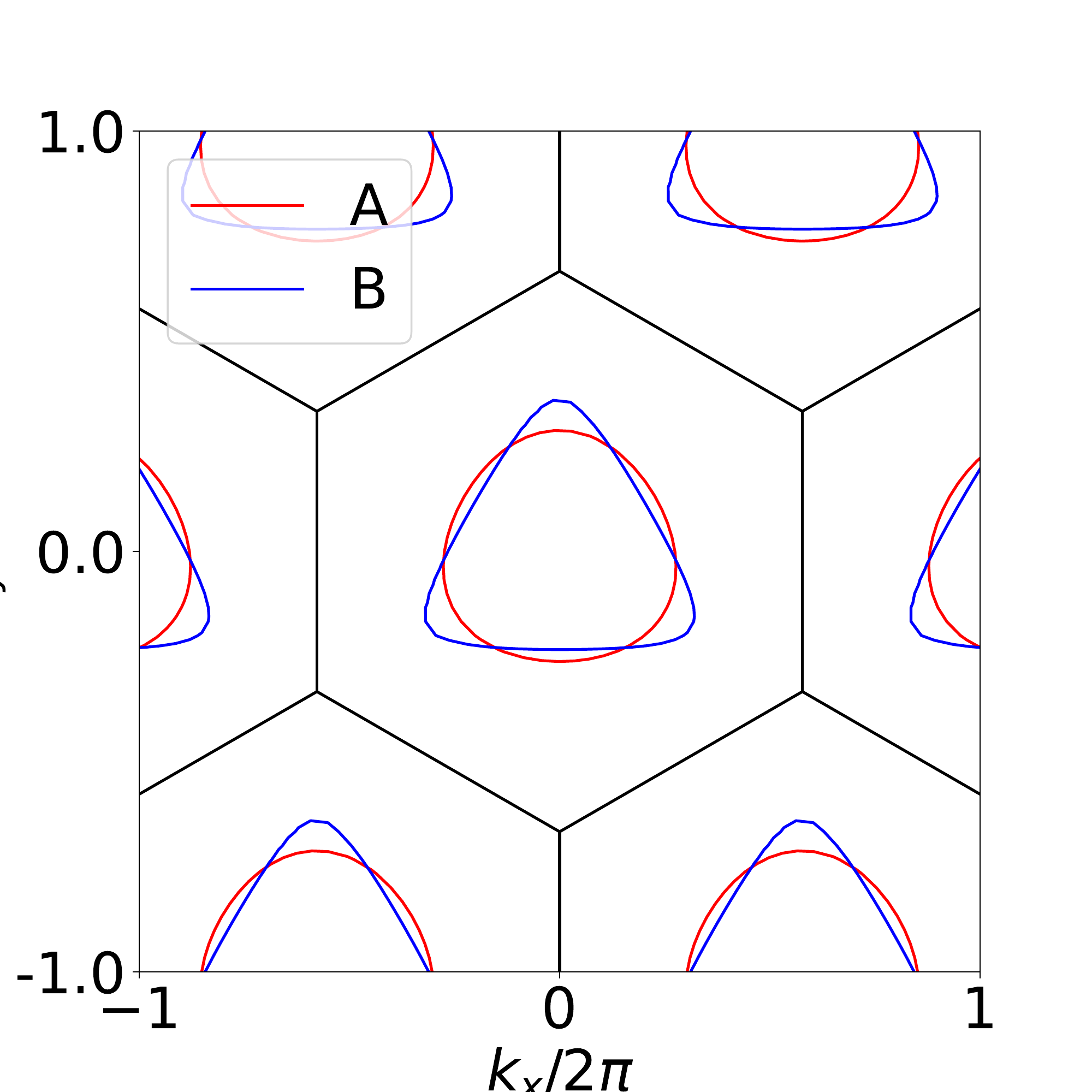}}}
  \subfloat[][]{\includegraphics[width=.2\textwidth]{{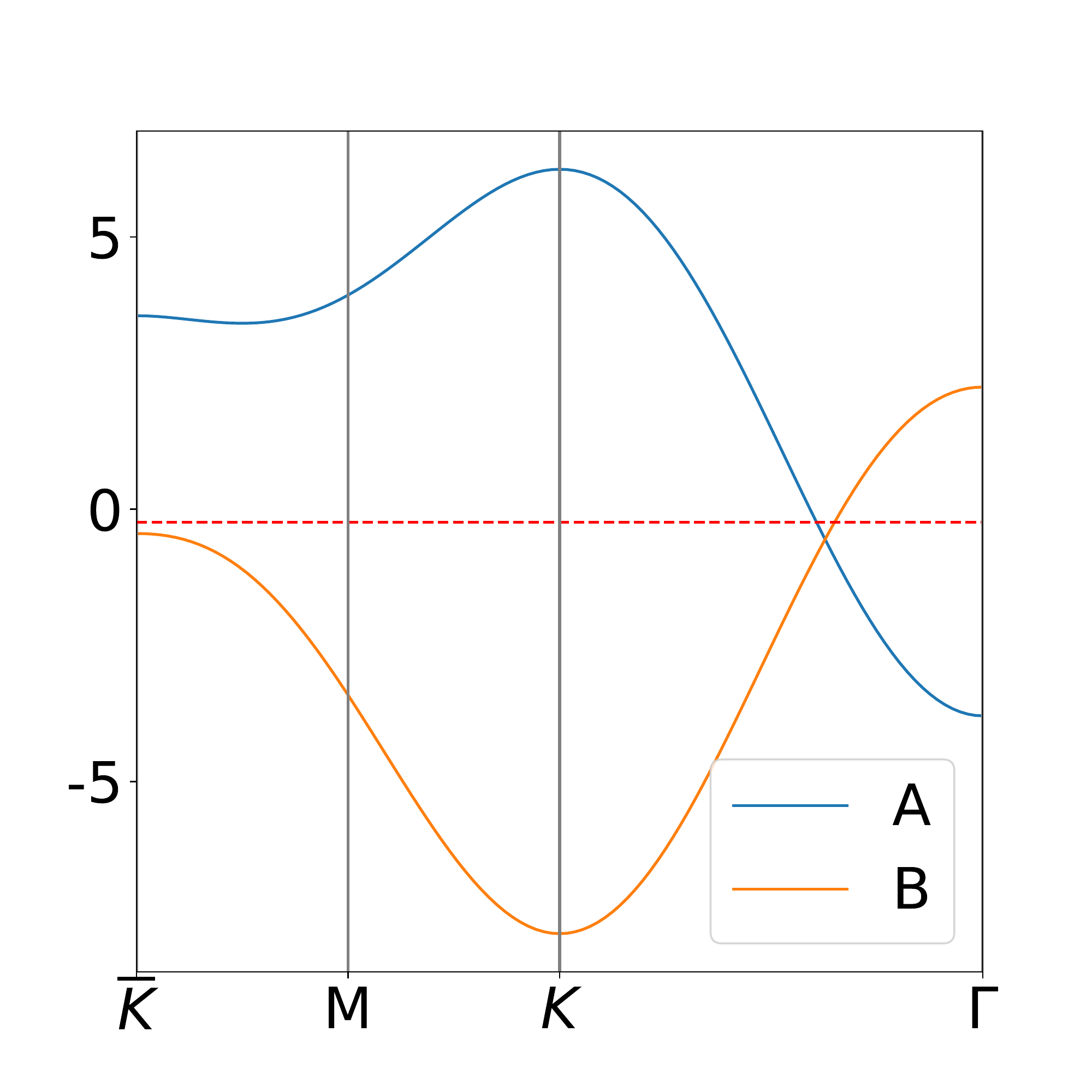}}}
  \caption{Band structure and fermi surface without pairing for $D=2.0$ (a)(b) $\phi_A=-\frac{7}{12}\pi$, $\phi_B=\frac{3}{4}\pi$, (c)(d) gauged to $\phi_A=\frac{1}{12}\pi$, $\phi_B=\frac{3}{4}\pi$. We choose the gauge under which the center of fermi pockets are at $\Gamma$. For this choice of $\phi_A, \phi_B$, the two fermi surfaces are perfectly nested after gauging, leading to a logarithmic divergence of exciton susceptibility. However, fermi liquid is still stable in this case for regions without full valley polarization.}
  \label{fermi_surface}
\end{figure}

In this section, we discuss the symmetry of exciton order parameters. Following the Schwinger boson theory, we only need to consider a spinless model for the charge degree of freedom. The exciton arises from pairing between $f_A$ and $h_B=f^\dagger_B$. This exciton order parameter
\be
\chi_{ij}  = \sum_{L,Q,\v k} \langle f^\dagger_{A,L\v K+\v k} f_{B,L\v K+Q\v K+\v k}\rangle e^{-i(L\v K+\v k)(\v r_i-\v r_j)+iQ \v K \v r_j}
\ee
induces hybridization between A-layer particle pocket at $L \v K$ and B-layer hole pocket at $(L+Q) \v K$.
We note in passing that Eq.5 in the main text is given by defining the Fourier transform
\be
    \chi_{Q,L} = \chi_{Q,L}(r_i, r_j) = \sum_{\v k} \langle \chi^{Q,L}_{\v k}\rangle e^{i\v k\v (r_j-r_i)} = \sum_{\v k}\langle f^\dagger_{A,L\v K+\v k} f_{B,L\v K+Q\v K+\v k}\rangle e^{i\v k\v (r_j-r_i)} 
\ee
To extract the intrinsic exciton pairing angular momentum, we follow the convention of first shifting the band bottom of A and band top of B to the vicinity of $\Gamma$ through a gauge transform, 
\begin{align}
    &f_{A,i}\rightarrow f_{A,i}e^{-i \delta L \v K\cdot \v r_i} \notag\\
    &f_{B,j}\rightarrow f_{B,j}e^{-i (\delta L+\delta Q) \v K\cdot \v r_j}
    \label{gaugetransform}
\end{align}
For $\phi_A = (-2/3+1/12)\pi$, $\phi_B = (2/3+1/12)\pi$, the particle and holon pockets before and after gauge transform is shown in Fig.\ref{fermi_surface} as an example.

We focus on the simple scenario with only one particle and one hole pocket. Then $\delta L$ and $\delta L+\delta Q$ correspond to the band bottom of the A-layer and band top of the B-layer in the free model (before exciton condensation). 
After the gauge transform Eq.\ref{gaugetransform}, exciton order should only couple two pockets at $\v \Gamma_M$, therefore it carries no momentum $Q'=Q-\delta Q=0$. Now we are ready to read off the exciton order $\chi(\v k)$ from the Fourier transform of the gauged $\chi_{ij}$.

To determine the large momentum $\delta L$ and $\delta Q$ for gauge transform, we note that the mean-field Hamiltonians for $f_A$ and $f_B$ are just two triangular lattice tight-binding models with only nearest-neighbor hopping, 
\begin{align}
    H_A &= -\sum_{\langle i,i'\rangle_A} \tilde{t}^A e^{i\tilde{\phi}^A_{ii'}} f^{A\dagger}_i f^A_{i'} +h.c. \notag \\
    H_B &= -\sum_{\langle j,j'\rangle_B} \tilde{t}^B e^{i\tilde{\phi}^B_{jj'}} f^{B\dagger}_j f^B_{j'} +h.c. 
\end{align}
whose band top/bottom positions can be easily read off from the hopping phases $\tilde{\phi}_A$, $\tilde{\phi}_B$. (Here we use a tilde to distinguish from the valley contrasting phases in the original spinful model, these phases are the effective hopping phases for the charges once the spin gets integrated out.) Namely, the A-band bottom is at $\v \Gamma_M$ for $\tilde{\phi}_A\in (-\pi/3, \pi/3)$, while the band top is at $\v \Gamma_M$ for $\tilde{\phi}_B\in (2\pi/3, 4\pi/3)$. Furthermore, under the gauge transform Eq.\ref{gaugetransform} the $\tilde{\phi}$ transform as
\begin{align}
    \tilde{\phi}^{ii'}_A &\rightarrow \tilde{\phi}^{ii'}_A + \delta L \v K\cdot (\v r_{i}-\v r_{i'})\notag \\
    \tilde{\phi}^{jj'}_B &\rightarrow \tilde{\phi}^{jj'}_B + (\delta L+\delta Q) \v K\cdot (\v r_{j}-\v r_{j'})
\end{align}
and the order parameter becomes
\be
    \tilde{\chi}_{ij} = \chi_{ij} e^{i \delta L\v K\cdot \v r_i-i(\delta L\v K+\delta Q\v K) \cdot \v r_j}
\ee

\begin{figure}[!ht]
  \centering
  \subfloat[][]{\includegraphics[width=.3\textwidth]{{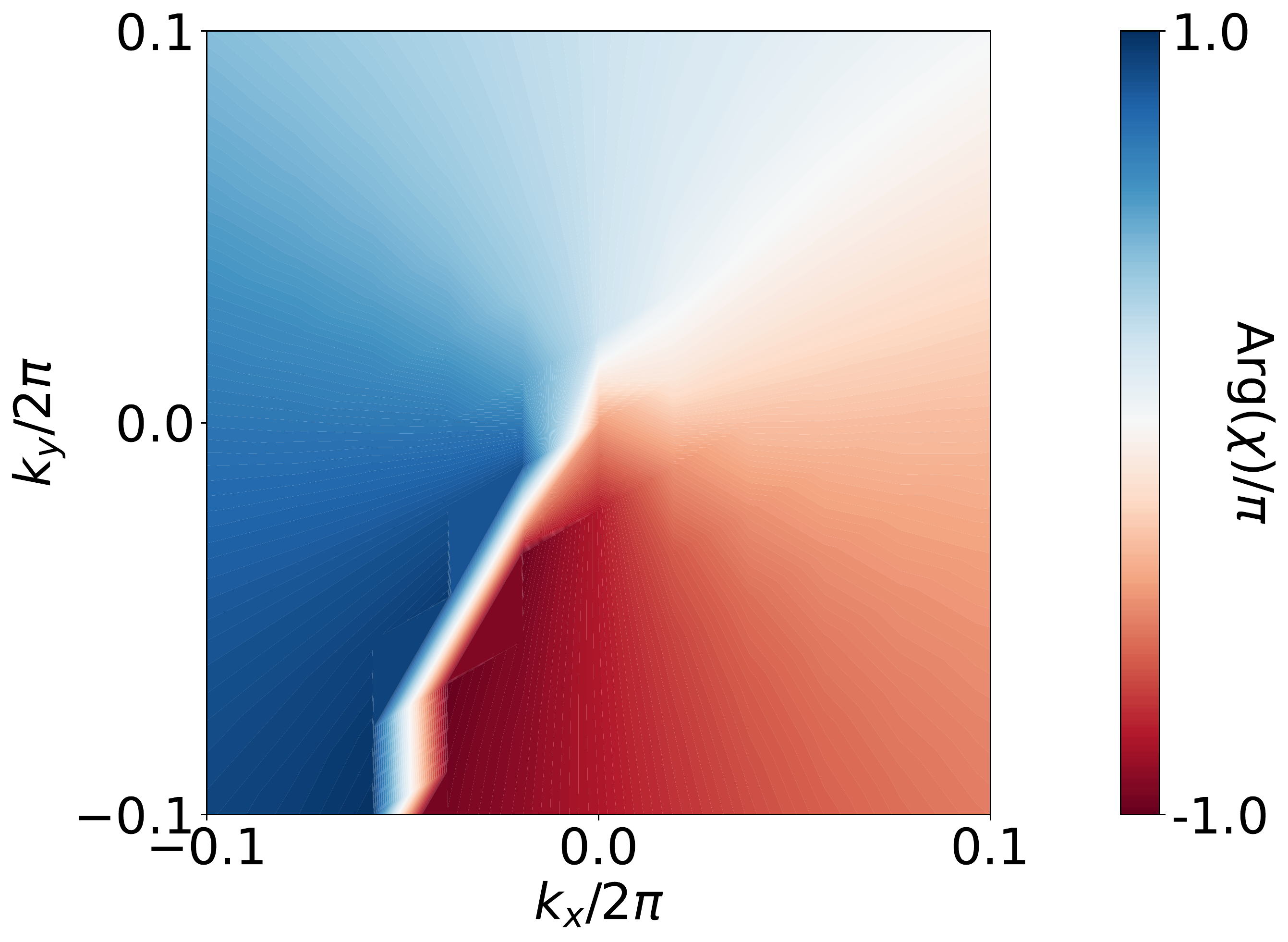}}}\quad
  \subfloat[][]{\includegraphics[width=.3\textwidth]{{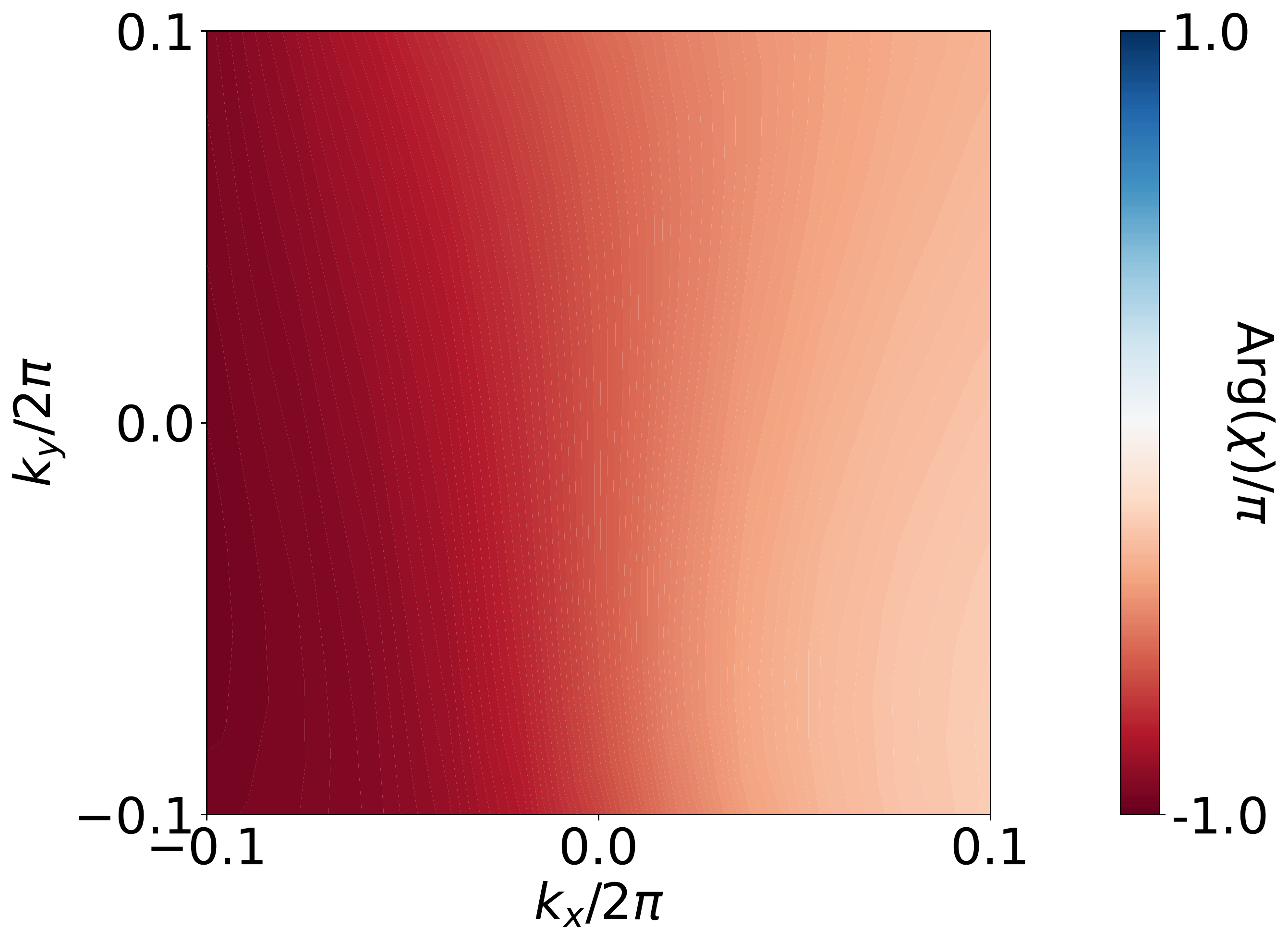}}}\\
  \caption{ Momentum space distribution of exciton. (a) $p+ip$ order of ECI at $D=0.0$, $V=6.0$ (b)$s+p$-wave exciton order of EI at $D=6.0$, $V=2.0$. A strong nematicity is observed in the latter example. The transition from ECI to EI happens when the vortex in $\chi(\v k)$ gets pushed out of the small Fermi surface region by $s$-wave component. Plotted here is the phase of the gauged exciton order. }
  \label{vortex}
\end{figure}
Therefore our procedure for extracting pairing symmetries will be
\begin{itemize}
\item (1) For given a spin configuration $F_i$, extract the effective hopping phase $\tilde{\phi}^{ij}_{A/B}=\mathrm{Arg}(\langle F_i|e^{i\phi^{ij}_{A/B} \sigma_z}|F_j\rangle)$
\item (2) Find the $(\delta Q,\delta L)$ that sends $\tilde{\phi}_{A/B}$ to the proper ranges mentioned above.
\item (3) Fourier transform the gauged exciton order $\tilde{\chi}_{Q'L'}=\tilde{\chi}_{ij} e^{i L'\v K\cdot \v r_i-i(L'\v K+Q'\v K) \cdot \v r_j} $ to read off its remaining momentum $Q'$ and angular momentum $L'$.
\item (4) Fourier transform to extract $\tilde{\chi}(\v k) = \sum_{ij}\tilde{\chi}_{ij} e^{i\v k(\v r_i-\v r_j)}  = \sum_{ij}\chi_{ij}  e^{i(L\v K+\v k)r_i-i(L\v K+Q\v K+\v k)\v r_j}$
\end{itemize}
Note that by gauging the pockets to $\v \Gamma_M$ following step (2), we have basically picked a gauge under which the center of mass of the exciton pair carries momentum $\v Q=0$.
In some cases, there can be coexisting $p$-wave and $s$-wave exciton orders. We extract each component through a Fourier transform for $\chi(\v k)$
\be
    \chi_l(|\v k|) = \int d\theta_{\v k} \chi(\v k) e^{-il\theta_{\v k}}
\ee
Alternatively, one can read off the strength of $p$/$s$ wave orders from the $(Q',L')$ Fourier components of $\chi_{ij}$.
In Fig.\ref{vortex} we show typical patterns of phase variation of exciton order $\chi$ in momentum space for the $p+ip$ ECI and $s+p$ nematic EI.

\section{Competition between the $s$ wave and $p$ wave exciton condensation}
In this section, we discuss why the $p$ wave is favored over the $s$ wave exciton condensation.
The nearest neighbor interaction is decoupled into
\be
    H_{V}
    = -V \sum_{\langle ij\rangle} \chi_{ij}^* c^\dagger_{i\sigma} f_{j\sigma} +h.c.
    = -V\sum_{\v k,l} \chi^*_l F^l_{\v k}  c^\dagger_{i\alpha} \sigma^\nu_{\alpha\beta}f_{j\beta} +h.c.
\ee
where $F^l_\v k = \sum_{\v \delta}^{NN} e^{il\theta_\delta}e^{-i\v k \v \delta}$, and $\chi_l =\frac{1}{3}\sum_{\v \delta}^{NN} \chi_{\v \delta}e^{-il\theta_{\v\delta}}$.
The self-consistent equation
\begin{align}
    \chi_{l} &= \frac{1}{3}\sum_{\v k,\omega} F^{l*}_{\v k}\langle c^\dagger_\v k f_\v k\rangle\\
    & = -\frac{1}{3}\sum_{\v k,\omega} G^c(\v k, i\omega) V\chi_{m} F^m_{\v k} \tilde{G}^f (\v k, i\omega)F^{l*}_{\v k}\\
    & = -\frac{V}{3}\sum_{\v k,\omega} G^c(\v k, i\omega) \tilde{G}^f (\v k, i\omega) F^{l*}_{\v k} F^m_{\v k} \chi_{m}
\end{align}
Here $G_c = (i\omega-\epsilon^A_\v k)^{-1}$ and $\tilde{G}_f = (i\omega-\epsilon^B_\v k -\Sigma_f(\v k,i\omega))^{-1}$. This is an eigenvalue problem. The off-diagonal($l\neq m$) contributions would be responsible for the $s+p$ wave exciton order observed. For notational simplicity, we will restrict ourselves to the diagonal part. Existence of a non-trivial solution for $\chi_l$ demands
\begin{align}
    \frac{1}{V}
    &= -\sum_{\v k, \omega} G^c(\v k, i\omega) \tilde{G}^f (\v k, i\omega) |F^{l}_{\v k}|^2/3\\
    &= -\sum_{\v k} \frac{f(\xi_{\v k,+})-f(\xi_{\v k,-})}{\xi_{\v k,+}-\xi_{\v k,-}}|F^{l}_{\v k}|^2/3\\
    &= - \int^{E_+}_{E_-} d^2\v k \frac{|F^{l}_{\v k}|^2/3}{\xi_{\v k,+}-\xi_{\v k,-}}\\
    &= - \int^{E_+}_{E_-} d^2\v k \frac{|F^{l}_{\v k}|^2}{\sqrt{\delta_\v k^2+V^2|\chi_l|^2|F^{l}_{\v k}|^2/3}}
\end{align}
In the third line, we have considered a simplest case with particle-hole symmetry $\epsilon^{A/B}_\v k = \epsilon_F\pm\delta_{\v k}/2$, namely the particle and hole pockets are perfectly nested. A short discussion of cases without perfect nesting will be presented later. Defining $\Phi_l(\v k) = \frac{1}{\sqrt{3}}V\chi_l F^l_\v k$, we find
\be
    \frac{1}{V}
    = - \int^{E_+}_{E_-} d^2\v k \frac{|F^{l}_{\v k}|^2}{\sqrt{\delta_\v k^2+4\Phi_l^2(\v k)}}
\label{sc}
\ee
This is analogous to the self-consistency equation for BCS instability. At zero temperature the logarithmic divergence of integral at $\Phi=0$ indicates that an infinitesimal $V$ would lead to exciton condensation. So the Fermi liquid is always unstable in the $V>0$ regime. From Eq.\ref{sc} we obtain the scaling of exciton order
\be
    \chi^l \sim \frac{\rho(\epsilon_F)}{V\eta^{l}(\epsilon_F)}e^{-\frac{1}{2V\eta^{l}(\epsilon_F)}}
\ee
where $\eta(\epsilon_F)$ is the weighted average of form factor in channel $l$ around the fermi surfaces
\be
    \eta^l(\epsilon_F) = \int_{FS}\frac{d\v k_F}{(2\pi)^2}\frac{|F^{l}_{\v k}|^2}{v_{\v k_F}}
\ee
here the weight is the density of states. This expression reduces to
\be
    \eta^l(\epsilon_F) \sim \rho(\epsilon_F) \int_{FS} d\v k_F |F^{l}_{\v k_F}|^2
    \label{F12}
\ee
in the small $k_F$ limit where the band structure has an approximate rotational symmetry. The angular momentum channel of the condensed exciton is efficiently predicted by comparing the parameter $\eta^l(\epsilon_F)$ across all angular momentum channels $l=0,1,-1$.

\begin{figure}[!ht]
  \centering
  \subfloat[][]{\includegraphics[width=.3\textwidth]{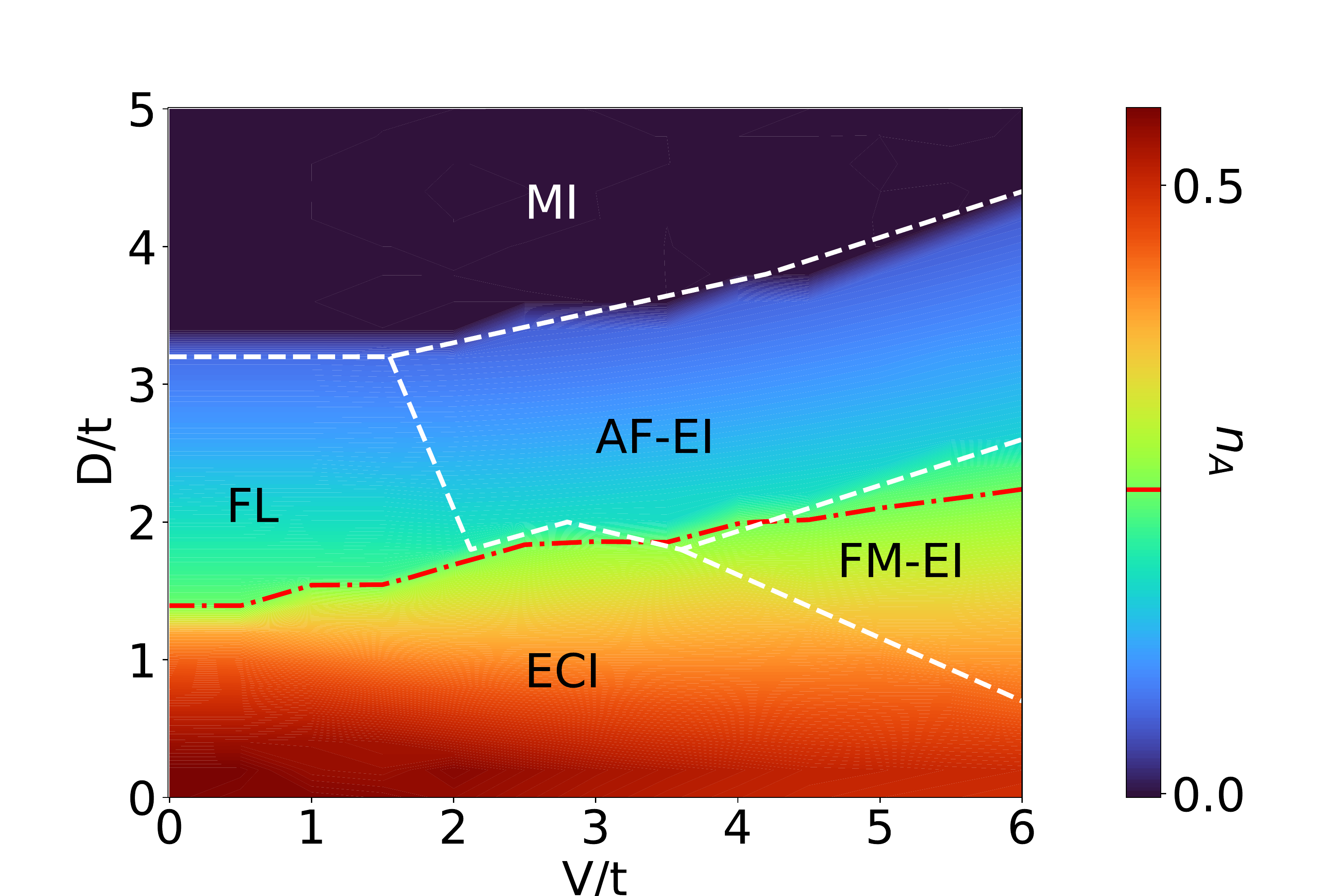}}\quad
  \subfloat[][]{\includegraphics[width=.3\textwidth]{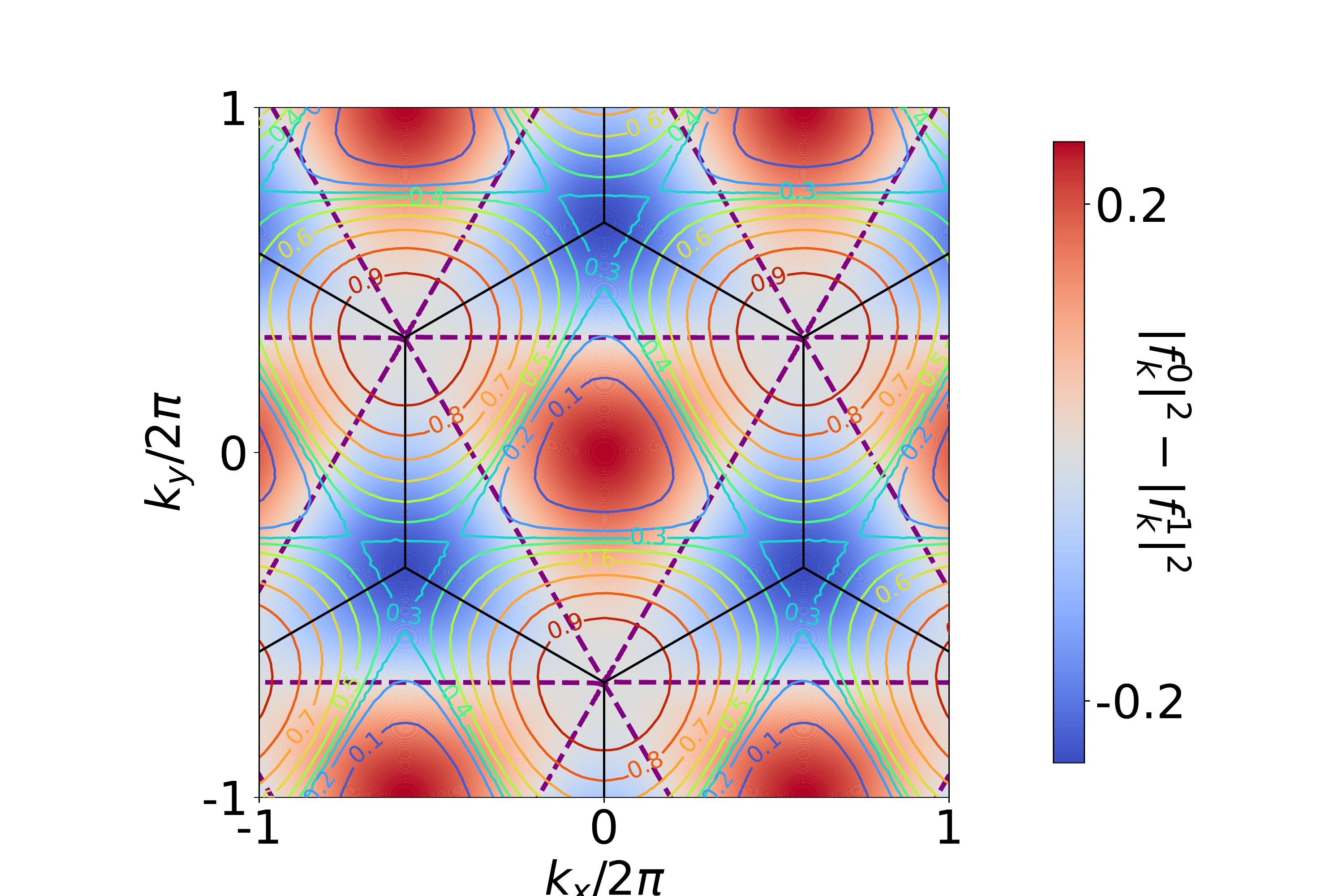}}
  \caption{(a) Exciton density. The contour in red corresponds to $n_{A(c)}=0.25$. It closely tracks the onset of $p+ip$ order. (b) Difference between form factors in $l=0$ and $l=1$ channels. The purple dashed contour marks momentum where two form factors are equal. Overlaid contours show the Fermi surface at a sequence of exciton densities $n_A$. Above $n_A=0.2$, the Fermi surface starts to pass through the area with a stronger $p$-wave form factor. At $n_A=0.3$ the Fermi surface is predominantly through the blue area. Somewhere in between these two densities, the pairing instability should change from $s$-wave to $p$-wave. This estimation is consistent with the critical density read off from (a)}
  \label{instability}
\end{figure}

In the following, we explore the cases without nesting between particle and hole pockets. For convenience we denote the susceptibility $\Pi_{\v k'} = \int d\omega G^A(\v k', \omega)G^B(\v k', \omega)$.
Generally speaking, at the transition point $\chi=0$, exciton susceptibility
\be
    \Pi_{\v k} \sim -\frac{f(\epsilon^A_{\v k})-f(\epsilon^B_{\v k})}{\epsilon^A_{\v k} - \epsilon^B_{\v k}}\
\ee
where $f(\epsilon) = (1+e^{\beta (\epsilon-\epsilon_F)})^{-1}$ is the Fermi-Dirac distribution.
The dominating contribution comes from the $\v k$ points with $\epsilon^A_{\v k}=\epsilon^B_{\v k}=0$.
For generic parameters $(\phi_A, \phi_B)$, the two fermi surfaces do not overlap, but the total filling $\nu_T=1$ dictates that they intersect at $3N$ momentum points, where $N$ is an even number. Assume $f^l_{\v k}$ vary slowly in the vicinity of this spot. The dispersion is
\be
    \epsilon^A(k) = u_A k_x + v_A k_y, \quad \epsilon^B(k) = u_B k_x + v_B k_y
\ee
for simplicity we will take $u_A = -u_B = u$ and $v_A = v_B = v$, assume $u/v=\cot\theta>1$, where $2\theta$ is the angle between two fermi surfaces
\be
    \int_0^\Lambda d^2\v k \Pi_{\v k} =4\int^\Lambda_0  dk_y\int^\Lambda_{\frac{v}{u}k_y} dk_x  \frac{1}{2u k_x}=2\int^\Lambda_0  dk_y  \ln \frac{u\Lambda}{vk_y} = 2\Lambda(1+\ln\frac{u}{v})
\ee
Each one of these spots contributes a finite instability. This means now it would require a finite strength of $V=V_c>0$ for the condensation to happen. A logarithmic divergence shows up as we approach the perfect nesting limit $\theta\rightarrow0$. And the $p_x\pm ip_y$ wave exciton gets favored as long as these intersections fall into the $|F^{\pm 1}_{\v k}|>|F^0_{\v k}|$ region.

For parameters closer to the experimental setup (see caption of Fig.\ref{realistic}), we show in Fig.\ref{fermi_surface} the Fermi pockets and band structure before exciton condensation happens. For the model used in the main text (parameter listed in the caption of Fig.\ref{pd}), there is an exact nesting of particle and hole pockets after gauging. In Fig.\ref{instability}(b) we overlay the Fermi surface of this model on top of the color map of the function $\Gamma_\v k = |F^0_\v k|^2-|F^1_\v k|^2$. This allows us to read off a critical density $n_{A(c)}\sim 0.25$ that is consistent with the transition between $s$ and $p+ip$ exciton order indicated by a red contour in Fig.\ref{instability}(a).

\section{Phase diagram for other values of $\Phi_A, \Phi_B$}
\begin{figure}[!h]
  \centering
  \includegraphics[width=.3\textwidth]{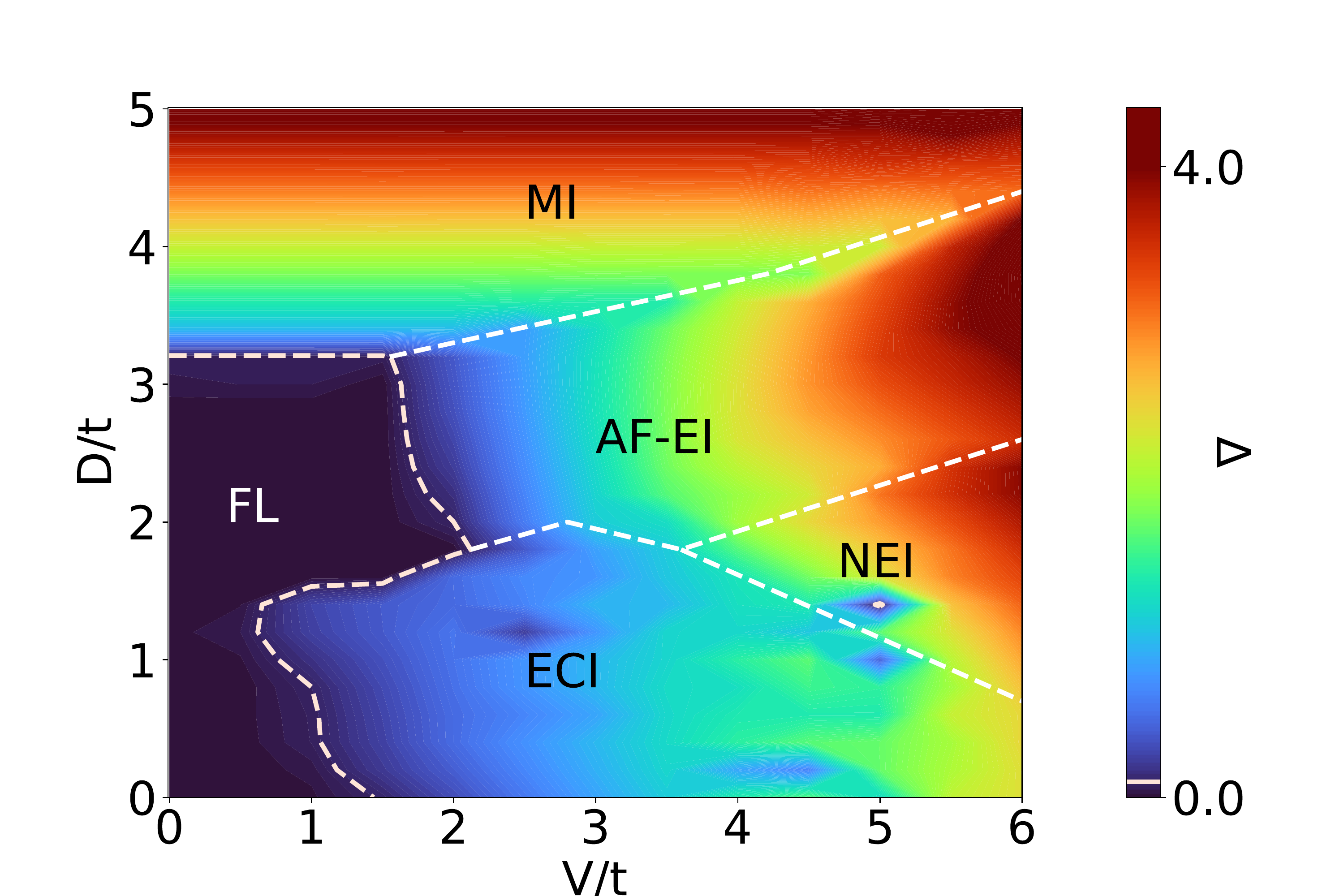}
  \includegraphics[width=.3\textwidth]{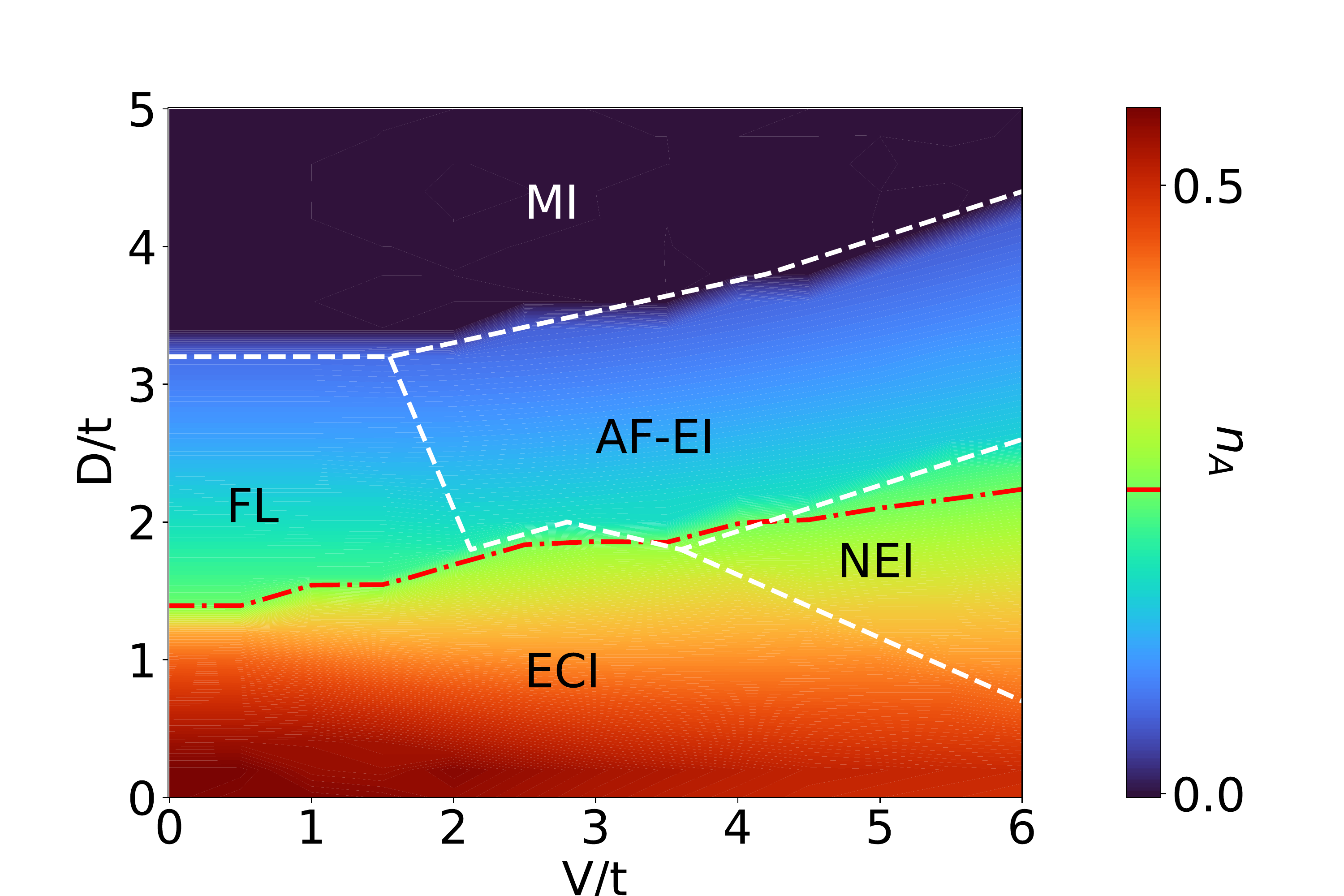}
  \caption{Schwinger boson mean field phase diagram in $D-V$ plane. (a) Colors show the single electron charge gap $\Delta_c$. We use the following parameters: $U=50$, $t_A=2$, $t_B=1$, $\phi_A = (-2/3+1/12)\pi$, $\phi_B = (2/3+1/12)\pi$, $V'=0.5V$. (b) shows the exciton density $n_A$. The dashed contour in red marks the $n_A=0.35$, which maps out the phase boundary between ECI and EI.}
  \label{realistic}
\end{figure}

Fig.\ref{realistic} we show the phase diagram obtained using parameters closer to the experimental system of AB-stacked WSe$_2$-MoTe$_2$ bilayer. The phase diagram Fig.\ref{realistic} is qualitatively similar to Fig.\ref{pd} in the main text. A $p+ip$ instability shows up for exciton density $n_A>0.35$.

\begin{figure}[!ht]
    \centering
    \includegraphics[width=0.3\linewidth]{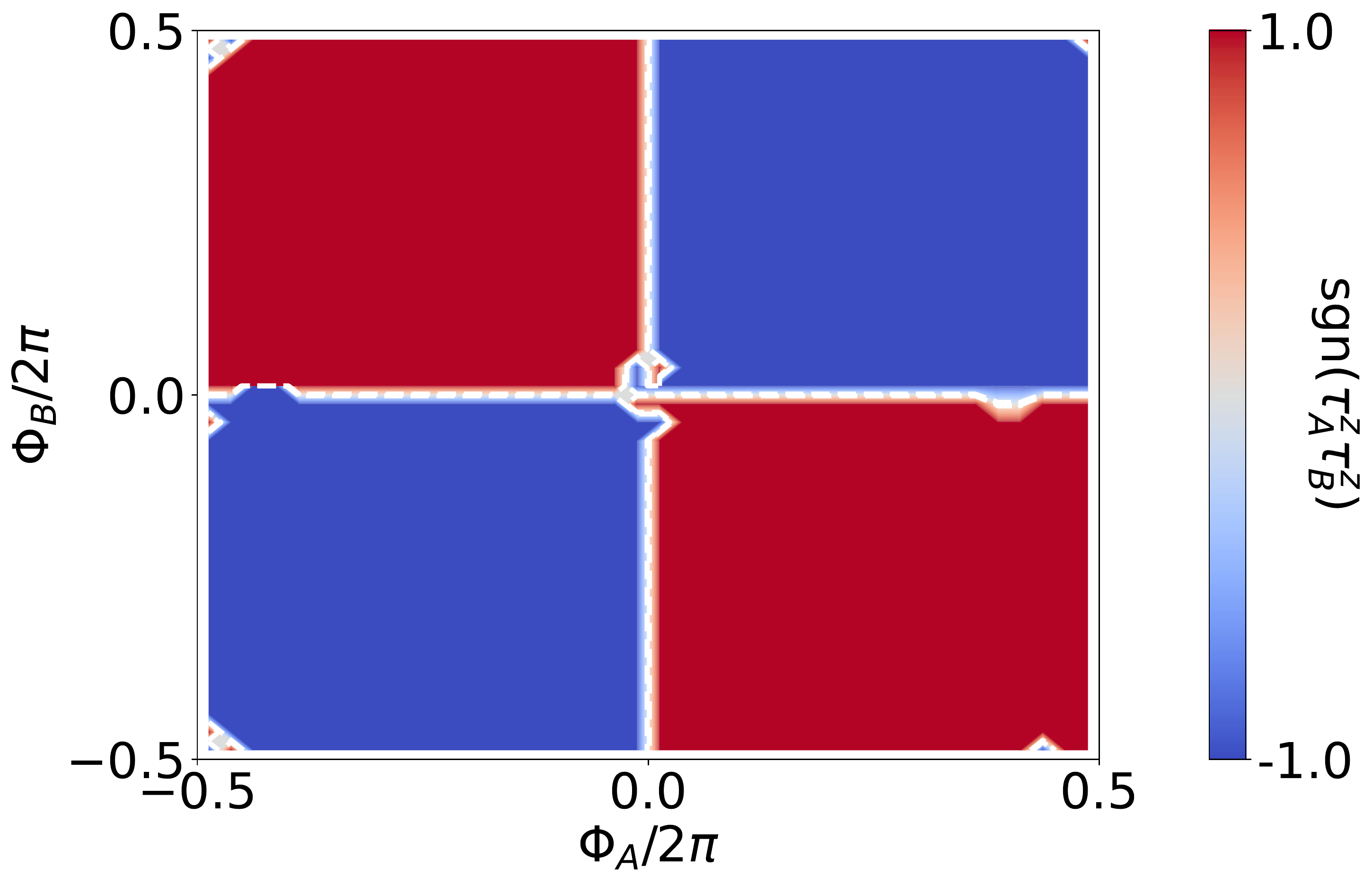}\quad
    \includegraphics[width=0.3\linewidth]{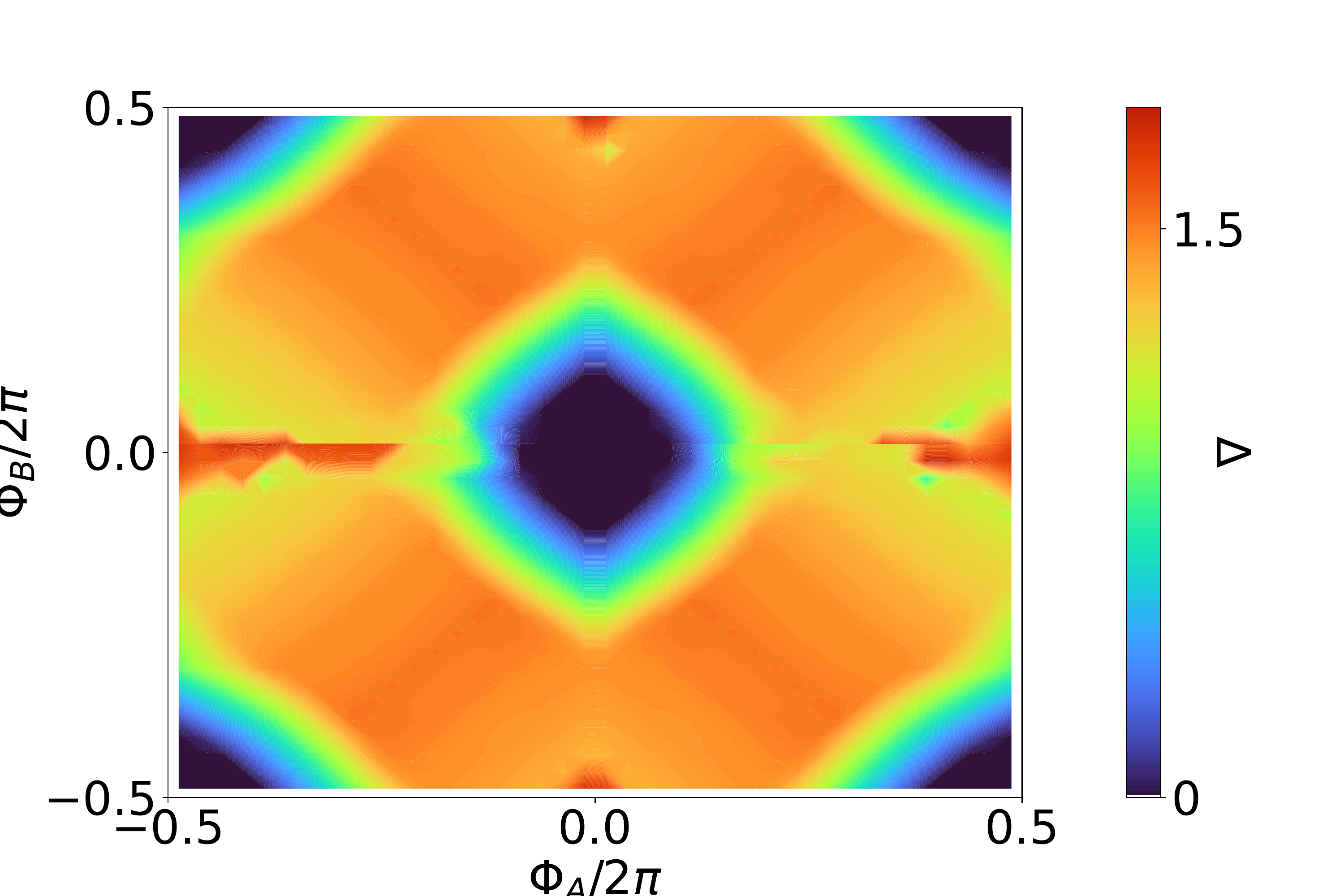}
    \caption{Hartree-Fock phase diagram for $V=4.0$, $D=0.5$. Other parameters are the same as in Fig.\ref{pd}. (a) Sign of $\langle \tau^z_A\tau^z_B \rangle$. (b) Charge gap. For time-reversal breaking phases, there is a two-fold ground state degeneracy. Our convention is to focus on the one with $\langle \tau^z_A\rangle \sim \langle S^z_A\rangle>0$. }
    \label{app_pd}
\end{figure}
In Fig.\ref{app_pd} we show AB valley alignment and charge gap for the same parameter space as in Fig.\ref{pd}. For $\Phi=n\pi$, the model has an enlarged SU(2)$\times$SU(2) symmetry. Therefore the valley-polarized and spin-polarized states are degenerate. A finite valley contrasting flux breaks the symmetry down to a time reversal, lifting this degeneracy and stabilizing one of the two phases on either side of $\Phi=0$. For this particular choice of $V/t=4$, we find Fermi liquid in the vicinity of $\Phi_A=\Phi_B=\pi$ and exciton Chern insulators in a wide range of $\Phi$. As we will find in Fig.\ref{app_SU2} of Appendix.\ref{APP_SU2}, with a stronger $V$ we get a nematic EI at $\Phi\sim0\;(\mathrm{mod} \pi)$.

\section{kinetic magnetism and canted spin order}
\begin{figure}
    \centering
    \includegraphics[width=0.5\linewidth]{{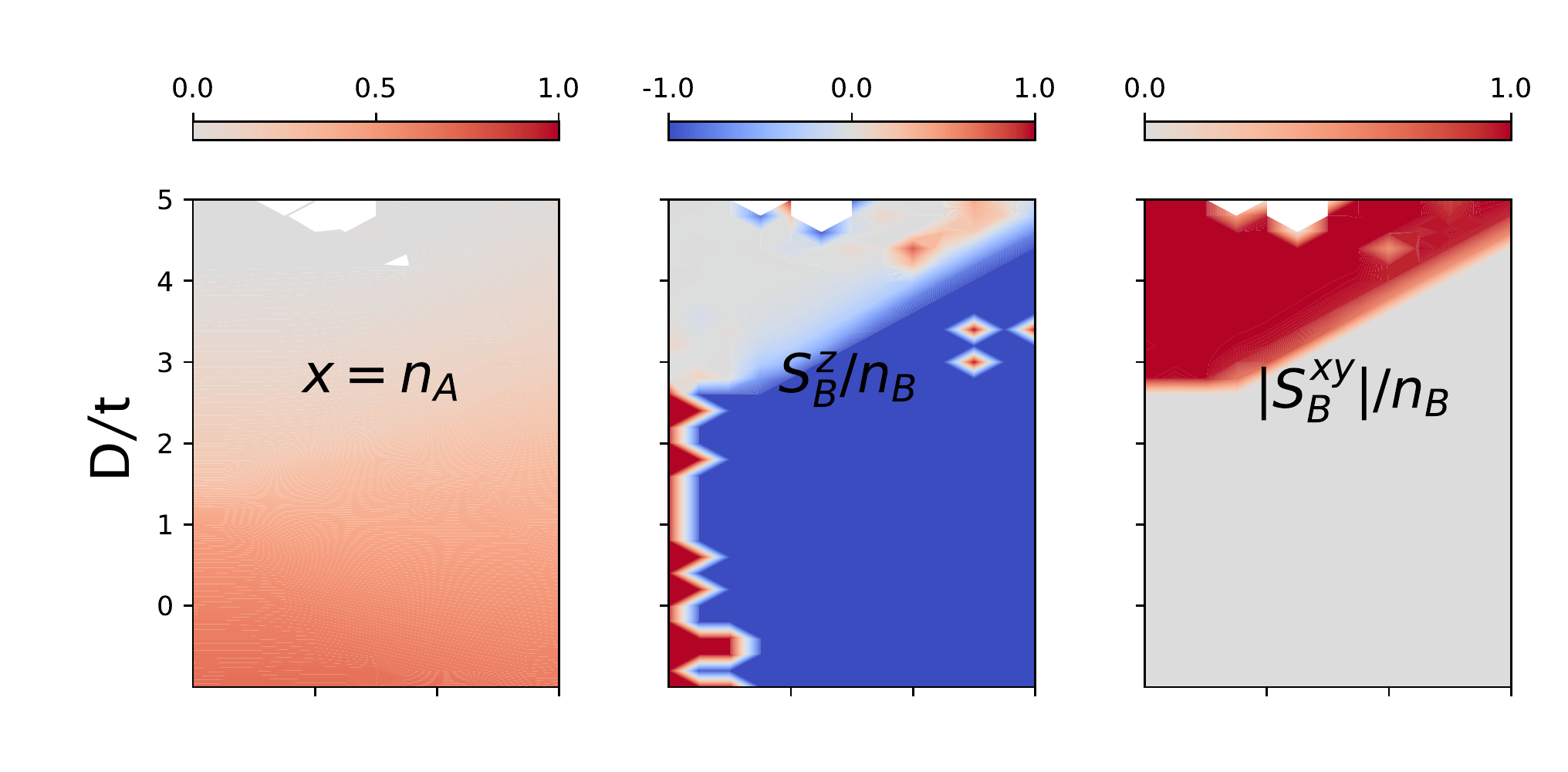}}
    \includegraphics[width=0.35\linewidth]{{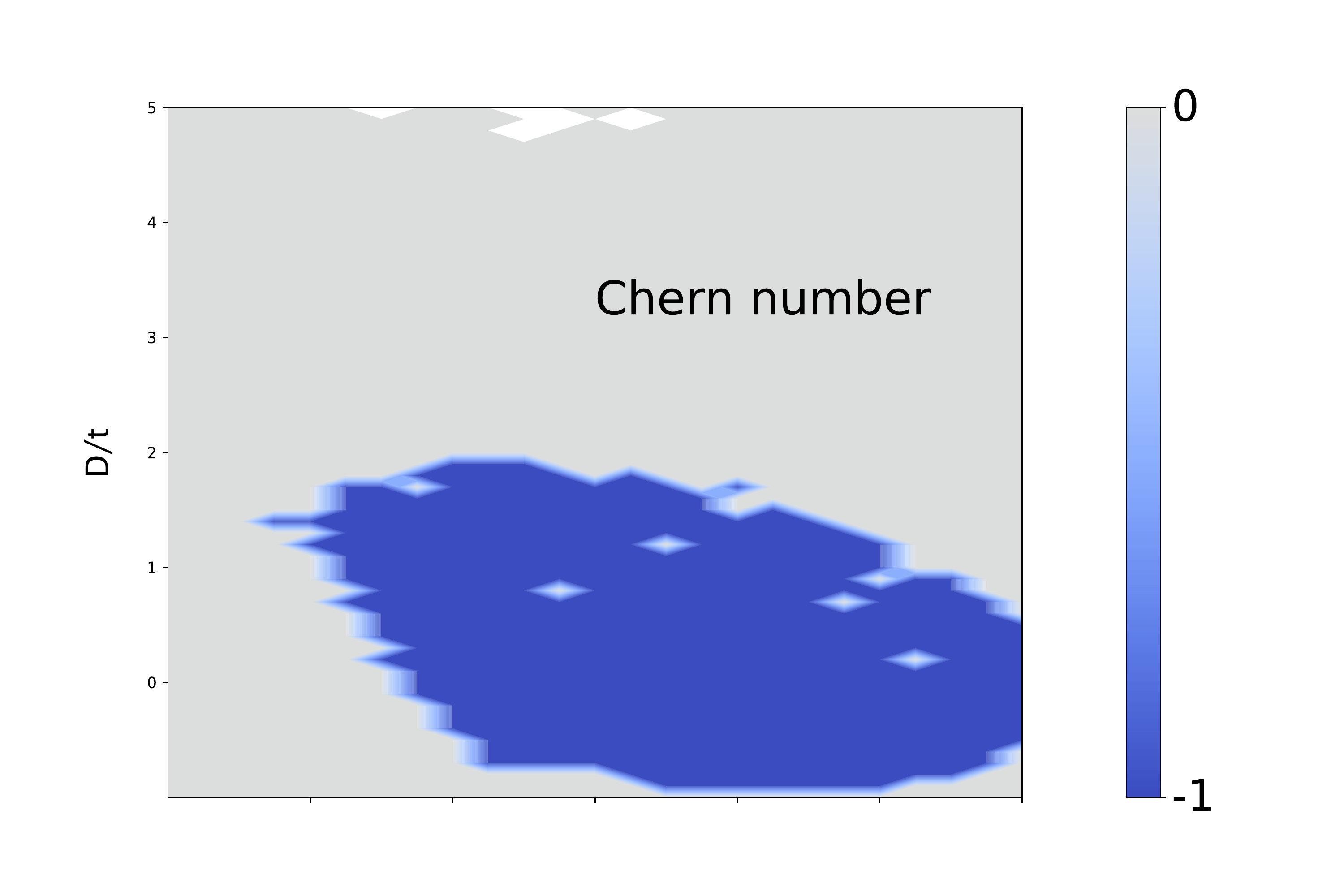}}\\
    \includegraphics[width=0.5\linewidth]{{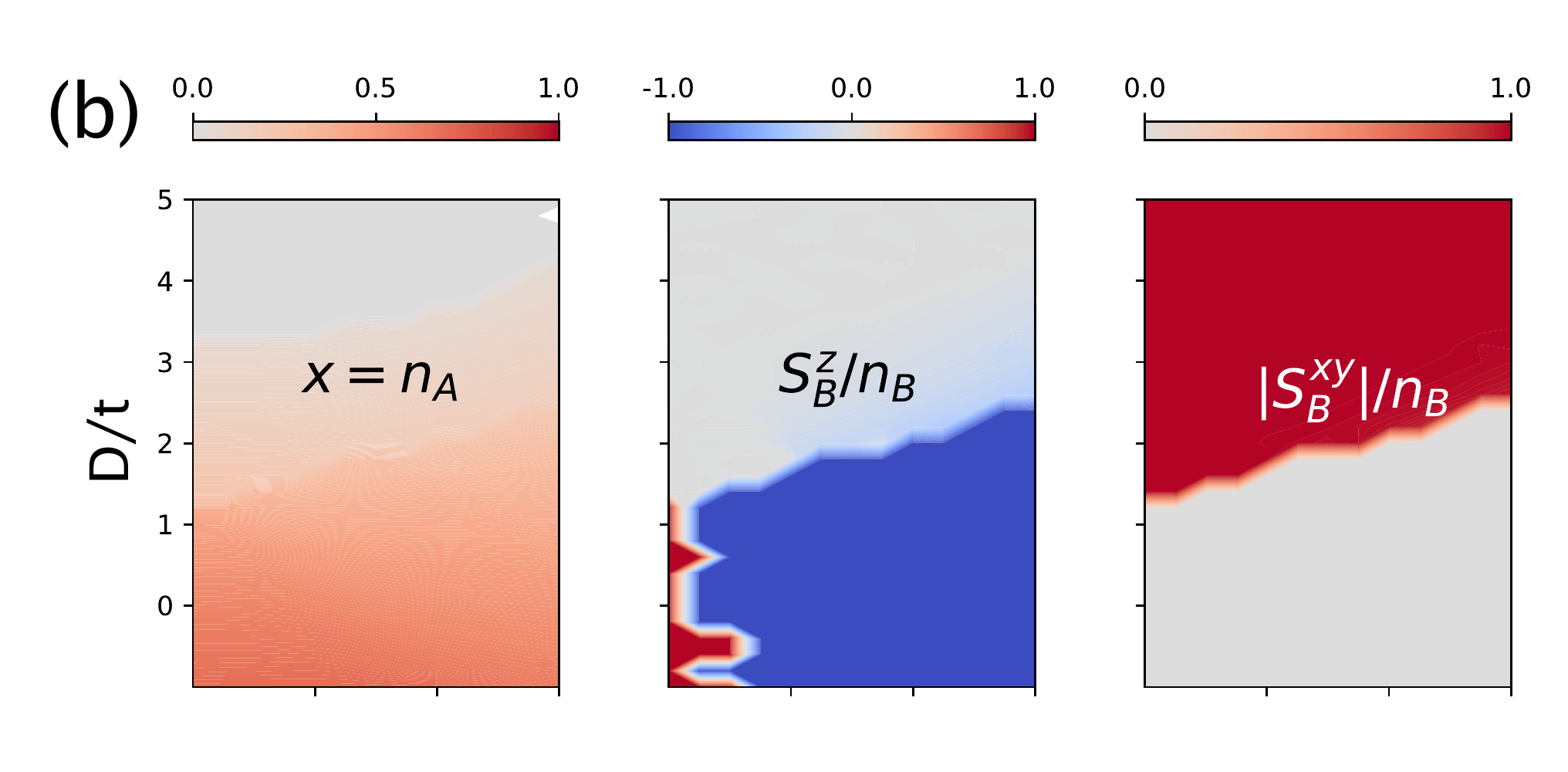}}
    \includegraphics[width=0.35\linewidth]{{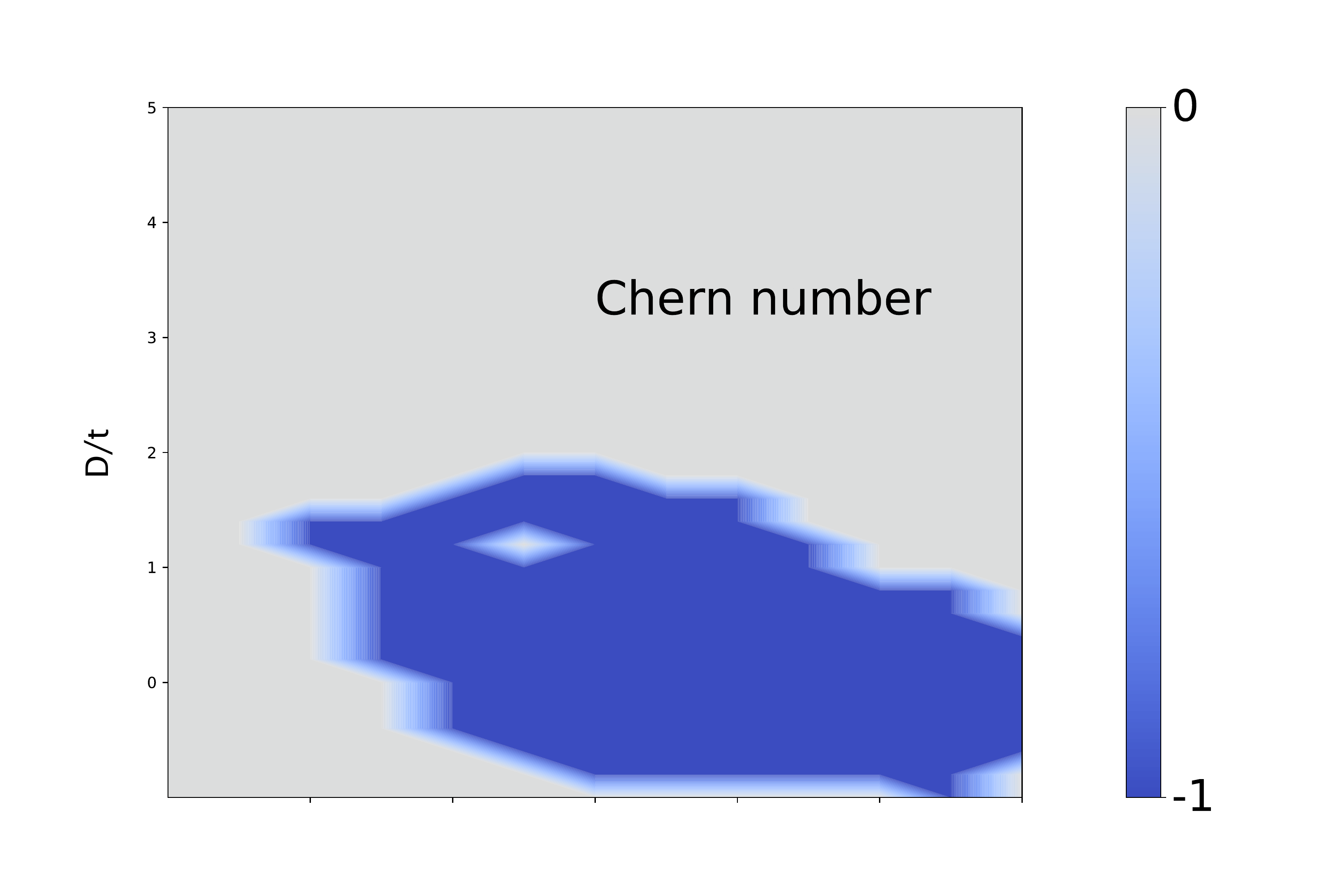}}\\
    \includegraphics[width=0.5\linewidth]{{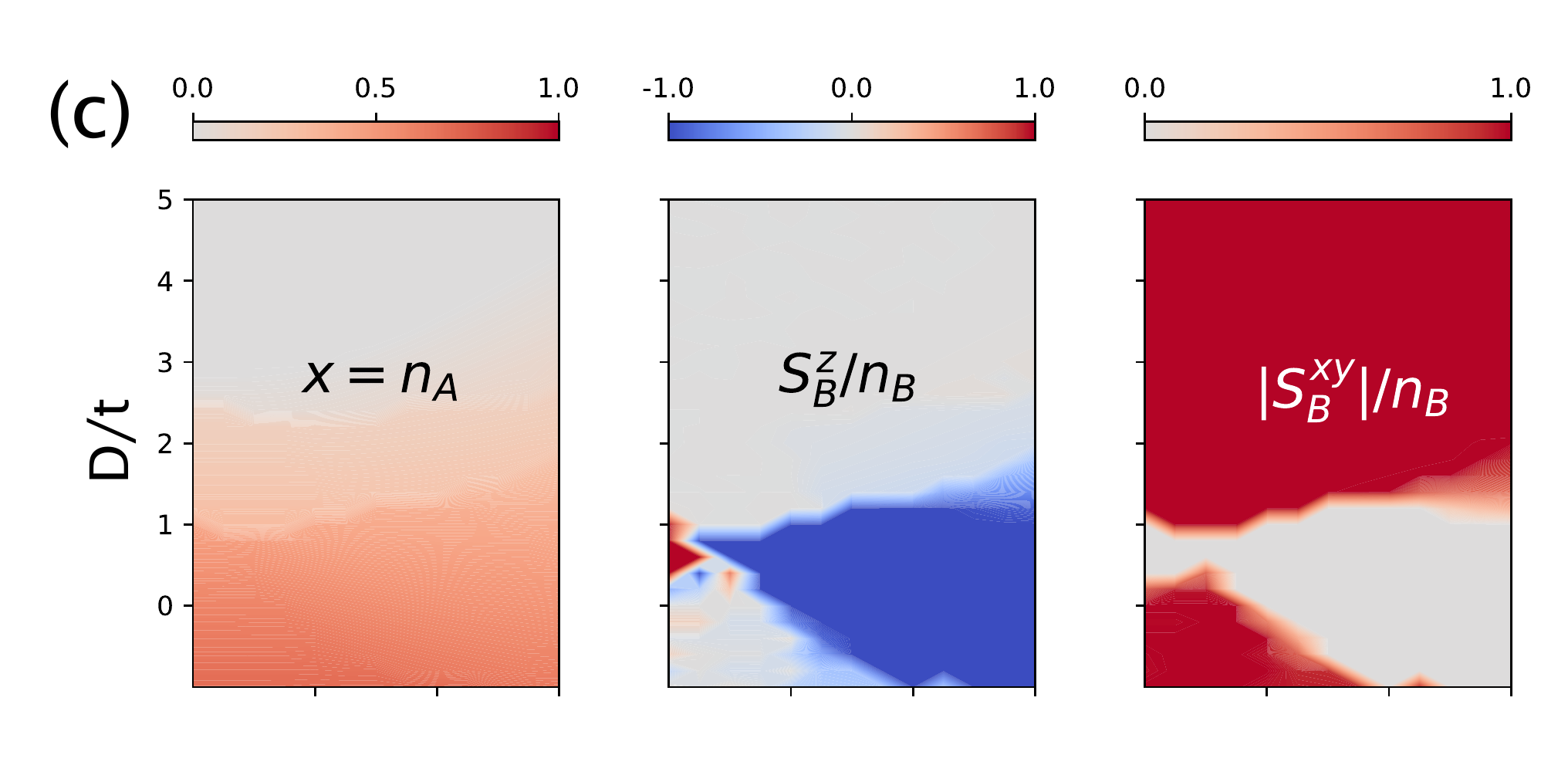}}
    \includegraphics[width=0.35\linewidth]{{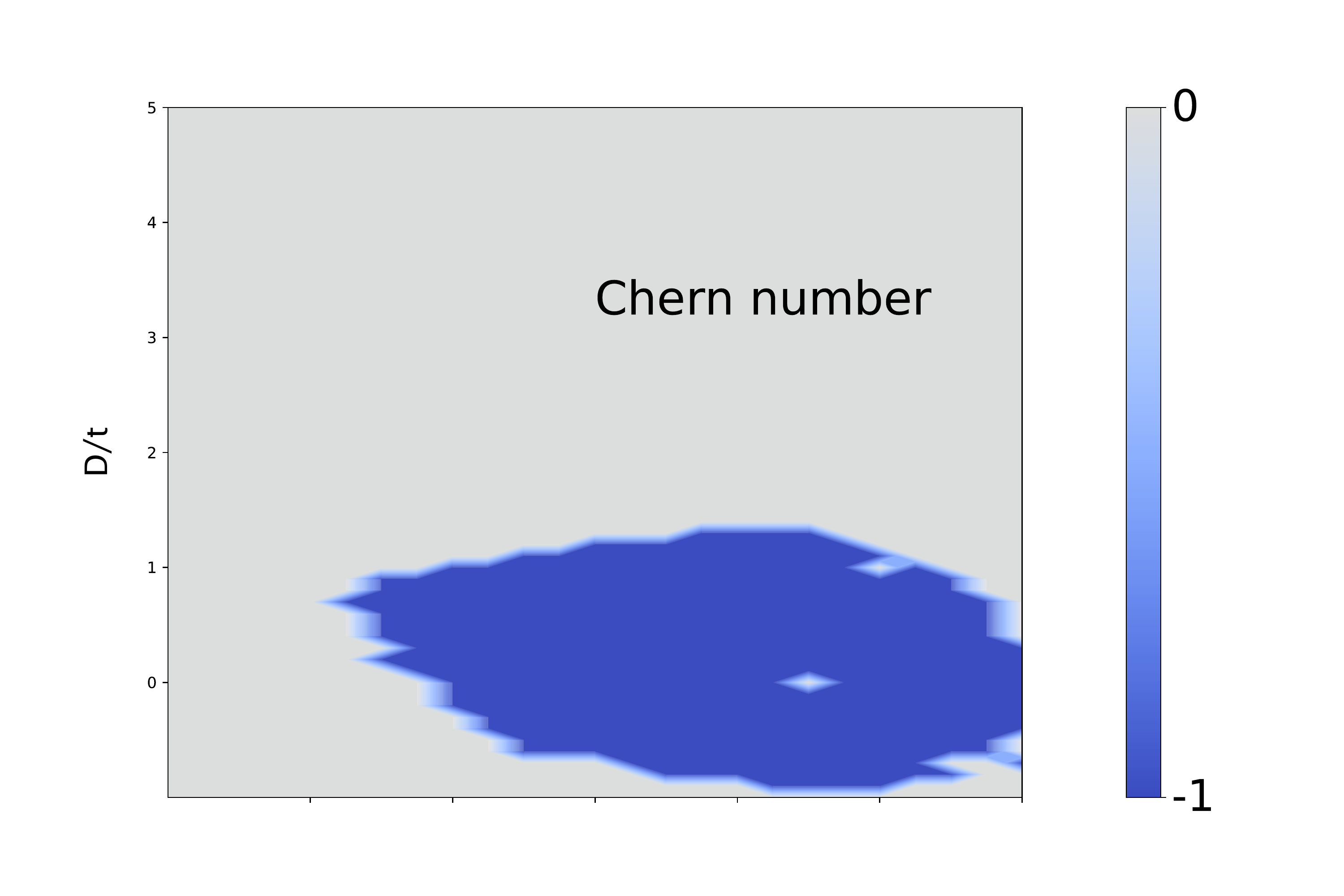}}\\
    \includegraphics[width=0.5\linewidth]{{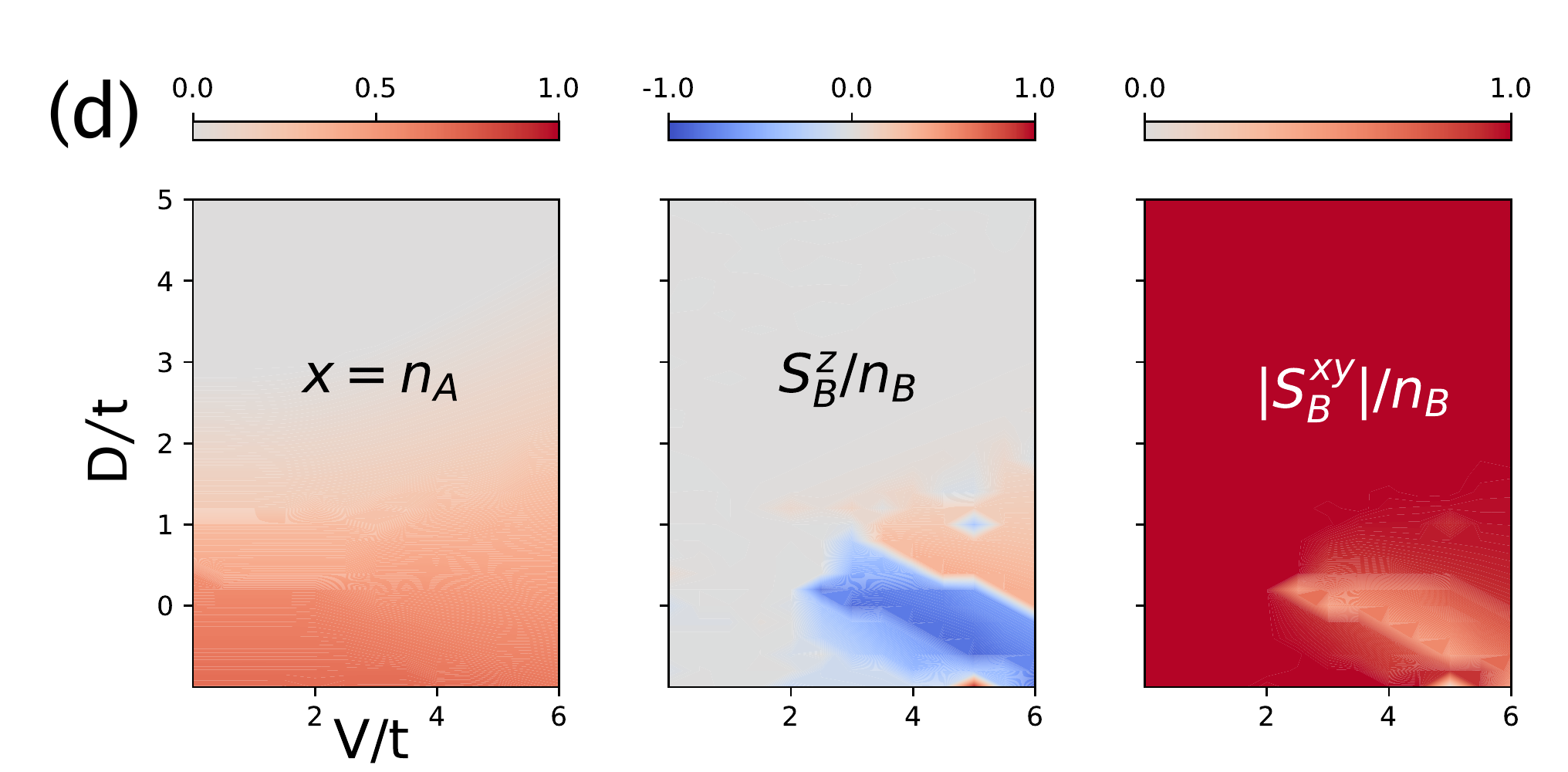}}
    \includegraphics[width=0.35\linewidth]{{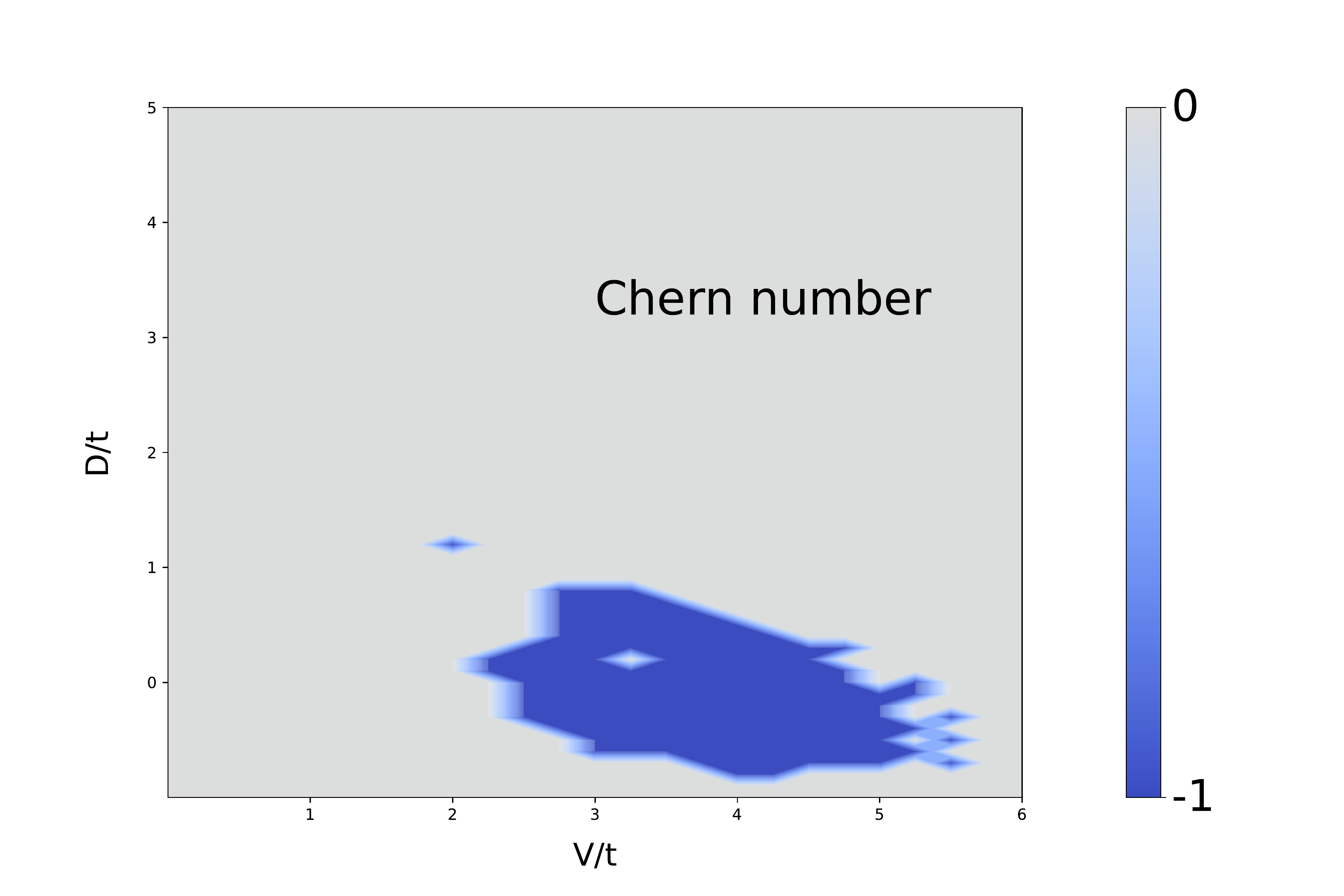}}\\
    \caption{Schwinger boson mean field phase diagram for $t_A=t_B=1$, $\phi_A=\pi/4$, $\phi_B=\pi/12$, $V'/V=0.5$. (a)(b)(c)(d)(e) for $J=0.001,0.02,0.08$ and $0.2$, respectively. }
    \label{tJpd2}
\end{figure}

In Fig.\ref{tJpd2} we show the evolution of the $D-V$ phase diagram calculated for the t-J model in Eq.\ref{tJham}. Focusing on the evolution spin order on the B sublattice as $J$ increases, we find the region with finite in-plane magnetic order expands. At $J=0.2$, which corresponds to $U/t\sim 20$, the spin order for QAH phase stops being fully polarized to $z$ direction and acquires a $120^\circ$ AFM component in $s_{xy}$ plane. Namely, in the B layer, we find a $120^\circ$ in-plane AFM order with canting along the $z$ axis.

 Obviously, there is a competition between the super-exchange $J$ term favoring the $120^\circ$ in-plane magnetic order and kinetic term $t_{BB}$ that stabilizes intra-layer valley polarization. This is reminiscent of Nagaoka ferromagnetism predicted for the doped square lattice Hubbard model at an extremely strong interaction limit, where kinetic energy dominates over super-exchange and leads to ferromagnetism. Interestingly, for the SU(2) symmetric t-J model on a triangular lattice, due to kinetic frustration, the kinetic energy also favors $120^\circ$ in-plane magnetic order, hence the $t$ and $J$ term collaborates and there is no magnetic transition as $t/J$ varies. In our case, however, the scenario is complicated by the absence of SU(2) symmetry and by having two sublattices. Eventually, we still observe kinetic ferromagnetism.
 
This canted spin order is consistent with recent experimental observation\cite{tao2022valley}. Here we are simply demonstrating the possibility for the exciton condensate to allow a canted order while retaining quantize Hall conductance. In the meantime, we believe the evidence provided by Ref.\cite{tao2022valley} alone is not strong enough to rule out other explanations such as phase separation caused by the disorder.
\section{$U(2)\times U(2)$ symmetric point with $\Phi_A=\Phi_B=0$}
\label{APP_SU2}
\begin{figure}[!ht]
    \centering
    \includegraphics[width=0.23\linewidth]{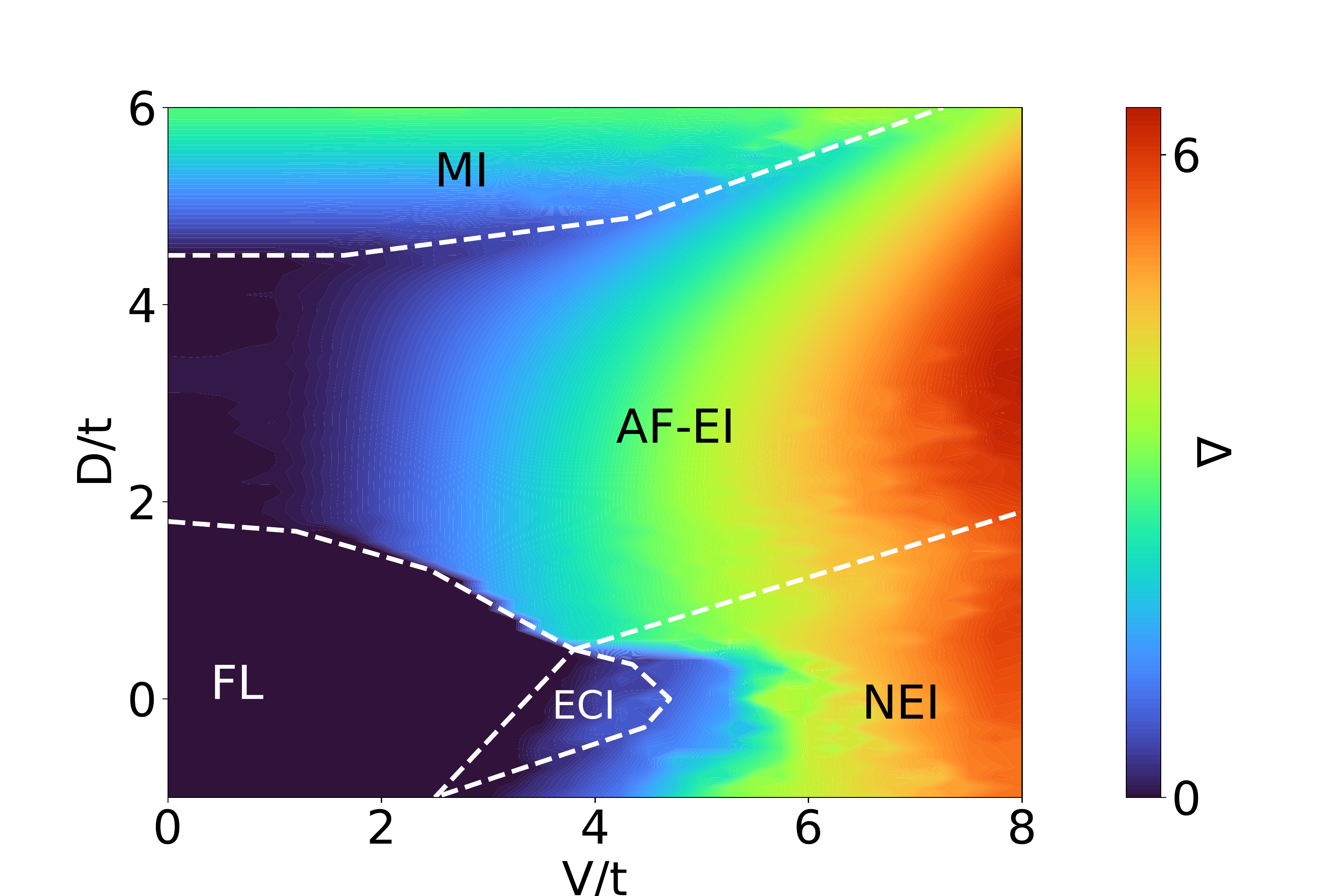}
    \includegraphics[width=0.23\linewidth]{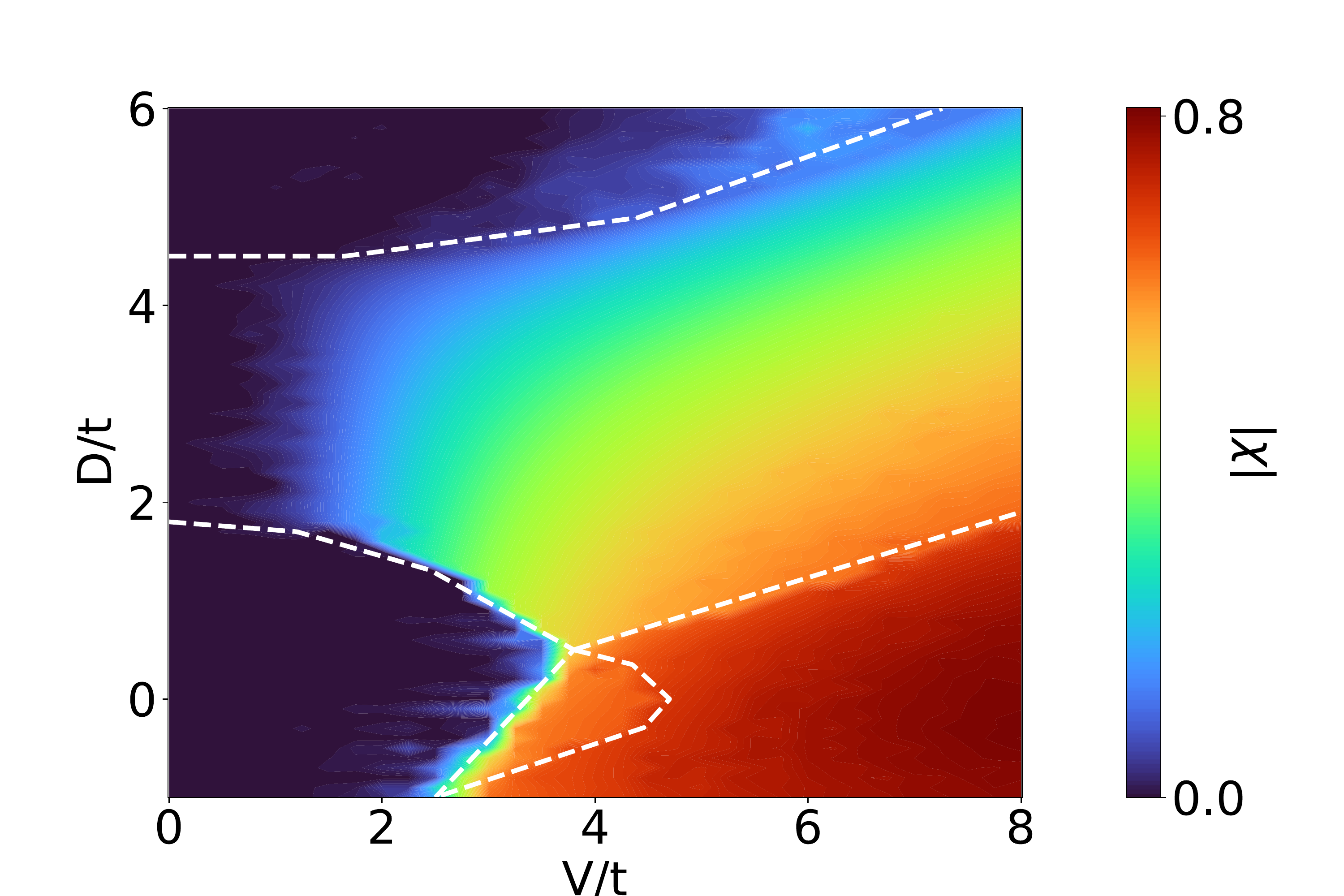} 
    \includegraphics[width=0.23\linewidth]{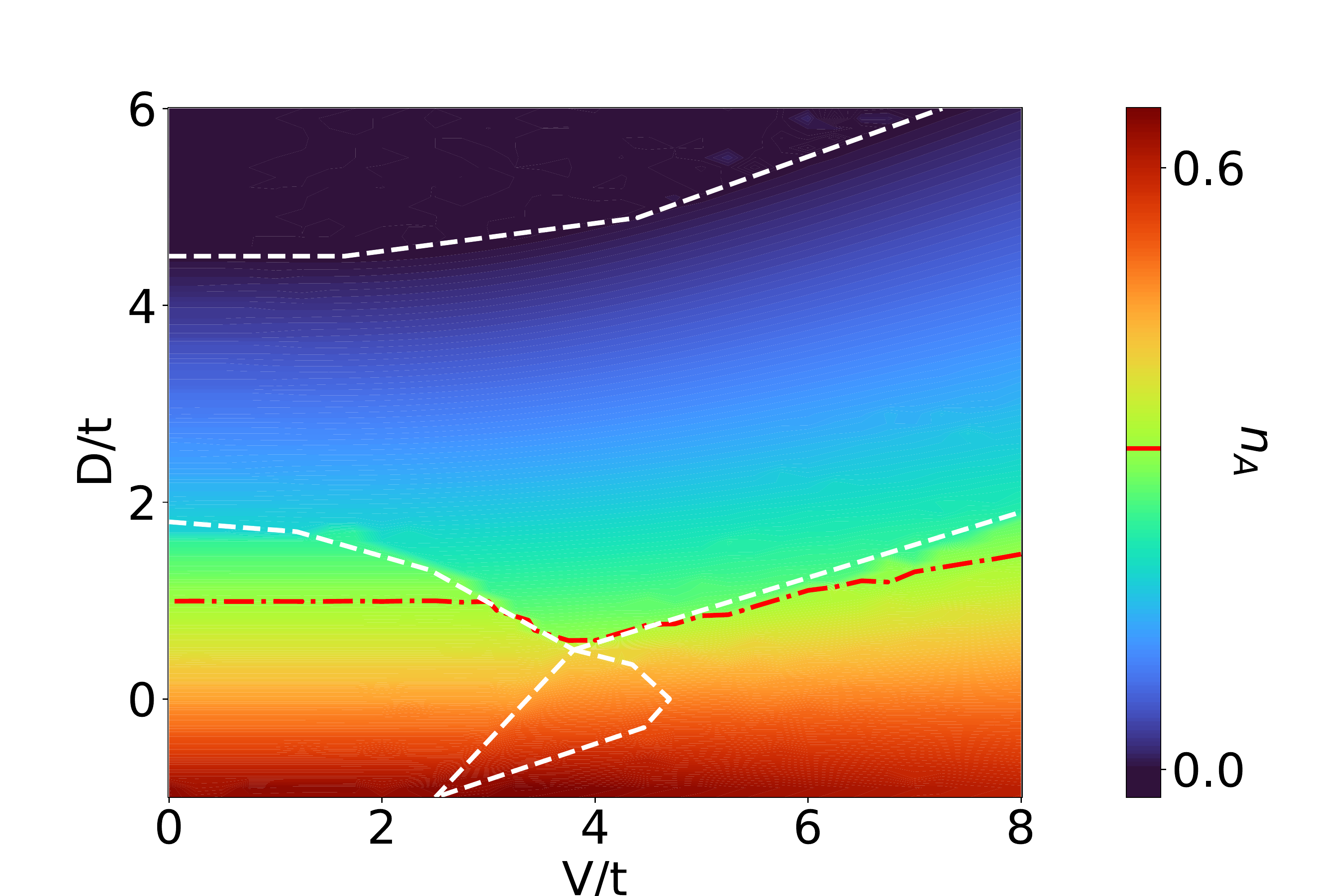}
    \includegraphics[width=0.23\linewidth]{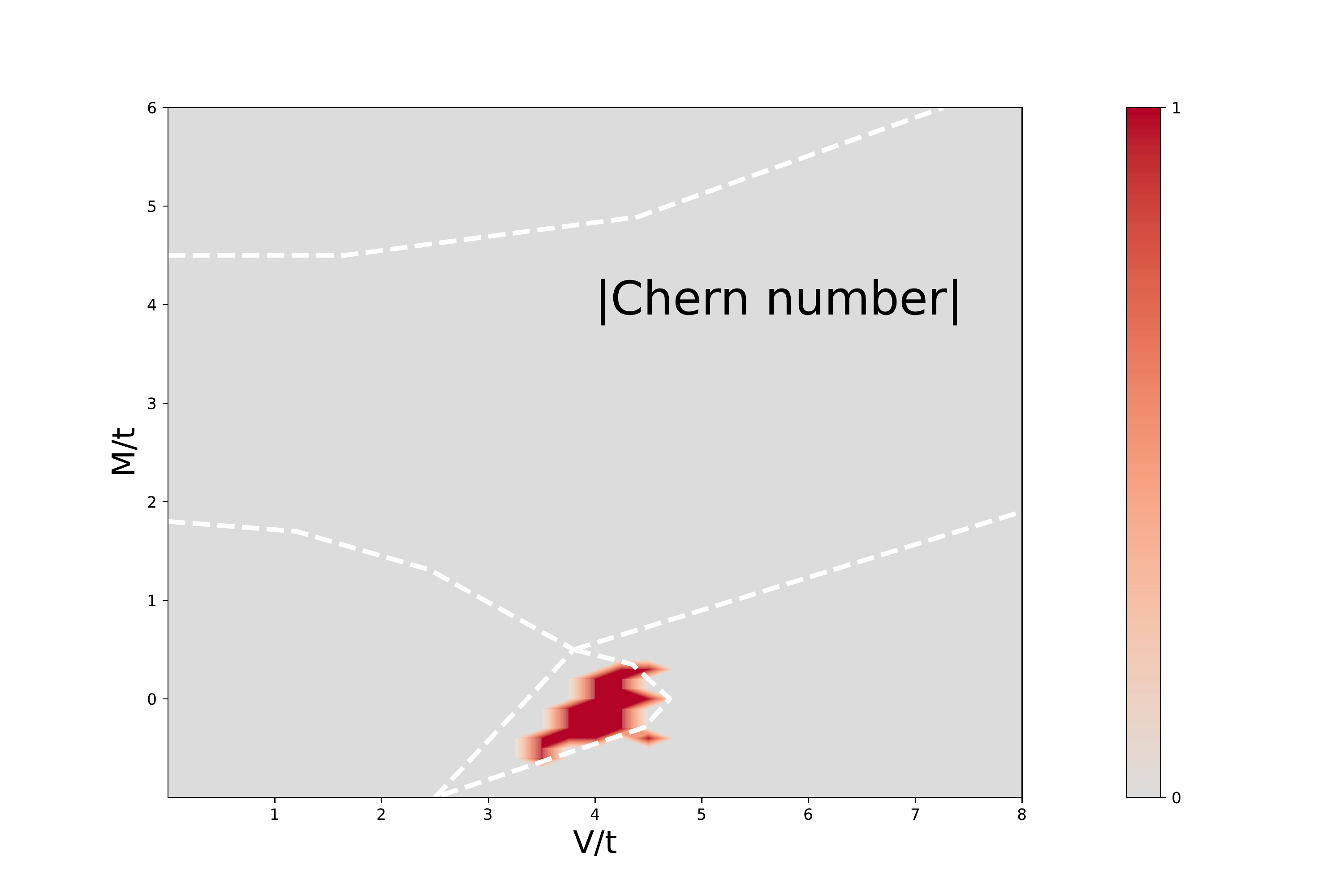}
    \caption{Schwinger boson mean field phase diagram in $D-V$ plane without valley contrasting flux. $t_{AB}=0$ and $t_A=t_B=1$. (a) Charge gap. (b) The magnitude of exciton order. (c) Exciton density. (d) Chern number.}
    \label{app_SU2}
\end{figure}

At $\Phi_A=\Phi_B=0$, there is a valley SU(2) symmetry in each layer. In Fig.\ref{app_SU2} we present the $D-V$ phase diagram for this case. We show in Fig.\ref{app_SU2}(b) the magnitude of order parameters. The $p+ip$ order now appears only in a narrow region around $V=4.0$, $D=0$.

We start our discussion from the low exciton density regime $n_A\sim0$. As before, we find the AF-EI phase with $s$ wave exciton order at small exciton density. But now the AF-EI phase is more robust and the NEI phase with p wave exciton order  starts only when $n_A> 0.35$. The reason is the following. With the $\tau_z$ polarized in both layers,  there are two degenerate holon pockets from B layer around $\v K_M$ and $\v K'_M$ points, and one electron pocket from A layer at $\v \Gamma_M$. In this case exciton order only hybridizes one of the hole pockets with the electron pocket and  is not effective to gap out the electron-hole pocket. This penalizes the energy of the FM phase. On the other hand, with the $120^\circ$ AF order in the B layer, there is only one hole pocket at $\Gamma_M$ and an s wave exciton order effectively gaps out the electron pocket and the hole pocket.  So the AF EI phase has better energy and occupies larger phase regions when $\Phi_A=\Phi_B=0$.




\begin{figure}[!ht]
    \includegraphics[width=0.3\linewidth]{{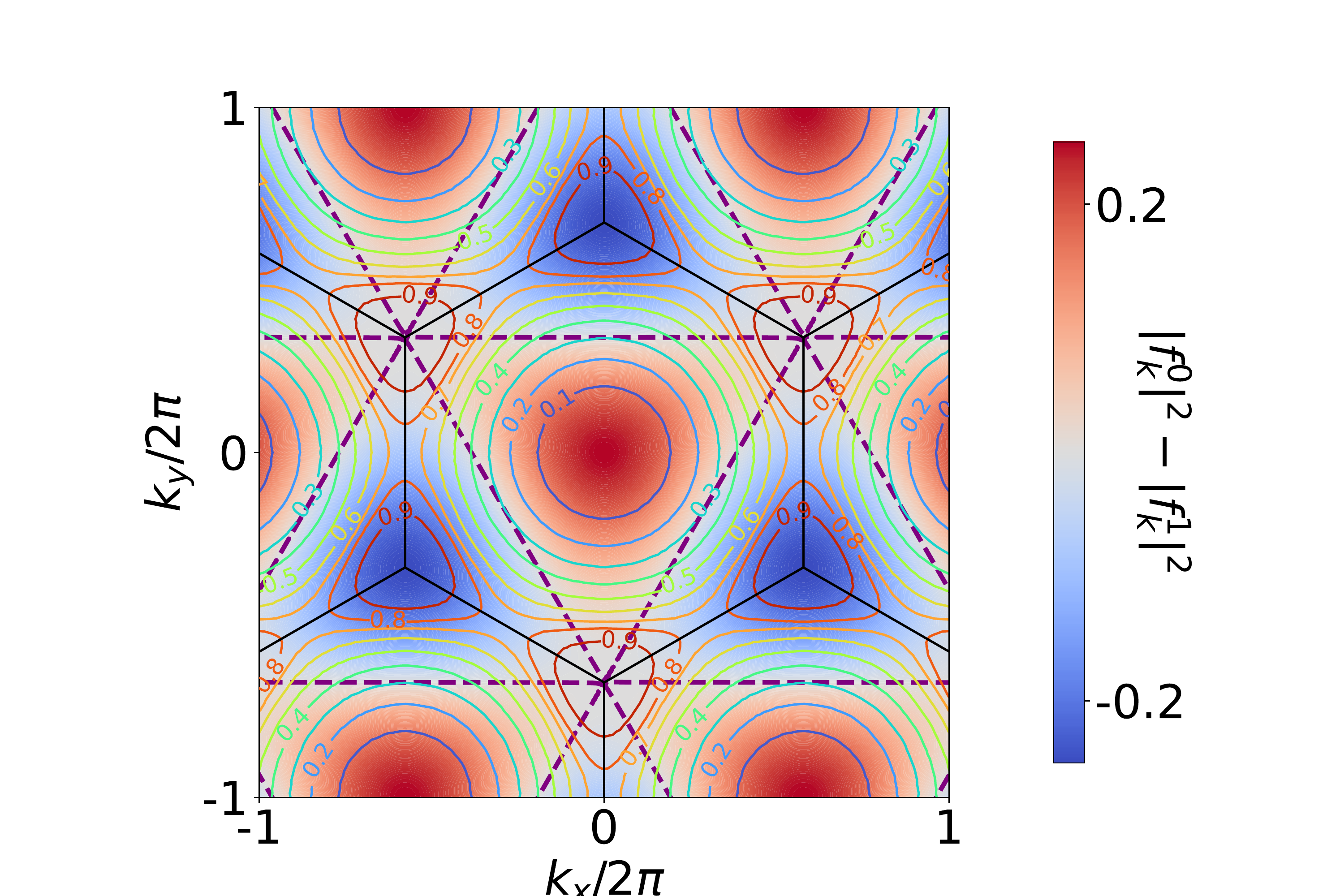}}
    \includegraphics[width=0.3\linewidth]{{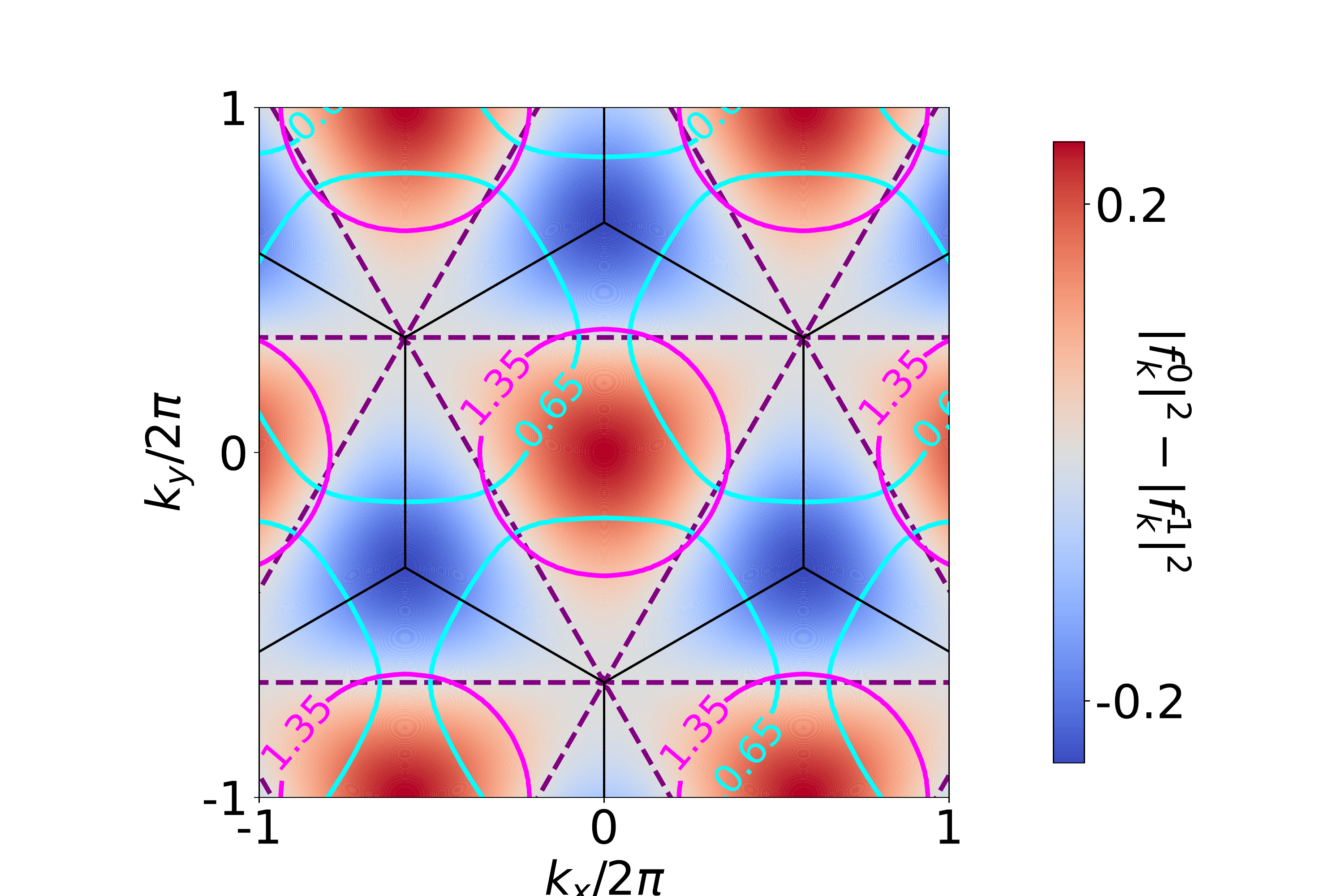}}
    \caption{(a) Form factor difference between $l=1$ and $l=0$ channels. The overlaid contours are Fermi surfaces at various fillings for a triangular lattice tight-binding model with no hopping phases. This plot helps determine pairing instability when the B layer is in $120^\circ$ magnetic order so that the holon and particle bands are perfectly nested. Numbers labeling these contours correspond to the $n_A$ associated with the two Fermi surfaces. (b) Gauged Fermi surfaces for $n_A=0.35$. The gauge choice is such that the two pockets are both centered at $\v \Gamma_M$. At this density, the intersections between two Fermi surfaces enter the region where $p+ip$ pairing is stronger than that of $s$ wave.}
    \label{instability2}
\end{figure}


On the other hand, consider the competing state with ferromagnetic order in B layer. In contrast to the AFM state, with an FM background, there is no $\pi$ Berry flux contributed by the spin configuration. As a result, the holon band gets inverted, and the Fermi surface now sits in the vicinity of $\v K_M$ and $\v K'_M$, where the $p$ wave form factors $F^{\pm1}_\v k$ dominate over that of $s$ wave (see orange contour in Fig.\ref{instability2}(a), or blue contour in Fig.\ref{instability2}(b) for the holon pocket after gauging). So we get the ECI phase with $\tau_z$ FM when $n_A>0.35$. However, the ECI phase is quite fragile and unstable to the nematic EI (NEI) phase for this parameter. To get a robust Chern insulator, a small valley contrasting flux $\Phi_A$ or $\Phi_B$ is needed.


\section{Competition between band Chern insulator and excitonic insulator with $t_{AB}$}
\begin{figure}[!ht]
    \centering
    \includegraphics[width=0.3\linewidth]{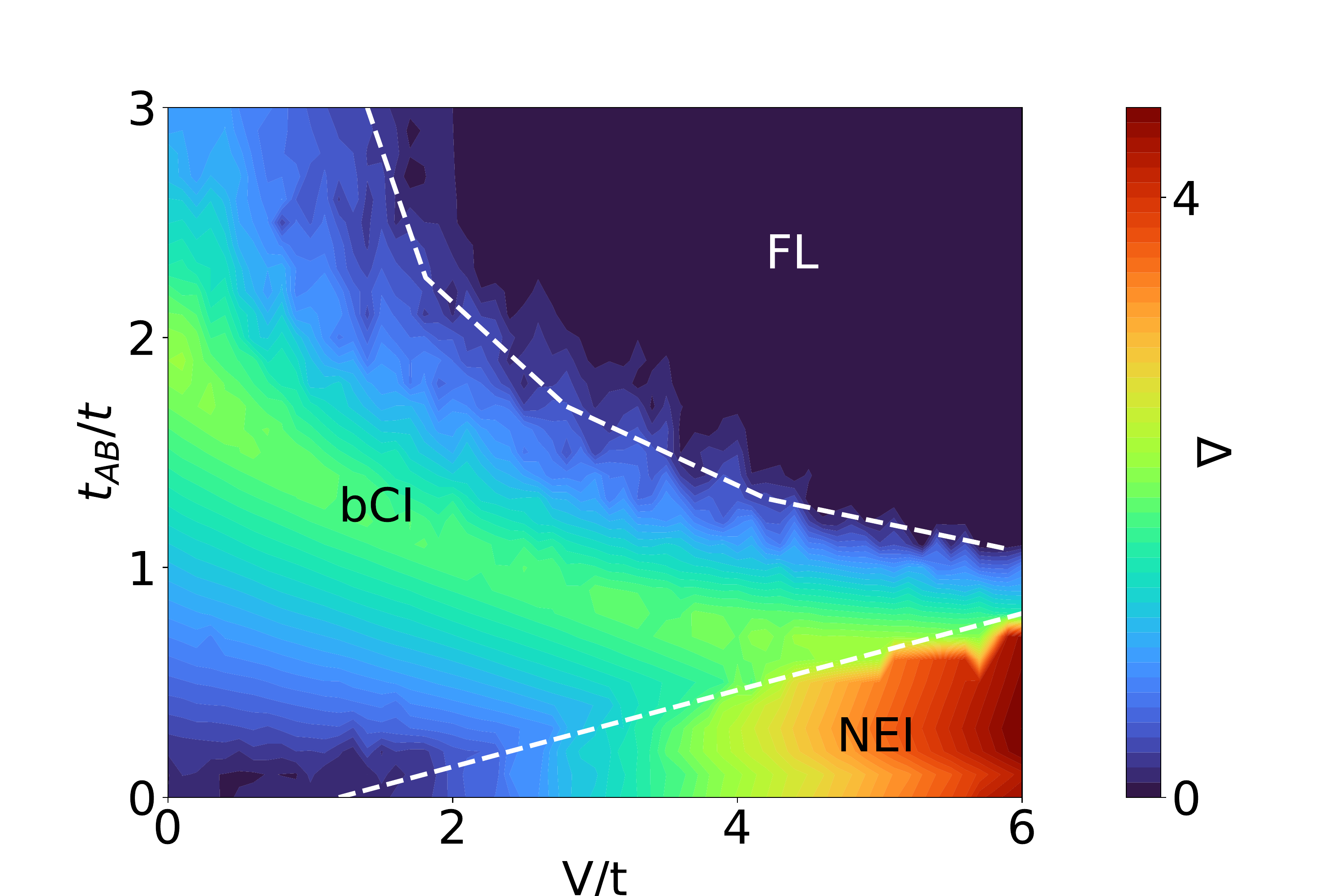}
    \includegraphics[width=0.3\linewidth]{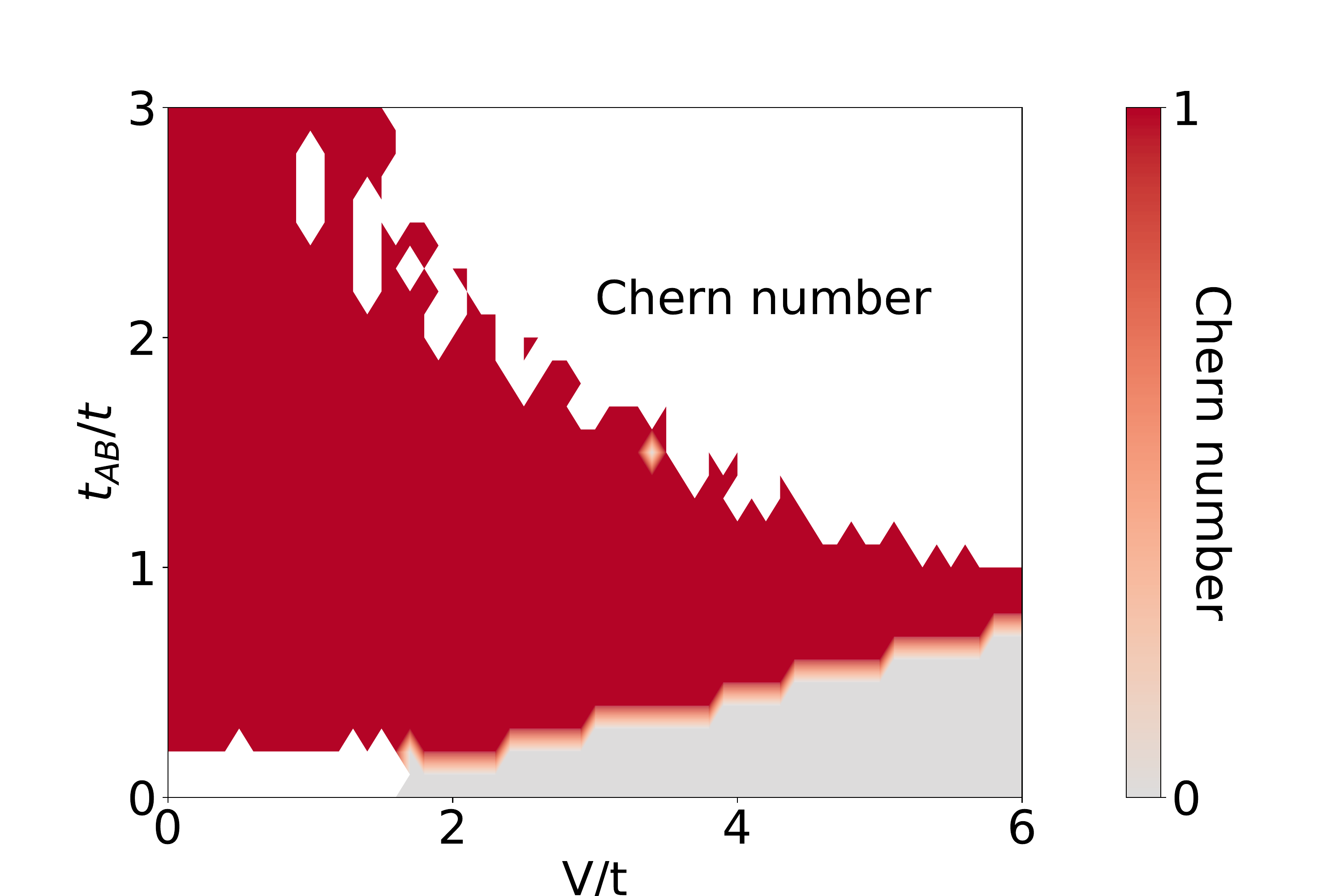} 
    \includegraphics[width=0.3\linewidth]{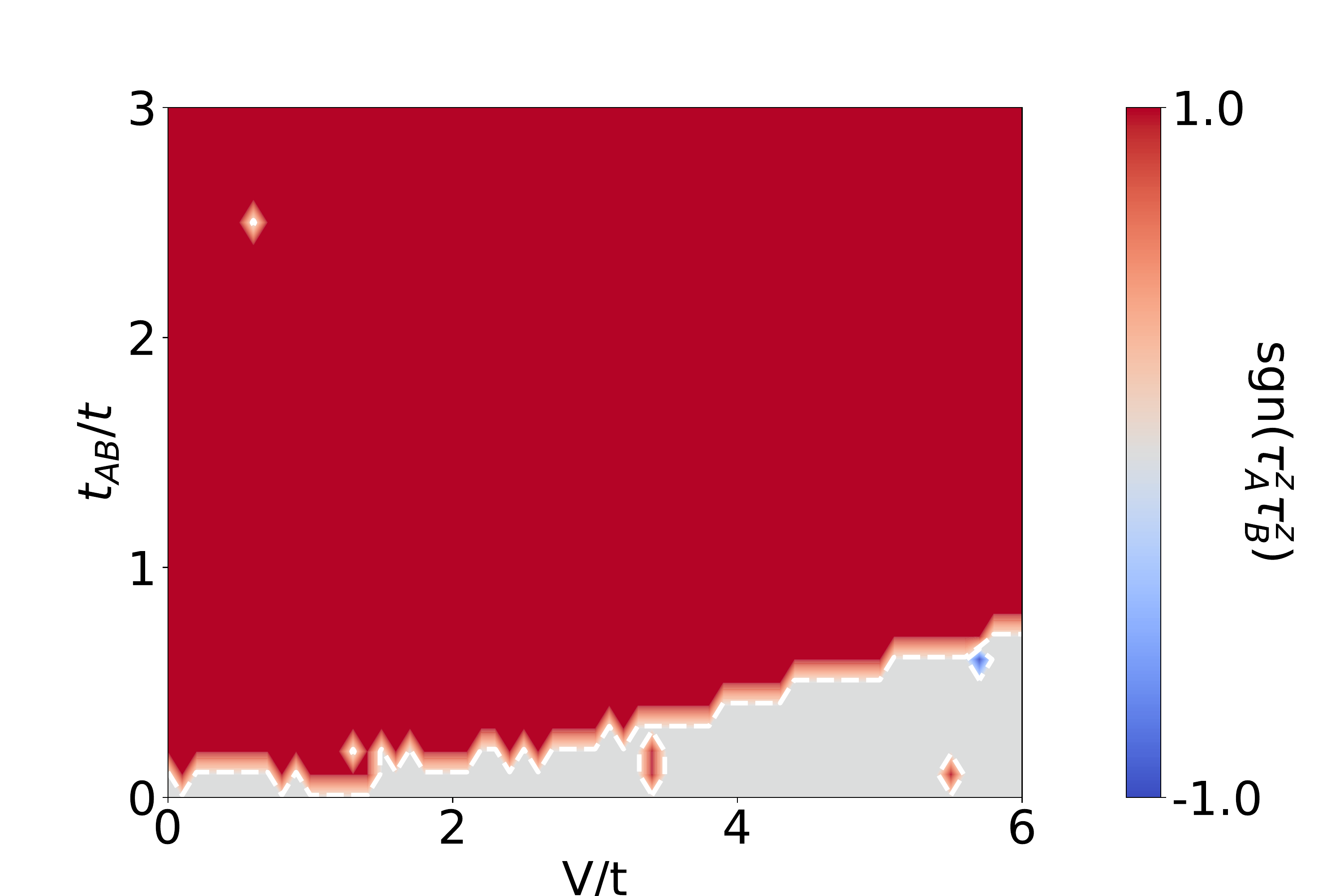}
    \caption{Hartree-Fock phase diagram in $t_{AB}-V$ plane. (a) Charge gap. (b) Chern number. (c) The sign of $\tau^z_A \tau^z_B$ when it is nonzero. This phase diagram is calculated using the same set of parameters as in Ref\onlinecite{devakul2022quantum}, $t_A=t_B=1$, $\Phi_A=2\pi$, $\Phi_B=0$, and $D=3$.}
    \label{app_CI2ECI}
\end{figure}

With a strong enough $t_{AB}$ to hybridize the two layers, a band Chern insulator (bCI) with canted $120^\circ$ antiferromagnetic order was found in a previous study\cite{devakul2022quantum}. Note here band Chern insulator may not be a good name because the phase apparently still requires strong intra-layer repulsion to generate the AF order. Here we mainly focus on the effective Hamiltonian $H_f$ of the charge part after the magnetic order is fixed.  Then the phase in Ref.~\onlinecite{devakul2022quantum} is realized in the large $t_{AB}$ but small $V$ limit. The hopping of $f$ is decided entirely by the original single electron hopping $t_{AB}$ and we call it band insulator to distinguish it from the excitonic insulator in this paper.    In Fig.\ref{app_CI2ECI} we investigate the transition from this band Chern insulator to our interaction-driven exciton insulators by turning on the repulsion $V$ and $V'$. In presence of a $t_{AB}$ term, the U(1) symmetry associated with layer charge conservation is explicitly broken. 

The effective hopping of the bCI phase is actually equivalent to the $p+ip$ exciton order in our ECI phase.  To simplify the discussion, we follow the Schwinger boson mean field theory, where spin magnetic order contributes to the charge part through the Berry phase. The mean field Hamiltonian for the charge part is again purely a spinless model:
\begin{align}
    H_c^{MF} =&- \sum_{\langle i,j \rangle_{A,B}} t_{AB} (F_i^{a\dagger} F^a_j) f^\dagger_i f_j + h.c.   \\ \nonumber&- \sum_{a=A,B} \sum_{\langle \langle ij \rangle \rangle_a} t_a (F_i^{a\dagger} e^{i \phi_{ij;a} \tau_z} F^a_j) f^\dagger_i f_j + h.c.   \\ \nonumber
    &- V\sum_{\langle ij \rangle} \chi_{ji} f^\dagger_i f_j + D\sum_{i \in A}  n_i -D \sum_{j \in B} n_j
\end{align}
And we will define the effective charge-hopping phase as
\be
    t^f_{ij;a}
    =t_a (F_i^{a\dagger} e^{i \phi_{ij;a} \tau_z} F^a_j)
    =\tilde{t}_a e^{i\tilde{\phi}_a}
\ee
along with effective flux $\tilde{\Phi}_a=3\tilde{\phi}_a$. $\chi_{ji}=\langle f^\dagger_i f_j \rangle$ gives the exciton order. The bCI has a full $\tau_z$ polarization in layer A. In layer B it has a $120^\circ$ order in $\tau_{xy}$ plane, along with a uniform $\tau^z$ component, which satisfies $\langle \tau^z_A\tau^z_B\rangle>0$. The Schwinger boson mean field ansatz is
\begin{align}
\label{gauge}
    F^A_i &= (1,0)^T\\
    F^B_j &= e^{\frac{i}{2}(\sigma^z+\mathit{\mathbb{1}})\v K \cdot \v r_j}(\cos{\frac{\theta}2},\sin{\frac{\theta}2})^T
\end{align}
where $\theta\in(0,\pi/2)$ is the angle of canting.

There is a competition between $t_{AB}$ and $V$. At $V=0$ and before layer hybridization, both the electron pocket of the A layer and the holon pocket of the B layer is centered at $\v K_M$ under the gauge choice of Eq.\ref{gauge}. The effect of $t_{AB}$ term is similar to that of mean field Hamiltonian $H_V = - V\sum_{\langle ij \rangle} \chi_{ji} f^\dagger_i f_j$ when $\chi_{Q,L}\neq0$ for $Q=L=0$, namely, a completely uniform distribution of exciton order on every bond. To match the discussion of Sec.~\ref{appendix:exciton order parameter}, we gauge the electron and hole pockets back to $\v \Gamma_M$ by a gauge transformation
\begin{align}
    &f_{A,i}\rightarrow f_{A,i}e^{-i \v K\cdot \v r_j} \notag\\
    &f_{B,j}\rightarrow f_{B,j}e^{-i \v K\cdot \v r_j}
    \label{gt}
\end{align}
under which the $t_{AB}$ term becomes
\be
    t_{AB} f^\dagger_{A,i} f_{B,j} \rightarrow t_{AB} e^{i\v K \cdot \v (\v r_j-\v r_i)} f^\dagger_{A,i} f_{B,j} = \tilde{t}_{AB} f^\dagger_{A,i} f_{B,j}
\ee
Now the new inter-layer tunneling $\tilde t_{AB}$ transforms non-trivially under $C_3$ rotations. At the mean-field level for $f$, this $\tilde{t}_{AB}$ is equivalent to the $p+ip$ exciton order in the ECI phase.  So the charge part of the bCI and our ECI has the same mean-field ansatz; thus, both have $C=1$. When increasing the $V$ term,  exciton order $\chi_{ij}$ decoupled from the interaction dominates over the $t_{AB}$ term. At low density, $s$ wave exciton condensation is favored. Hence the effective hopping now has both the $p+ip$ component from the external $t_{AB}$ term and $s$ component from exciton order. Consequently, we expect an NEI with $s+p$ exciton order. This NEI has $120^\circ$ magnetic order in both layers with the same chirality and no $\tau_z$ polarization. This is demonstrated in Fig.\ref{app_CI2ECI}(c). The strongest bond connects the two AB sites with parallel valley $\tau_{xy}$. So this NEI is connected to the $s$ wave AF-EI phase we found earlier at $t_{AB}=0$.  For the parameter $\Phi_A=2\pi, \Phi_B=0$ here, the ECI phase is quite fragile as shown in the previous section.  But if we choose a different flux $\Phi$, the bCI phase can be driven into the ECI phase by increasing $V$.  Note although both the bCI and the ECI have the same $p+ip$ order in the mean-field level of the electron and holon $f$, their magnetic orders are different and there must still be a phase transition between them.

\begin{figure}[!ht]
    \centering
    \includegraphics[width=0.9\linewidth]{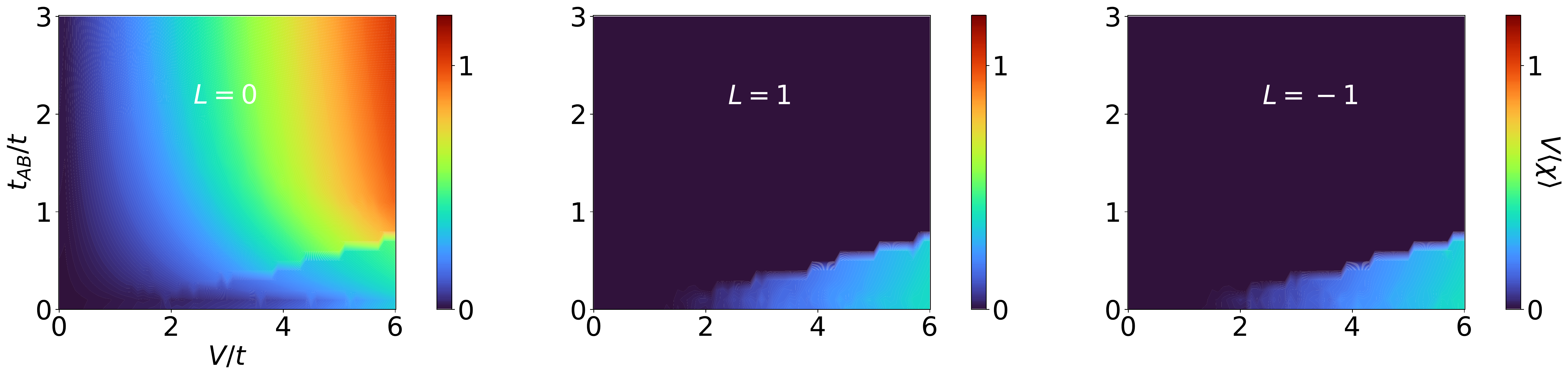}
    \caption{Order parameter in each angular momentum channel. This is the result before gauging. After the gauge transform (see text), the angular momentum quantum number $L$ gets shifted by $1$.}
    \label{app_CI2ECI_chiL}
\end{figure}

The bCI phase by Ref.~\onlinecite{devakul2022quantum} and our ECI phase have the same origin of the non-zero Chern number. But there are some essential differences: (1) The effective hopping in the ECI phase is spontaneously generated from inter-layer Coulomb interaction while the hopping of the bCI phase inherits from single electron hopping $t_{AB}$. (2) The $p\pm ip$ exciton order of the ECI phase does not rely on the $120^\circ$ magnetic order in the B layer. Actually, in our ECI  phase,  both layers are fully valley polarized.  On the other hand, $120^\circ$ order is essential for the bCI phase.  (3) The bCI picture is appropriate only when the inter-layer interaction $V$ is small. In the real system the inter-layer tunneling $t_{AB}$ is small, but the repulsion $V$ is quite large. So we should be in the limit $t_{AB}\ll t_A\ll V$. Thus the effective hopping of the charge $f$ is dominated by the exciton order $V \chi_{ij}$, which could be $s$ wave or $p\pm ip$ depending on parameters. In the small $t_{AB}$, large $V$ regime appropriate to realistic systems, the bCI is unstable, and a Chern insulator phase, if exists, should originate from exciton order as in our ECI phase.

\section{Excitonic topological insulator at $\nu_T=2$}

 Let us also have some discussions for the total filling $\nu_T=n_A+n_B=2$. When $D$ is large, the ground state should be a band insulator with two electrons per triangular site on the B layer while the A layer is empty. Again electrons are transferred to the A layer when $D$ is decreased. Experimentally a topological insulator with helical edge modes is observed at $\nu_T=2$\cite{li2021quantum} for intermediate values of $D$. Presumably,  two Chern bands with $C=\pm 1$ are occupied. Each Chern band can be described by a mean-field theory similar to our discussion at $\nu_T=1$ and can still arise from the excitonic mechanism. Both intra-valley and inter-valley exciton condensation are possible, leading to either a quantum valley Hall (QVH) insulator or a quantum spin Hall (QSH) insulator QVH insulator can crossover to a band insulator from pure single particle physics. QSH insulator has additional inter-valley-coherent (IVC) order and maybe a parent state of superconductivity upon doping\cite{po2018origin,kozii2020superconductivity,chatterjee2021inter}.  We hope future experiments can distinguish these two different states.

\end{document}